%% file: main.tex
\documentclass[apj,twocolumn]{openjournal}
\usepackage{amsmath}
\usepackage{booktabs}
\usepackage{placeins}
\input{commands}
  \graphicspath{{AllResultsSeptember/AllFiguresSeptember/}}
  \makeatletter
  \def\input@path{{AllResultsSeptember/AllTablesSeptember/}}
  \makeatother
\input{AllResultsSeptember/AllTablesSeptember/CommandList_September}
\usepackage{float}
\defcitealias{2021ApJ...910...18C}{C21}
\defcitealias{2023ApJ...953....1C}{C23}
\defcitealias{2023AJ....166...76S}{S23}
\newcommand{\booV}{Bo\"otes~V}
\newcommand{\lmi}{Leo~Minor~I}
\newcommand{\dtwo}{DELVE~2}
\usepackage{rotating}
\usepackage{booktabs}

\usepackage[dvipsnames]{xcolor} 
\usepackage[breaklinks,colorlinks,citecolor=blue,urlcolor=blue,linkcolor=blue]{hyperref}

\usepackage{orcidlink}
\usepackage{subfigure}

\renewcommand{\arraystretch}{1.2}  
\usepackage{listings}
\usepackage{color}
\definecolor{dkgreen}{rgb}{0,0.6,0}
\definecolor{gray}{rgb}{0.5,0.5,0.5}
\definecolor{mauve}{rgb}{0.58,0,0.82}
\definecolor{golden}{rgb}{0.86,0.65,0.01}
\usepackage{soul}
\usepackage{amsmath}
\usepackage{xspace}
\usepackage{graphicx}
\usepackage{enumitem}
\usepackage{amssymb}
\usepackage{xifthen}
\usepackage{hyperref}
\usepackage[normalem]{ulem}
\usepackage{multirow}

\begin{document}

\title{Spectroscopic Confirmation of the Milky Way Satellites Bo\"otes~V and Leo Minor I \\ and the LMC Satellite DELVE 2: Three Small Ultra-Faint Dwarf Galaxies}

\author{W.~Cerny\orcidlink{0000-0003-1697-7062}$^{1}$}
\author{M.~Geha\orcidlink{0000-0002-7007-9725}$^{1}$}
\author{J.~D.~Simon\orcidlink{0000-0002-4733-4994}$^{2}$}
\author{T.~S.~Li\orcidlink{0000-0002-9110-6163}$^{3}$}
\author{A.~B.~Pace\orcidlink{0000-0002-6021-8760}$^{4}$}
\author{R.~P.~van~der~Marel\orcidlink{0000-0001-7827-7825}$^{5,6}$}
\author{N.~Kallivayalil\orcidlink{0000-0002-3204-1742}$^{4}$}
\author{A.~P.~Ji\orcidlink{0000-0002-4863-8842}$^{7,9}$}
\author{A.~Drlica-Wagner\orcidlink{0000-0001-8251-933X}$^{7,8,9,10}$}
\author{D.~Crnojevi\'c\orcidlink{0000-0002-1763-4128}$^{11}$}
\author{G.~Limberg\orcidlink{0000-0002-9269-8287}$^{7}$}
\author{N.~E.~D.~Noël\orcidlink{0000-0002-8282-469X}$^{12}$}
\author{A.~H.~Riley\orcidlink{0000-0001-5805-5766}$^{13}$}
\author{D.~J.~Sand\orcidlink{0000-0003-4102-380X}$^{14}$}
\author{G.~S.~Stringfellow\orcidlink{0000-0003-1479-3059}$^{15}$}
\author{Y.~Choi\orcidlink{0000-0003-1680-1884}$^{16}$}
\author{P.~S.~Ferguson\orcidlink{0000-0001-6957-1627}$^{17}$}
\author{C.~E.~Mart\'inez-V\'azquez\orcidlink{0000-0002-9144-7726}$^{18}$}
\author{P.~Massana\orcidlink{0000-0002-8093-7471}$^{19}$}
\author{G.~E.~Medina\orcidlink{0000-0003-0105-9576}$^{3,20}$}
\author{B.~Mutlu-Pakdil\orcidlink{0000-0001-9649-4815}$^{21}$}
\author{J.~D.~Sakowska\orcidlink{0000-0002-1594-1466}$^{22}$}
\author{C.~Y.~Tan\orcidlink{0000-0003-0478-0473}$^{7,23}$}
\author{A.~K.~Vivas\orcidlink{0000-0003-4341-6172}$^{24}$}
\author{J.~Bruce\orcidlink{0009-0000-9203-1653}$^{25}$}
\author{(DELVE Collaboration)}

\affiliation{$^{1}$Department of Astronomy, Yale University, New Haven, CT 06520, USA; \textcolor{blue}{williamcerny@yale.edu}}
\affiliation{$^{2}$Observatories of the Carnegie Institution for Science, 813 Santa Barbara St., Pasadena, CA 91101, USA}
\affiliation{$^{3}$Department of Astronomy and Astrophysics, University of Toronto, 50 St. George Street, Toronto, ON, M5S 3H4, Canada}
\affiliation{$^{4}$Department of Astronomy, University of Virginia, 530 McCormick Road, Charlottesville, VA 22904, USA}
\affiliation{$^{5}$Space Telescope Science Institute, 3700 San Martin Drive, Baltimore, MD 21218, USA}
\affiliation{$^{6}$Center for Astrophysical Sciences, The William H.\ Miller III Department of Physics \& Astronomy, Johns Hopkins University, Baltimore, MD 21218, USA}
\affiliation{$^{7}$Kavli Institute for Cosmological Physics, University of Chicago, Chicago, IL 60637, USA}
\affiliation{$^{8}$Fermi National Accelerator Laboratory, P.O.\ Box 500, Batavia, IL 60510, USA}
\affiliation{$^{9}$Department of Astronomy and Astrophysics, University of Chicago, Chicago, IL 60637, USA}
\affiliation{$^{10}$NSF-Simons AI Institute for the Sky (SkAI), 172 E.\ Chestnut St., Chicago, IL 60611, USA}
\affiliation{$^{11}$Department of Physics \& Astronomy, University of Tampa, 401 West Kennedy Boulevard, Tampa, FL 33606, USA}
\affiliation{$^{12}$Department of Physics, University of Surrey, Guildford GU2 7XH, UK}
\affiliation{$^{13}$Lund Observatory, Division of Astrophysics, Department of Physics, Lund University, SE-221 00 Lund, Sweden}
\affiliation{$^{14}$Department of Astronomy/Steward Observatory, 933 North Cherry Avenue, Room N204, Tucson, AZ 85721-0065, USA}
\affiliation{$^{15}$University of Colorado Boulder, Boulder, CO 80309, USA}
\affiliation{$^{16}$NSF National Optical-Infrared Astronomy Research Laboratory, 950 North Cherry Avenue, Tucson, AZ 85719, USA}
\affiliation{$^{17}$DiRAC Institute, Department of Astronomy, University of Washington, 3910 15th Ave NE, Seattle, WA 98195, USA}
\affiliation{$^{18}$NSF NOIRLab, 670 N. A'ohoku Place, Hilo, Hawai'i 96720, USA}
\affiliation{$^{19}$NSF NOIRLab, Casilla 603, La Serena, Chile}
\affiliation{$^{20}$Dunlap Institute for Astronomy \& Astrophysics, University of Toronto, 50 St. George Street, Toronto, ON M5S 3H4, Canada}
\affiliation{$^{21}$Department of Physics and Astronomy, Dartmouth College, Hanover, NH 03755, USA}
\affiliation{$^{22}$Instituto de Astrofísica de Andalucía (IAA-CSIC), Glorieta de la Astronom\'ia s/n, E-18008 Granada, Spain}
\affiliation{$^{23}$Department of Physics, University of Chicago, Chicago, IL 60637, USA}
\affiliation{$^{24}$Cerro Tololo Inter-American Observatory/NSF NOIRLab, Casilla 603, La Serena, Chile}
\affiliation{$^{25}$Department of Astronomy, Indiana University, Swain West, 727 E. 3rd Street, Bloomington, IN 47405, USA}
\shortauthors{Cerny et al. (DELVE Collaboration)}

\shorttitle{Spectroscopic Confirmation of the Ultra-Faint Dwarf Galaxies \booV{}, \lmi{}, and \dtwo{}}

\begin{abstract}
We present the first multi-object spectroscopy of \booV{}, \lmi{}, and \dtwo{} -- three small ($20 \rm \ pc$~$\leq  r_{1/2} \leq  26$~pc), ultra-faint $(-3.2 \leq M_V \leq  -2.1$) stellar systems in the outer Galactic halo ($\rm 71 \ kpc \leq$ $D_\odot \leq  102 \rm \ kpc$) with heretofore ambiguous classifications.  Leveraging at least two epochs of Keck/DEIMOS or Magellan/IMACS observations targeting each satellite, we identify \BooVNmemclear{}, \LeoMinorINmemclear, and \DELVEtwoNmemclear{}  spectroscopic members in the three systems, respectively.  We tentatively resolve a \DELVEtwovdisp{} velocity dispersion for \dtwo{} but place upper limits of \BooVvdisp{} and \LeoMinorIvdisp{} for \booV{} and \lmi{}. We find that all three systems are very metal-poor, with mean iron abundances $-2.85 \rm \leq\overline{[Fe/H]}  \leq -2.15$. Both \booV{} and \lmi{} display unambiguous metallicity spreads (\BooVfehdisp{} and \LeoMinorIfehdisp{}), while \dtwo{} displays a modestly significant spread (\DELVEtwofehdisp{}) across three stars. Based primarily on these metallicity measurements, we confidently classify \booV{} and \lmi{} as dwarf galaxies and determine that \dtwo{} is most likely a dwarf galaxy as well. We explore the orbit of each satellite in a joint Milky Way + LMC potential, confirming for the first time that \dtwo{} is a long-term bound satellite of the LMC. Finally, we review the empirical Milky Way satellite classification landscape and find that nearly every spectroscopically characterized satellite with azimuthally averaged half-light radius $r_{1/2} > 15$~pc and surface brightness $\mu_{V,1/2} > 26.5 \rm \ mag \ arcsec^{-2}$ is likely a dwarf galaxy. This result can guide initial probabilistic classifications for satellites discovered by \textit{Rubin}, \textit{Roman}, and \textit{Euclid}. \newline
\end{abstract}


\section{Introduction}
\label{sec:intro}
The Milky Way (MW) is predicted to host a population of $\sim$120--600 satellite galaxies within its virial radius \citep{2008ApJ...688..277T,2009AJ....137..450W,2009ApJ...696.2179K,2014ApJ...795L..13H, 2018MNRAS.479.2853N,2020ApJ...893...48N,2022MNRAS.516.3944M,2024MNRAS.529.3387A}, the vast majority of which are expected to be \textit{ultra-faint dwarfs}: ancient, very metal-poor, and highly dark-matter-dominated galaxies devoid of neutral gas at the present day \citep[e.g.,][and references therein]{2019ARA&A..57..375S, 2025arXiv250206948D,2026arXiv260210202G}. By virtue of their minimal baryonic contents, these extreme galaxies are valuable laboratories for testing the microphysics of dark matter and its distribution on the smallest spatial scales. In particular, the luminosity function and radial distribution of MW satellite galaxies provide strong constraints on models which predict a suppression in the matter power spectrum \citep[e.g.,][]{2010MNRAS.404L..16M,2019ApJ...878L..32N,2021JCAP...08..062N,2022PhRvD.106l3026D,2026ApJ..1000...88L}. In addition, the internal stellar kinematics of these resolved systems can be used to test dark matter models which predict different velocity dispersions and density profiles for the least-massive galaxies \citep[e.g.,][]{2016ApJ...824L..31B,2017PhRvL.119d1102K,2020ApJ...893...21S,2021ApJ...912L...3H,2024JCAP...05..043T,2025arXiv251204156D,2026arXiv260304508F}. 

\par
\begin{figure}
    \centering
    \includegraphics[width=0.5\textwidth]{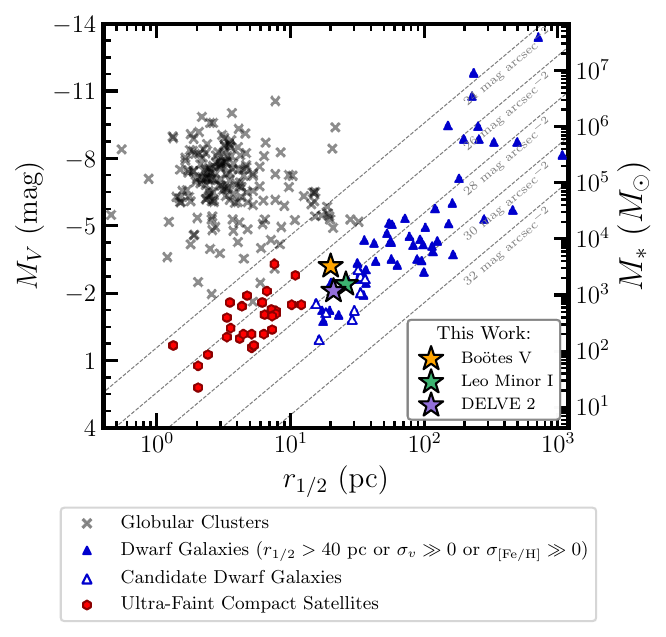}
    \caption{Absolute magnitude ($M_V$) vs. azimuthally averaged half-light radius ($r_{1/2}$) for three populations of MW satellites: globular clusters (black $\times$s), dwarf galaxies (blue triangles), and ultra-faint compact satellites (red hexagons). \booV{}, \lmi{}, and \dtwo{} are shown as orange, green, and purple stars, lying at intermediate physical sizes.  See Appendix \ref{sec:datareferences} for details on the underlying references represented in this figure.}
    \label{fig:sizeluminosity}
\end{figure}

 As wide-field sky surveys have improved in sensitivity, dedicated matched-filter searches have pushed the frontier of MW satellite discovery to previously unimaginable stellar masses and physical sizes \citep[e.g.,][]{2007ApJ...654..897B,2015ApJ...813..109D,2016ApJ...832...21H,2016ApJ...820..119K,2023ApJ...953....1C,2024ApJ...961...92S,2026ApJ..1000...46T}. A recurring challenge in characterizing these boundary-pushing MW satellites -- and thereby unlocking the range of constraints that these systems can provide -- has been separating true dark-matter-dominated galaxies from baryon-dominated star clusters \citep[e.g.,][]{2011AJ....142..128W,2016MNRAS.458L..59M,2016MNRAS.460.3384V,2019MNRAS.490.1498L,2022MNRAS.510.3531B,2024ApJ...976..256S,2026arXiv260217652C}. This is because, at the faint end ($M_V \gtrsim -4$, or equivalently, $L_V \lesssim 3500 \rm \, L_\odot$), ultra-faint dwarfs lying on the empirical mass-size trend exhibit typical half-light radii $ r_{1/2} \lesssim 30 \rm \, pc$ -- comparable to, or smaller than, the sizes of the most extended MW globular clusters (see e.g., the compilations of \citealt{2012AJ....144....4M,2018ApJ...860...66M,2025OJAp....8E.142P} and \figref{sizeluminosity}). Because of this overlap, the photometric datasets  used to discover these systems are often insufficient to determine their physical classifications.  Instead, detailed spectroscopy of the resolved stars in these systems is typically required to establish their true natures.
\par Multi-object spectroscopy offers two complementary probes of an ultra-faint satellite's dark matter content and, by extension, its nature as a star cluster or dwarf galaxy. The first and most direct probe relies on stellar kinematics:  the line-of-sight velocity dispersion across member stars can be used to infer a satellite's dynamical mass under the assumption of virial equilibrium and thereby estimate a mass-to-light ratio that directly quantifies the amount of dark matter present (see \citealt{2022NatAs...6..659B} for a review). Past observational work examining the MW's ultra-faint satellite galaxies has revealed that they exhibit high mass-to-light ratios of $\mathcal{O}(100$--$1000)$~\MLunit{} within their half-light radii \citep[e.g.,][]{2007ApJ...670..313S,2007MNRAS.380..281M}. This contrasts sharply with globular clusters, which typically exhibit $M/L_V \approx 1$--3~\MLunit{} \citep[e.g.,][]{1991A&A...252...94M,1993ASPC...50..357P,2000ApJ...539..618M,2007A&A...469..147R,2020PASA...37...46B}.  
\par While the dynamical identification of a satellite's dark matter halo is the gold standard for classification, resolving the velocity dispersions of the faintest dwarf galaxies -- expected to be $\sim$1--3~\kms{} \citep[e.g.,][]{2024PhRvD.110l3013E,2025arXiv251115808A} -- has proven challenging due to the velocity precision floors limiting most current multi-object spectrographs ($\sim$0.6--2~\kms; \citealt{2021ApJ...920...92J,2023ApJS..268...19W,2026ApJ...999..140G}). This challenge has motivated the use of stellar metallicity dispersions as a second probe of satellite classifications (see e.g., \citealt{2007MNRAS.380..281M,2011AJ....142..128W,2013ApJ...770...16K,2023ApJ...958..167F, 2024ApJ...968...21H, 2025ApJ...979..164C} for illustrative examples).  This probe relies on the fact that the host dark matter halos of faint galaxies provide potential wells deep enough to enable the retention of supernova ejecta, allowing self-enrichment across generations of star formation \citep{2012AJ....144..183L,2012AJ....144...76W} and thereby contrasting them against baryon-dominated star clusters. The discriminating power of this diagnostic is empirically supported by the observation that nearly all MW satellite galaxies exhibit substantial metallicity dispersions  of $\sigma_{\rm [Fe/H]} \approx 0.2$--$0.6$~dex (see e.g., \citealt{2023ApJ...958..167F,2023ApJS..268...19W,2026arXiv260210202G}), in contrast to globular clusters which exhibit typical spreads of $\sigma_{\rm [Fe/H]} \lesssim 0.05$~dex \citep[e.g.,][]{2009A&A...508..695C,2019ApJS..245....5B}.

In this work, we present new multi-object spectroscopic observations to study the stellar kinematics and metallicity distributions of \booV{}, \lmi{}, and \dtwo{} -- three MW satellites discovered with the second data release of the DECam Local Volume Exploration Survey (DELVE; \citealt{2022ApJS..261...38D})  by \citet{2021ApJ...910...18C,2023ApJ...953....1C} (hereafter referred to as \citetalias{2021ApJ...910...18C} and \citetalias{2023ApJ...953....1C}). \booV{} was also contemporaneously discovered by \citealt{2023AJ....166...76S} (hereafter \citetalias{2023AJ....166...76S}) in the Ultraviolet Near Infrared Optical Northern Survey (UNIONS; \citealt{2025AJ....170..324G}), who presented a medium-resolution longslit spectrum of the system's brightest star. These three satellites are exemplars of the faint stellar systems at  $r_{1/2} \approx 20$--30~pc for which classification based on size and luminosity alone has historically proven challenging. Our goal in this work is to resolve this ambiguity for \booV{}, \lmi{}, and \dtwo{}, and use these results to clarify the broader MW  satellite classification landscape.

\par The remainder of this article is organized as follows. In \secref{data}, we introduce our new observations and subsequently describe our procedures for data reduction and spectroscopic measurements. In \secref{membership}, we combine these measurements with \Gaia astrometry and ground-based photometry to isolate member stars in each satellite. We then use these member samples to probe the satellites'  dynamical masses  (\secref{dynamics}) and metallicity distributions (\secref{chemistries}). In \secref{orbits}, we derive new mean proper motions for each satellite and investigate their orbital histories. We also explore the connection between \dtwo{} and the Large Magellanic Cloud (LMC). In \secref{disco}, we synthesize our measurements to probe whether each system is more likely a dwarf galaxy or a star cluster. We then review the broader classification landscape for ultra-faint satellites at the dawn of several major sky surveys. We summarize our findings in \secref{summary}.

\input{AllResultsSeptember/AllTablesSeptember/T1_ObsTable_September}

\section{Observations, Data Reduction, and Spectroscopic Measurements}
\label{sec:data}
\subsection{Keck Observations of \booV{} and \lmi}
We obtained medium-resolution spectra of stars in the field of \booV{} and \lmi{} with the DEep Imaging Multi-Object Spectrograph (DEIMOS; \citealt{2003SPIE.4841.1657F}) on the Keck II telescope on Maunakea, Hawai`i. Our observations used the instrument's highest-resolution grating (1200G) and OG550 order-blocking filter, yielding $R \approx 6000$ across a wavelength range $\sim$6500--9000 $\rm \AA$ that includes $\rm H\alpha$ at  $6563 \ \rm \AA$ and the Calcium Triplet (CaT) lines at $8498, 8542, 8662~\rm \AA$. Each satellite was observed on two separate runs (\tabref{obstable}): \lmi{} was observed on a two-night run spanning 2022 June 4--5 ($1\times1200$s and $2\times1500$s on the former night, $1\times1800$s on the latter) and a single-night run on 2023 April 23 ($6\times1200$s), while \booV{} was observed on the nights of 2023 April 23 ($7\times1200$s) and 2023 June 11 ($7\times1200$s).  Internal XeNeArKr arcs and  quartz flats were taken at the beginning of each night.\footnote{The raw DEIMOS spectra are available in the Keck Observatory Archive (\url{https://koa.ipac.caltech.edu/}) under programs Y115 and Y167 (PIs: Cerny, Geha).}
\par We observed a distinct slit mask for each satellite on each run, i.e., a total of four masks. These four masks each used slits of width $0.7''$ and minimum lengths $4.5$--$5\arcsec$. Stellar targets for each mask were selected based on the available DECam photometry (for both satellites) as well as the Gemini-North/GMOS-N photometric catalogs from \citetalias{2023ApJ...953....1C} for \booV{} only. Parallaxes and proper motions from \Gaia DR3 \citep{2016A&A...595A...1G,2023A&A...674A...1G} were further used to exclude foreground interlopers, and we particularly prioritized the high-probability proper motion members cataloged in Appendix A of \citetalias{2023ApJ...953....1C}. The second iteration of masks also benefited from our initial membership determinations based on the first run of observations for each satellite.
\subsubsection{DEIMOS Data Reduction, Velocity Measurements, and Equivalent Width Measurements }
\par We reduced the raw 2D DEIMOS spectra using the \texttt{PypeIt} package \citep{2020JOSS....5.2308P} following the procedures described in Section 2.3 of \citet{2026ApJ...999..140G}. \texttt{PypeIt} performs standard reduction operations including flat fielding, bias subtraction, sky subtraction, and extraction of the 1D spectra.  Each run for a given satellite was reduced independently; the two nights of the 2022 June 4--5 run used the same \lmi{} mask and DEIMOS flexure compensation reference and thus were treated as an effective single night of observation.
\par We measured stellar radial velocities with  \texttt{dmost} \citep{2026ApJ...999..140G}, a specialized suite of codes for the analysis of DEIMOS spectra taken with the instrument's 1200G grating. The  \texttt{dmost} package derives velocities by forward-modeling the observed spectra with synthetic stellar templates from the PHOENIX library \citep{2013A&A...553A...6H} as well as a telluric absorption template from \texttt{TelFit} \citep{2014AJ....148...53G}. The latter is used to correct for the wavelength shift induced by the miscentering of stars within their slits. The best-fitting templates are selected through $\chi^2$ minimization, and posterior distributions for the velocity and telluric corrections are derived through a Markov Chain Monte Carlo (MCMC) approach using the \texttt{emcee} \citep{2013PASP..125..306F} sampler. Velocities and initial statistical uncertainties ($\epsilon_{\rm v, MCMC}$) are derived based on the MCMC posteriors, and poor fits are automatically excluded based on the skew and kurtosis of the posteriors. Finally, \texttt{dmost} applies a per-epoch systematic uncertainty floor of $1.1 \rm \km \ s^{-1}$ along with a 1.4$\times$ rescaling of the MCMC statistical errors, i.e., $\epsilon_v = \sqrt{(1.4\epsilon_{\rm v, MCMC})^2 + 1.1^2}$.  If a majority of individual exposures did not yield a velocity, \texttt{dmost} instead coadds first and then measures the velocity following a similar set of procedures but with a modified systematic prescription (see \citealt{2026ApJ...999..140G}).

\par We also determined the Equivalent Widths (EWs) of the three CaT lines from our DEIMOS spectra using updated routines implemented within \texttt{dmost} (to be described more extensively in a future contribution). Each CaT line was fit with a Gaussian-plus-Lorentzian model within a Bayesian framework. Estimates of the summed CaT EW  across the three lines ($\sum \rm EW_{\rm CaT} \equiv EW_{8498} + EW_{8542} + EW_{8662}$) were obtained through evaluation of the analytical integrals, with statistical uncertainties derived by marginalizing across the fit parameter posterior distributions. The final uncertainty was obtained by applying a scaling-plus-floor systematic prescription similar to that used for the velocities.  This procedure was applied to the coadded 1D spectrum for each source from a given mask, and the mask-level $\rm \sum EW_{CaT}$ measurements were combined with inverse-variance weighting if a source was observed more than once. For this work, we adopted a per-epoch minimum threshold of $S/N > 7 \rm \, pixel^{-1}$ for DEIMOS EW measurements and excluded fits with $\rm \sum EW_{CaT}$ uncertainties $>1 \rm \, \AA$.

\subsection{Magellan Observations of DELVE~2}
We targeted \dtwo{} ($\delta \approx -68\degree$) with the Inamori-Magellan Areal Camera and Spectrograph (IMACS; \citealt{10.1117/12.670573,Dressler_2011}) on the 6.5-m Magellan Baade telescope at Las Campanas Observatory, Chile.  We observed the system on five separate runs split across eight total nights in 2021--2024 (see \tabref{obstable}).  All observations used the same 1200~$\ell \rm \ mm^{-1}$ grating, which provides $R\approx11000$ across a wavelength range $7500$--$9000 \ \rm \AA$. We observed a single slitmask comprised of 62 $0.7'' \times 5''$ slits covering stellar targets selected based on Dark Energy Camera (DECam; \citealt{2015AJ....150..150F})  broadband photometry from DELVE DR2 and \Gaia EDR3 astrometry \citep{2016A&A...595A...1G,2021A&A...649A...1G,2021A&A...649A...2L}. We obtained HeNeArKr arc and quartz flat frames after every 2--3 science exposures. 
\par We encountered poor seeing and/or clouds for the majority of our \dtwo{} runs, resulting in unusably low signal-to-noise for all but the brightest sources. Fortunately, the most recent run (in December 2024) enjoyed much better conditions and thus these observations carry the most weight in our analyses that follow. This variability in conditions explains the significant difference in the number of sources with recovered velocities per run ($N_*$) reported in \tabref{obstable}.
\subsubsection{IMACS Data Reduction, Velocity Measurements, and Equivalent Width Measurements }
\par We reduced our 2D IMACS spectra following steps similar to those described in \citet{2017ApJ...838...11S}. We first mapped slits across the detector plane and performed initial wavelength calibration based on the official Carnegie Observatories System for MultiObject Spectroscopy pipeline (COSMOS; \citealt{Dressler_2011,2017ascl.soft05001O}). Then, an adapted version of the DEEP2 reduction pipeline \citep{2012ascl.soft03003C} originally designed for Keck/DEIMOS was used to perform flat fielding, sky subtraction, and extraction of the 1D spectra. Each run was reduced independently, with all nights of a given run reduced together (see \tabref{obstable}).

\par Stellar velocities were measured from our 1D IMACS spectra from each run using a lightly updated version of the MCMC-based template fitting code first described in \citet{2017ApJ...838....8L}. Each spectrum was fit against three empirical spectral templates: the metal-poor red giant branch (RGB) star HD122563, the relatively more metal-rich RGB star HD26297, and the blue horizontal branch (BHB) star HD161817. The template fit yielding the lowest $\chi^2$ was selected for our nominal velocity fit. The Fraunhofer A-band region of each spectrum was then separately fit against an empirical telluric template, providing a velocity shift that enables us to account for the miscentering of stars in their slits. In both cases, statistical uncertainties were derived based on the fit posteriors after sigma clipping. Finally, we added a $1.0$~\kms{} systematic floor in quadrature based on the repeatability of velocity measurements across nights  \citep{2017ApJ...838...11S}. For stars observed on multiple runs, we allowed this uncertainty component to average down across runs; this diverges from the original treatment by \citet{2017ApJ...838...11S}, but was motivated by the stability of our velocity measurements of \dtwo{}'s brightest star and more closely resembles our DEIMOS procedure. We imposed a minimum $S/N > 2.5 \rm \, pixel^{-1}$ for velocities and further flagged stars with poor velocity fits based on the skew and the kurtosis of the fit posterior (as with DEIMOS).
\par We also derived CaT EWs from our IMACS spectra by fitting Gaussian-plus-Lorentzian profiles using Levenberg-Marquardt minimization with \texttt{mpfit} \citep{2009ASPC..411..251M} following the procedure from \citet{2017ApJ...838....8L}.  We imposed a minimum $S/N > 5 \rm \, pixel^{-1}$ for IMACS EWs (a less stringent threshold than for DEIMOS EWs given IMACS' higher spectral resolution) and visually inspected the EW fits for all member candidates. In a small number of cases where exactly one CaT line yielded a poor fit or was otherwise unusable (e.g., due to a chip gap or contamination by sky lines), we used the 
linear relations from \citet{2024ApJ...961..234H} to predict the EW of that line from the other two. In all cases, we assumed a $0.2 \rm \ \AA$ systematic uncertainty on $\sum \rm EW_{\rm CaT}$ measurements from each epoch \citep{2017ApJ...838....8L}. 

\subsection{Stellar Metallicities}
We derived stellar metallicities for RGB members in each satellite using the recent CaT EW--[Fe/H] calibration from \citet{2026MNRAS.546ag019N} closely following the procedures described in \citet{2026arXiv260217652C}. We adopted the form of the \citet{2026MNRAS.546ag019N} calibration that uses the summed EW across the three CaT lines  ($\sum \rm EW_{\rm CaT}$)  as well as absolute $V$-band magnitudes ($V_0$) as a surface-gravity proxy. We estimated $V_0$ for each star by transforming the DELVE photometry to $V$-band magnitudes using relations from \citet{Abbott_2021} and then subtracting the distance modulus of each satellite (from \tabref{propertiestable}). Finally, we derived posterior distributions for the metallicity of each star by propagating the two inputs through the \citet{2026MNRAS.546ag019N} relation, accounting for the uncertainty contributions from the photometry (assuming a floor of $0.02$ mag), the distance modulus of each satellite, $\sum \rm EW_{\rm CaT}$, and the calibration's coefficients.
\par Our final metallicity estimates were computed based on the median and 16th/84th percentiles of the marginalized [Fe/H] posterior after clipping nonphysically low metallicity samples ($\rm [Fe/H] < -6$). For the DEIMOS metallicities only, we further added a 0.1 dex systematic term in quadrature to the statistical metallicity uncertainties following \citet{2026ApJ...999..140G}. Matching prior analyses, we did not add the same term to the IMACS-derived metallicity uncertainties, but we note that the larger EW systematic uncertainty assumed for the IMACS measurements results in effectively the same $\sim$0.1~dex [Fe/H] uncertainty floor as for DEIMOS.

\section{Stellar Membership}
\label{sec:membership}
To determine which stars in our spectroscopic catalogs are true members of each galaxy, we applied a semi-automated, process-of-elimination procedure incorporating color--magnitude information from DELVE DR2/DR3, astrometric information from \textit{Gaia}, and radial velocity measurements from our new spectroscopy. We began by  filtering our spectroscopic catalog for each satellite in color--magnitude space based on a $\tau = 13.5$~Gyr, $Z=0.0001$ PARSEC isochrone\footnote{PARSEC: PAdova and Trieste Stellar Evolution Code v1.2S \citep{2012MNRAS.427..127B,2014MNRAS.444.2525C,2014MNRAS.445.4287T}.} set to the distance modulus reported for each system (\tabref{propertiestable}). We specifically performed 1D filtering in color assuming a $(g-r)_0$  tolerance of 0.2 mag about the isochrone RGB and MS. For the HB, we instead applied similar filtering in $g_0$ assuming a wider tolerance of 0.4 mag.  We then removed clear foreground stars in the MW disk based on \Gaia DR3 astrometry (where possible), as suggested by resolved parallaxes ($\varpi - 3\sigma_\varpi > 0$) or proper motions of $|\mu| >7~\rm mas \, yr^{-1}$ in either component. 
\par After these initial selections, we iteratively derived estimates of the weighted-mean radial velocity and proper motion of each satellite from stars within $r_{\rm ell} < 2a_{1/2}$ of each satellite's centroid. In each iteration, we excluded stars at any radii that deviated by $>\sqrt{(15\rm \ km \ s^{-1})^2+\epsilon_v^2}$ in velocity or by $\chi^2 > 9$ in proper motion relative to these automatically determined estimates. The velocity tolerance of $\sqrt{15^2+\epsilon_v^2}$ was chosen to be inclusive enough to not reject true members given the expected intrinsic velocity dispersions of faint dwarf galaxies ($\sim$1--5~\kms{}); in particular, simple Monte Carlo simulations suggest $<2\%$ of members would be missed by this velocity selection for all intrinsic dispersions $\leq5$~\kms{} given our typical observational uncertainties of $\sim$1--5~\kms{}.\footnote{See also \citet{GhodsiYengejeh_2026}, who find a 1.5\% miss rate assuming our limiting case of $\sigma_v = 5$~\kms{} even when incorporating stellar binaries with binary fraction $f_b = 0.4$.} These automated velocity and proper motion selections were repeated until the iteration converged. Finally, we  inspected the spatial positions, metallicities, and kinematics of all remaining stars and made any necessary final judgments for edge cases.
\par In the subsections below, we detail the application of this procedure to the three satellites. Our final member samples are presented in Tables \ref{tab:members_Bootes_V}--\ref{tab:supplemental_Bootes_V} (\booV{}), \tabref{members_Leo_Minor_I} (\lmi{}), and \tabref{members_DELVE_2} (\dtwo{}), and extended tables including non-member stars will become available on Zenodo (see \secref{dataavail}). Throughout, we enumerate individual member stars in each galaxy in order of increasing $g$-band magnitude (e.g., \texttt{BooV\_1} is the brightest star in \booV{}, \texttt{BooV\_2} the second brightest, etc.).

\begin{figure*}
    \centering
    \includegraphics[width=0.9\textwidth]{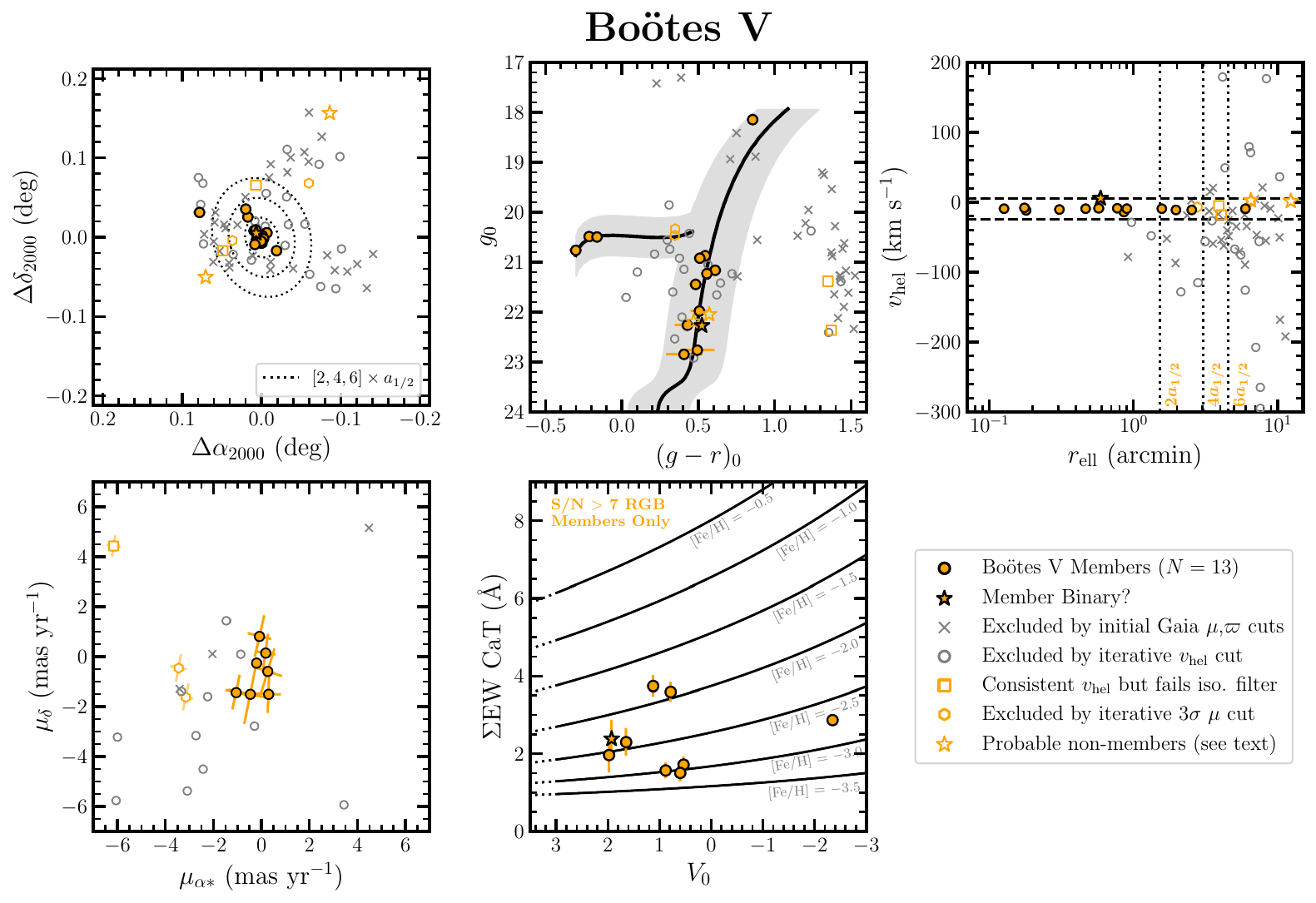}
    \caption{Membership diagnostics for \booV{} based on our Keck/DEIMOS data. (Top Left) Spatial distribution of spectroscopic targets. Contours of [2,4,6]$ \times a_{1/2}$ are plotted as dotted ellipses. (Top Center) Color--magnitude diagram of the same targets. A $\rm \tau = 13.5~Gyr, \rm [Fe/H] = -2.2$ isochrone is shown in black, with our photometric member selection shown in grey shading. (Top Right) Radial velocity vs. projected elliptical radius. (Bottom Left) Proper motion vector-point diagram. (Bottom Center) CaT EW vs. absolute $V$-band magnitude for $S/N > 7 \rm \, pixel^{-1}$ members only. Contours of constant metallicity based on the \citet{2026MNRAS.546ag019N} calibration are shown in black.   In all panels, uncertainties are only shown for stars passing our velocity selection (i.e., those in orange) for visual clarity.  We identify \BooVNmemclear{} clear members as well as one suspected member binary. \label{fig:booVmembership}}
\end{figure*}

\begin{figure*}
\centering
    \includegraphics[width=0.9\textwidth]{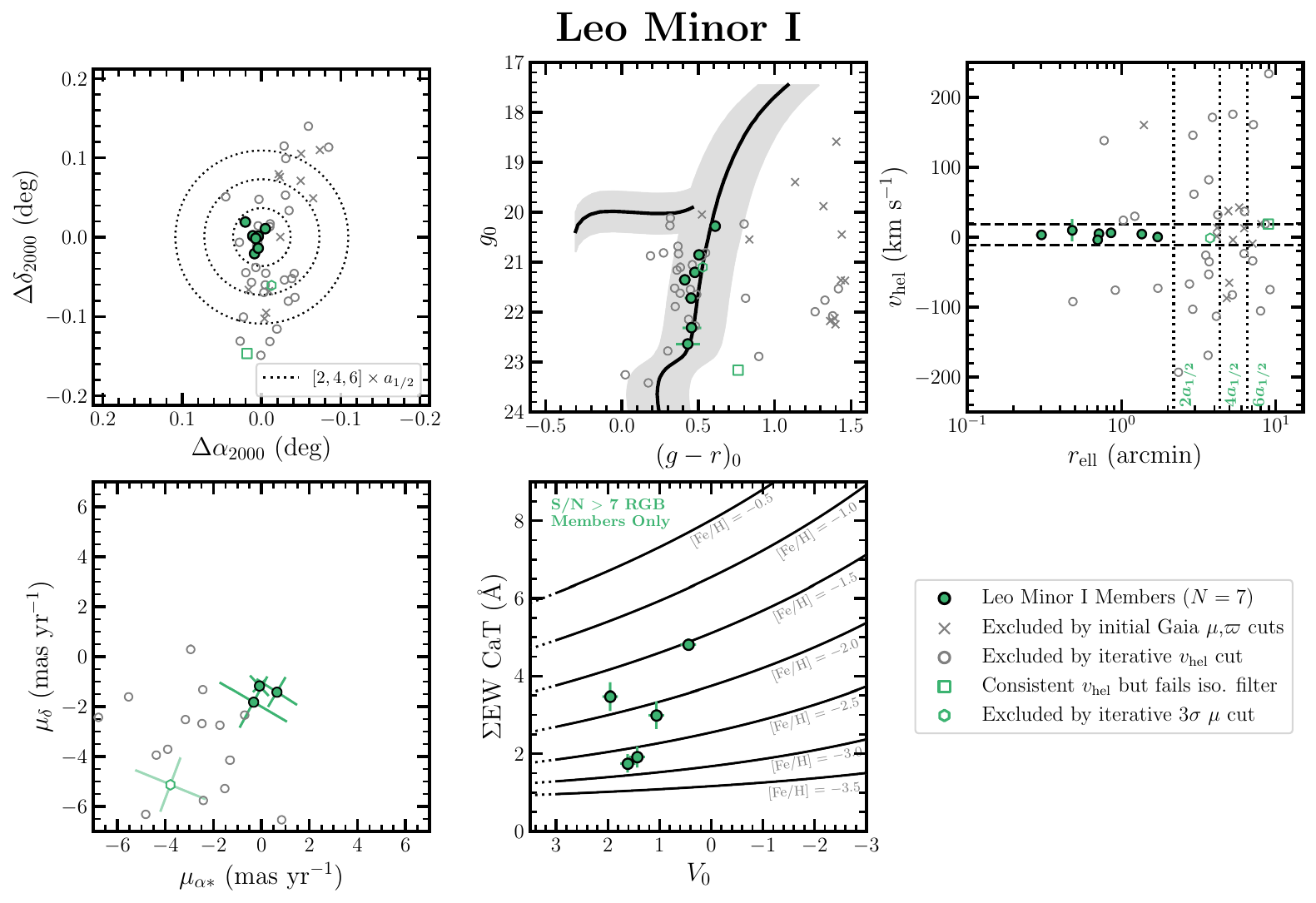}
    \caption{The same as \figref{booVmembership}, except for our Keck/DEIMOS observations of \lmi{}. We identify \LeoMinorINmemclearWord clear members, all falling on the RGB. \label{fig:lmimembership}}
        \includegraphics[width=0.9\textwidth]{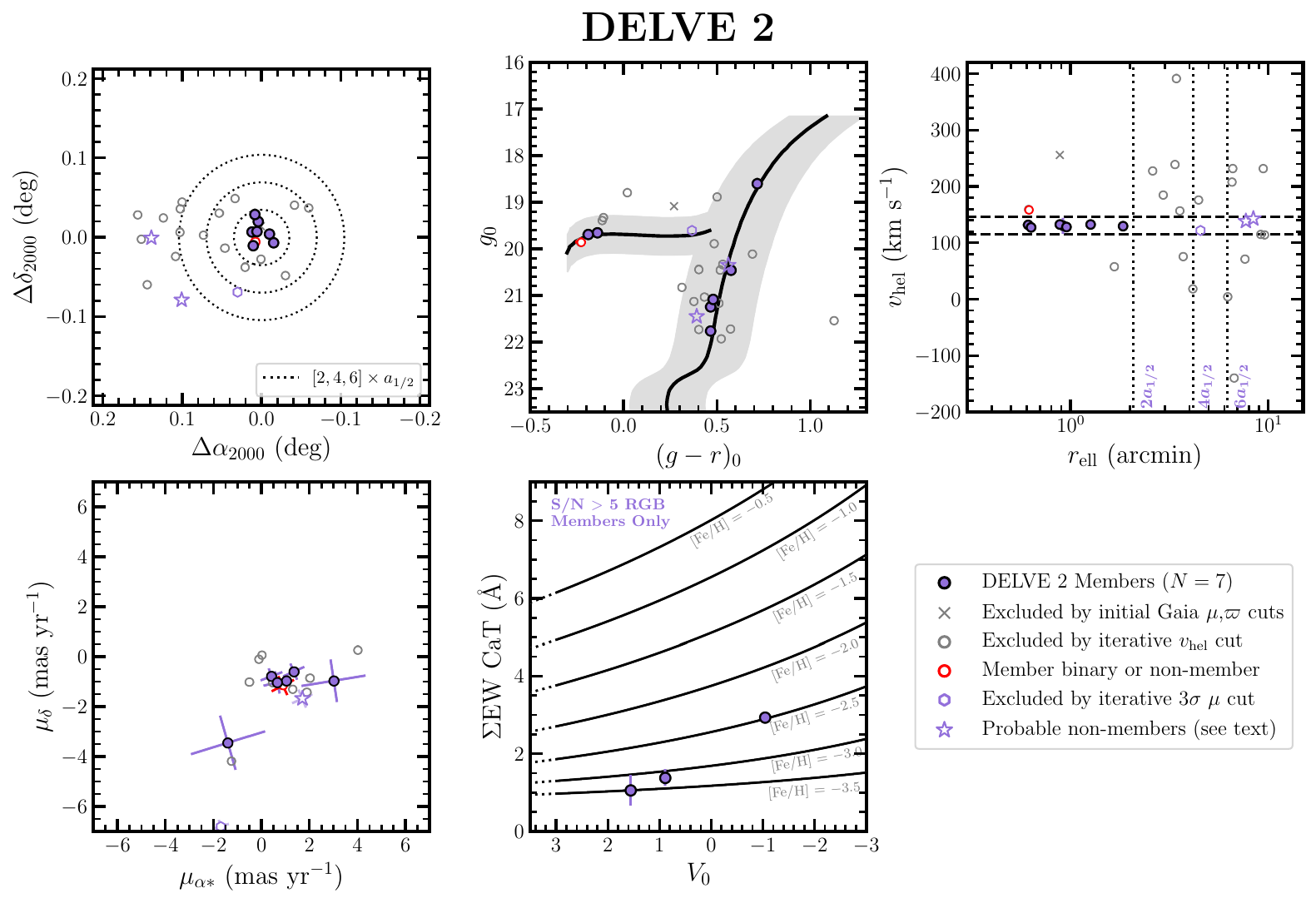}
    \caption{The same as \figref{booVmembership}, but for our Magellan/IMACS observations of \dtwo. We identify \DELVEtwoNmemclearWord clear members and flag one additional star on the horizontal branch with uncertain membership status. }
    \label{fig:delve2membership}
\end{figure*}

\subsection{\booV}
\label{sec:booV}
For \booV{} (\figref{booVmembership}), our initial proper motion and parallax selections reduced the pool of  spectroscopic targets from an initial \BooVNunique{} stars to \BooVNpassesinitial{} stars, of which \BooVNpassesInitialAndIso{} also pass our isochrone selection. Our automated velocity and proper motion loop converged in two iterations to initial mean estimates of $\overline{v_{\rm hel}} = \BooVitermeanvhel{}$ and ($\mu_{\alpha_*}$, $\mu_{\delta}$) = $(\BooVitermeanpmra{}, \BooVitermeanpmdec{})$ based on members within $r_{\rm ell} < 2a_{1/2}$.
Selecting stars within $\pm \sqrt{15^2 + \epsilon_{v}^2}$~\kms{} about this mean velocity and within $3\sigma$ in proper motion subsequently reduced this to a sample of \BooVNpassesbothiterative{} viable spectroscopic members.
\par While \BooVNmemclear{} of these \BooVNpassesbothiterative{} automatically selected candidates are high-confidence members based on their very similar velocities ($-14~\kms{} < v_{\rm hel} < -8~\kms{}$), we identified three edge cases with a lower likelihood of membership: one toward the  center of the system  ($r_{\rm ell}< 2a_{1/2}$) and two in the distant outskirts  (each $r_{\rm ell} > 6a_{1/2}$). The inner candidate's central position, photometric consistency with \booV{}'s red giant branch, and low metallicity ($\rm [Fe/H] = -2.31 \pm 0.30$, assuming it is a giant at the distance of \booV) support its membership, but its velocity, $v_{\rm hel} = 5.6 \pm 5.2$~\kms{}, makes it a clear outlier relative to the confident members. We suspect this star is a member binary and excluded it from our primary velocity dispersion fit in \secref{dynamics}. The remaining two edge cases, on the other hand, display \textit{both} a large spatial separation and a modest velocity offset. Given that the empirical MW foreground distribution is centered at a similar velocity to \booV{} (as seen in \figref{booVmembership}), these stars are more likely to be contaminants than true distant members. We therefore excluded them from our sample, leaving a final total of \BooVNmemclear{} clear members of \booV{} in addition to one candidate binary member pending further confirmation.  

\par \subsection{\lmi}
Relative to \booV{}, determining the stellar membership for \lmi{} (\figref{lmimembership}) was more straightforward and required no subjective decisions. Our initial color--magnitude diagram selection and \textit{Gaia} filtering reduced our starting pool of \LeoMinorINunique{} stars to \LeoMinorINpassesInitialAndIso{}. After two iterations, our automatic member selection converged to initial mean estimates of $(v_{\rm hel}, \mu_{\alpha*}, \mu_\delta) = (\LeoMinorIitermeanvhel, \LeoMinorIitermeanpmra, \LeoMinorIitermeanpmdec)$. Using these properties to set our velocity and proper motion selection windows as described above, we isolated a sample of \LeoMinorINpassesbothiterative{} stars, all of which lie on the RGB. The brightest three have self-consistent \textit{Gaia} proper motions, with a mean signal that is somewhat distinct from the proper motions of the non-member stars at $g\lesssim 21$ observed with DEIMOS.
\par Notably, the brightest star in our sample (\texttt{LMiI\_1}/\textit{Gaia} DR3 731478564536994432) is more metal rich than the vast majority of stars known in ultra-faint dwarfs: \LeoMinorIfehbrightest{}. While this could be taken as evidence for non-membership, this star's central spatial position ($r_{\rm ell} \approx 1.2a_{1/2}$) and proper motion consistency with the next two fainter members strongly support its membership. In addition, the EW of the Mg I 8807 $\rm \AA$ line for this star, $0.23  \pm  0.05 \rm\  \AA$, is consistent with the locus of giants based on the dwarf-giant separation criterion of $\rm EW_{\rm MgI} < 0.3 \rm \ \AA$ from \citet{2012A&A...539A.123B}. We retain this star, keeping our final sample of members at \LeoMinorINmem{} stars.

\pagebreak
\par \subsection{\dtwo} Given IMACS' lower sensitivity and the poor observing conditions we encountered, our spectroscopic observations of \dtwo{} are much shallower compared to those of \booV{} and \lmi{} from DEIMOS. Our spectroscopic catalog contains just \DELVEtwoNunique{} stars with measured velocities, of which \DELVEtwoNpassesInitialAndIso{} pass our isochrone filter and initial astrometric selections. Applying our iterative membership selection loop, we determined an approximate mean motion of $(v_{\rm hel}, \mu_{\alpha*}, \mu_\delta) = (\DELVEtwoitermeanvhel{}, \DELVEtwoitermeanpmra{}, \DELVEtwoitermeanpmdec{})$ and identified an initial sample of \DELVEtwoNpassesbothiterative{} stars. \DELVEtwoNmemclearWordUppercase{} of these \DELVEtwoNpassesbothiterative{}  stars fall within $2a_{1/2}$ and are unambiguous members of \dtwo{}, while the status of the remaining two stars at $>6a_{1/2}$ is more ambiguous.
\par We argue that these latter two distant stars, like the two distant stars in \booV{}, are most likely non-members. The brighter of these stars falls closely on the isochrone, but its proper motion deviates by $\sim$2.9$\sigma$ relative to the mean set by the more-confident inner members -- narrowly within our automated selection, but enough to raise doubts about its membership in combination with its distant spatial position. The other distant star is too faint for \textit{Gaia} but is a color outlier relative to the remaining RGB members; we again suspect it to be a non-member despite it falling within our filters. 
\par Lastly, we highlight one additional uncertain case (\textit{Gaia} DR3 4697408102458426624; shown as an unfilled red circle in \figref{delve2membership}) that is a BHB star located within \dtwo's half-light radius. This star was rejected by our automated membership procedure because its velocity ($v_{\rm hel} = 158.5 \pm 2.9$~\kms{}) is separated by $\sim$28~\kms{} from \dtwo's mean velocity. However, the likelihood of an interloper BHB falling within \dtwo's half-light radius and at the appropriate distance/magnitude and proper motion is small. One explanation for this star is that it is in a binary system, though explaining its significant velocity offset would likely require a short period and/or a highly eccentric orbit. An alternative possibility is that this star is a variable star such as an RR Lyrae, and that the observed velocity offset could be the result of the star's radial pulsations. However, our analysis of  15 epochs of archival DECam  photometric data (split across the $g,r,i,z$ bands) demonstrated evidence for only very small variation (e.g., $< 0.05$ mag in $g$), inconsistent with the RR Lyrae scenario (and possibly arising purely from photometric zeropoint uncertainties). Lacking a clear line of evidence supporting its membership, we continued to exclude this star (as well as the other two distant candidates described above)  from our member sample, keeping our final \dtwo{} member sample at \DELVEtwoNmemclearWord{} stars.

\subsection{Identifying Spectroscopic Binaries}
\label{sec:binaries}
\par Stellar binaries are common in low-metallicity dwarf galaxies \citep[e.g.,][]{2018AJ....156..257S,2019MNRAS.487.2961M,2023A&A...677A..95A,2026ApJ..1000...33Q}, and can strongly bias measurements of stellar velocity dispersions if left unaddressed \citep{2010ApJ...722L.209M,2022ApJ...939....3P,2023ApJ...956...91W,2025MNRAS.543.1120G,2026arXiv260303129A}. To search for unresolved spectroscopic binaries in \booV{}, \lmi{}, and \dtwo{} -- all of which were observed on at least two well-spaced observing runs --  we conducted a $\chi^2$ test for each star in our spectroscopic datasets  in which we evaluated the null hypothesis that the velocity was a constant value equal to the weighted mean velocity across runs, i.e., the velocity reported in the \textit{Combined} row for a given star in our member tables.  
This yielded $p$-values $  p_{\rm const} \equiv \rm 1- CDF(\chi^2, N_{\rm epoch}-1) $; we flagged all stars with  $p_{\rm const} < 0.05$ as likely velocity variables (see  Tables \ref{tab:members_Bootes_V}--\ref{tab:members_DELVE_2}) for our primary analyses.
\par For \booV, these tests identified no binaries below the $p_{\rm const} <0.05$ threshold in our DEIMOS data. We attribute our non-detection of binaries to the fact that just under half of our member sample ($6$ of \BooVNmem) have velocities from each of our two epochs and because the time baseline between these epochs was relatively short ($\sim$49 days). However, we reiterate that one marginal member candidate could be a large-amplitude member binary (see \secref{booV}). For \lmi{}, we identified two binaries based on high-significance variability: \texttt{LMiI\_3}, with $p_{\rm const}=0.011$, and \texttt{LMiI\_4}, with $p_{\rm const}=0.001$. Both of these stars varied by $>10$~\kms{} across our two epochs separated by $\sim$323 days. One additional star, \texttt{LMiI\_6}, varied at the $p_{\rm const} = 0.065$ level across our two epochs, but in this case we cannot clearly reject the null hypothesis of no variability. Lastly, for \dtwo{}, we found no evidence for $p_{\rm const}<0.05$ velocity variation in any of our \DELVEtwoNmemclearWord{} members. However, two stars fell just above this $p_{\rm const}=0.05$ threshold:  \texttt{DEL2\_2} ($p_{\rm const} = 0.089$) and \texttt{DEL2\_4} ($p_{\rm const} = 0.069$).
\par The binarity of these latter three stars with $0.05 < p_{\rm const} < 0.1$ (one in \lmi{} and two in \dtwo{}) remains unclear: we cannot reject the null hypothesis of no velocity variation, but the observed velocity differences in our data are well outside of the $1\sigma$ velocity uncertainties. Given the small spectroscopic sample sizes available, we retained these stars in our kinematic samples but later quantified the impact of excluding them on our kinematic constraints in \secref{vdispsystematics}.

\subsubsection{Additional Measurements for \booV{} Stars}
\par The binarity analysis described above was limited to our own DEIMOS and IMACS measurements. However, three member stars that we identified in \booV{} have pre-existing spectroscopic measurements from the literature: the brightest RGB star (\texttt{BooV\_1}/\Gaia DR3 1478121382994560256) and two of the BHB stars (\texttt{BooV\_2}/\Gaia DR3 1478121722292128768 and \texttt{BooV\_4}/\Gaia DR3 1478121584853174144). Specifically, \texttt{BooV\_1} was observed at medium resolution both by \citetalias{2023AJ....166...76S} using Gemini/GMOS-N and by the  Large Sky Area Multi-Object Fiber Spectroscopic Telescope (LAMOST) survey \citep{2012RAA....12.1197C,2019yCat.5164....0L}. In addition, this star and two BHB members were each observed by the Dark Energy Spectroscopic Instrument (DESI) Milky Way Survey \citep{2023ApJ...947...37C,2026OJAp....955260K}. These efforts provide $1$--3 additional velocity epochs per star, which we summarize in \tabref{supplemental_Bootes_V}. 
\par The addition of these external velocity measurements generally does not provide new evidence of binarity for any stars, though their larger uncertainties (often $\epsilon_v > 10$~\kms{}) make it challenging to rule out small velocity variation relative to our DEIMOS measurements.  A notable exception is the brightest star \texttt{BooV\_1}, for which the inclusion of a precise DESI DR1 radial velocity ($\epsilon_v \approx 3$~\kms{}) diminishes support for the null hypothesis of no variability to $p_{\rm const}=0.055$. This does not meet our adopted threshold for rejecting the null hypothesis of no variation, but the $\sim$8~\kms{} difference relative to our DEIMOS measurements motivates further long-term monitoring. We treat this star similarly to the other $0.05 < p_{\rm const} < 0.1$ cases, presenting our kinematic constraints with and without this star in \secref{vdispsystematics}.
\par In addition to providing velocity measurements, these prior spectroscopic efforts produced independent metallicity estimates for the brightest RGB member (\texttt{BooV\_1}); we summarize these in \tabref{supplemental_Bootes_V} as well. We find good consistency between our own metallicity estimate for \booV{}'s brightest giant (\BooVfehbrightest{}) and the similar CaT estimate from \citetalias{2023AJ....166...76S}, $\rm [Fe/H] = -2.85 \pm 0.10$ (based on the \citealt{2010A&A...513A..34S} calibration). DESI DR1 provides a modestly higher metallicity for this star, $\rm [Fe/H] = -2.5 \pm 0.19$, that is still consistent with our own within the joint 1$\sigma$ uncertainty.

\begin{figure*}
    \centering
    \includegraphics[width=0.3\linewidth]{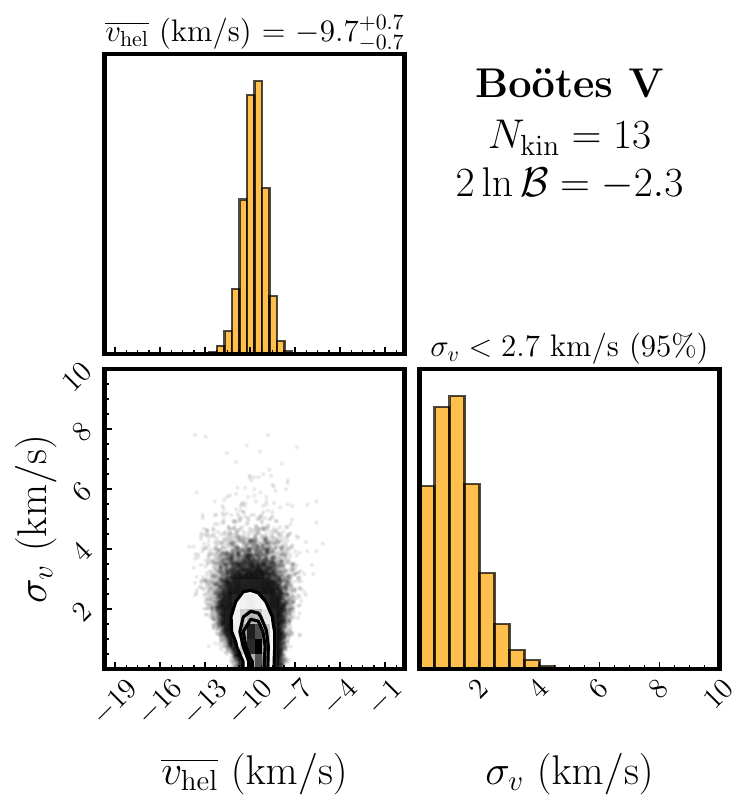}
    \    \includegraphics[width=0.3\linewidth]{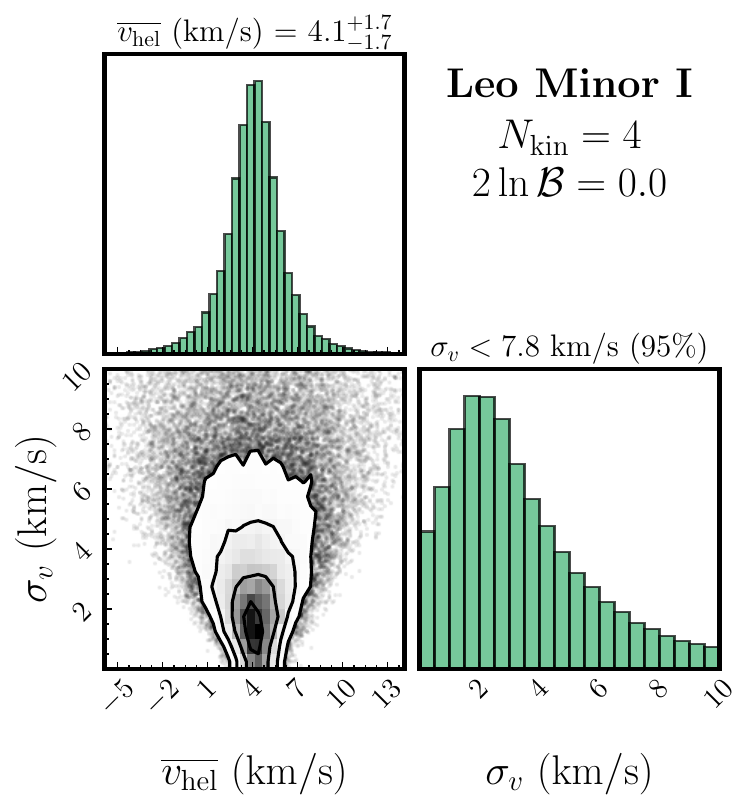}
     \includegraphics[width=0.3\linewidth]{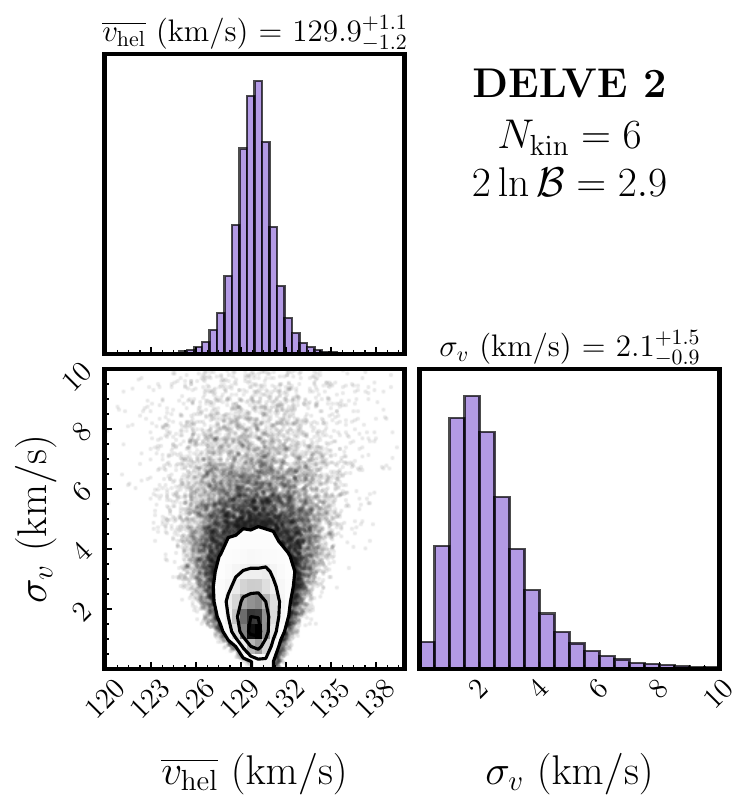}
    \caption{Posterior distributions for the  mean velocities ($\overline{v_{\rm hel}}$) and velocity dispersions ($\sigma_v$) for each satellite from our  \texttt{dynesty} nested sampling fits. For \booV{} (left) and \lmi{} (center), we find marginalized velocity dispersion posteriors with significant posterior mass down to our lower prior bound of $\sigma_v = 0$, as reflected in the Bayes factor comparison against zero-spread models ($2\ln \mathcal{B} \leq 0$); we therefore report upper limits in each case. For \dtwo{} (right), we find a small, modestly significant resolved dispersion with an extended posterior tail to higher values. }
    \label{fig:vdispposteriors}
\end{figure*}

\section{Stellar Kinematics, Dynamical Masses,
and Dark Matter Contents}
\label{sec:dynamics}
\subsection{Velocity Dispersion Fits}
We estimated the mean velocity $\overline{v_{\rm hel}}$ and line-of-sight stellar velocity dispersion $\sigma_v$ of \booV{}, \lmi{}, and \dtwo{} from our spectroscopic member samples through two-parameter Bayesian fits with the identical setup described in \citet{2026arXiv260217652C}. The velocity distribution of each satellite was modeled as a Gaussian distribution with a spread comprised of both an intrinsic dispersion component ($\sigma_v$) and a component associated with velocity measurement errors (see Eq. 8 of \citealt{2006AJ....131.2114W}). We assumed a uniform prior of $[-500, 500]~\kms{}$ on the mean velocity of each satellite and a uniform prior of $[0,10]~\kms{}$ on their velocity dispersions. Posterior probability distributions for the two model parameters  were derived  by performing dynamic nested sampling \citep{2004AIPC..735..395S,10.1214/06-BA127,2019S&C....29..891H} with the \texttt{dynesty} Python package \citep{2020MNRAS.493.3132S,2023zndo...3348367K}.  For our primary fits, we used the velocities and velocity errors for all members listed in Tables \ref{tab:members_Bootes_V}--\ref{tab:members_DELVE_2}, taking the velocity value from the \textit{Combined} rows if multiple velocity measurements for a given star were available. We excluded members with $\epsilon_v > 10~\kms{}$ as well as the small number of suspected binaries identified  in \secref{binaries}. 
\par To evaluate the significance of our dispersions and objectively decide when to quote resolved dispersions vs. upper limits, we computed logarithmic Bayes factors\footnote{Commonly, Bayes factors are reported in the form $\mathcal{B} \equiv e^{\Delta \ln \mathcal{Z}}$. We quote results in the form $2\ln\mathcal{B}$ because this closely resembles frequentist likelihood ratio test statistics (as discussed in \citealt{kassrafferty}). Approximate frequentist $z$-scores can be computed from our estimates according to $z \approx$  $\sqrt{|2\ln\mathcal{B}|}$.} $2\ln\mathcal{B} \equiv 2\left(\ln \mathcal{Z_\sigma} - \ln \mathcal{Z_{\rm \sigma =0}}\right) $ which quantify the evidence for our default dispersion model ($\ln \mathcal{Z_\sigma})$  relative to the evidence for the data given a single-parameter model in which the velocity dispersion was fixed to $\sigma = 0$~\kms{} ($\ln \mathcal{Z_{\rm \sigma =0}}$). Positive logarithmic Bayes factors $2\ln\mathcal{B}> 2$ indicate a clear preference for the model with the intrinsic dispersion left free, i.e., a resolved velocity dispersion, while $2\ln\mathcal{B} < -2$ implies that the simpler one-parameter model with $\sigma=0$~\kms{} is preferred, i.e., an unresolved dispersion. We refer to logarithmic Bayes factors between these regimes as marginally resolved ($0  < 2\ln\mathcal{B} < 2$) or marginally unresolved  ($-2 < 2 \ln\mathcal{B} \leq 0$).
\begin{figure*}
    \centering
    \includegraphics[width=\textwidth]{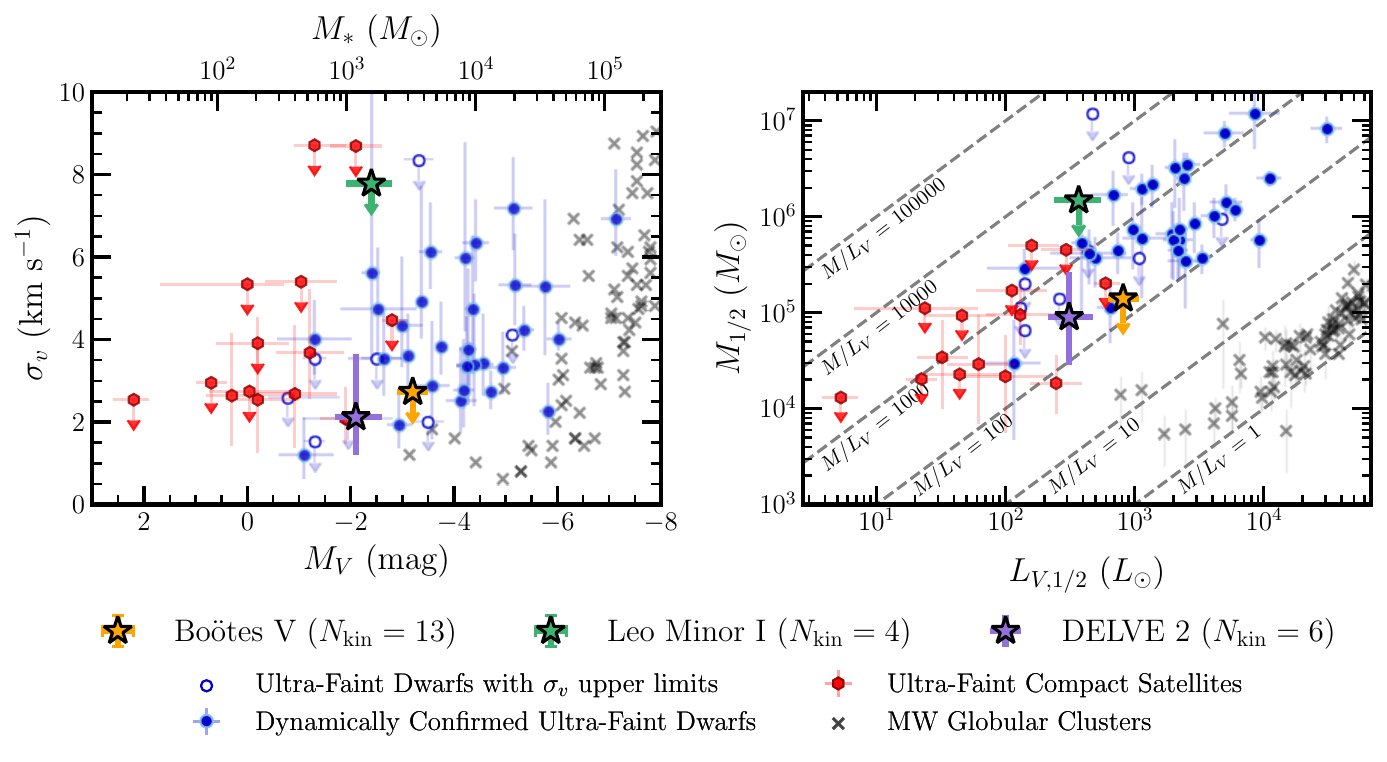}
    \caption{(Left) Velocity dispersion ($\sigma_v$) vs. absolute magnitude ($M_V$) for \booV{}, \lmi{}, and \dtwo{} compared to the populations of ultra-faint MW satellite galaxies (blue circles; filled if dynamically confirmed, and unfilled if upper limits; from the LVDB), globular clusters (black crosses; \citealt{2018MNRAS.478.1520B}), and ultra-faint compact satellites of ambiguous classification (red hexagons; \citealt{2026arXiv260217652C}). (Right) Dynamical mass within the half-light radius ($M_{1/2}$) vs. $V$-band luminosity within the half-light radius ($L_{V,1/2}$). The diagonal dashed lines correspond to mass-to-light ratios spanning $M/L_{V,1/2} = 1$~\MLunit{} to $M/L_{V,1/2} = 10^5$~\MLunit.   In both panels, \booV{}, \lmi{}, and \dtwo{} are broadly consistent with the empirical locus of MW ultra-faint dwarfs. }
    \label{fig:masses}
\end{figure*}

\subsubsection{Primary  Fit Results}
The results of our primary kinematic fits are presented in \tabref{propertiestable} and visualized in \figref{vdispposteriors}. In the latter, we present corner plots of the posterior distributions for both $\overline{v_{\rm hel}}$ and  $\sigma_v$ derived from our fits to each satellite's velocity distribution.  For \booV{} (left panel), fitting our \BooVNkin{} kinematic members resulted in a marginalized velocity dispersion posterior with significant mass down to the prior bound at zero, as also reflected in our Bayes factor estimate (\BooVvdispbayes{}); we therefore report an upper limit of \BooVvdisp{} at the 95\% credible level. For \lmi{} (center panel), our kinematic member sample is too small (\LeoMinorINkin) to be meaningfully constraining on its dispersion, though the mean velocity is well constrained. Our Bayes factor test leaves ambiguous whether the free-dispersion model is preferred or not (\LeoMinorIvdispbayes{}); we choose to report an upper limit in this case: \LeoMinorIvdisp{}. Finally, for \dtwo{} (right panel; \DELVEtwoNkin), we resolve a small dispersion of \DELVEtwovdisp{} that is modestly statistically significant (\DELVEtwovdispbayes{}). 

\par In the left panel of \figref{masses}, we compare our results to the observed dispersions of the MW's ultra-faint dwarf galaxies as well as the population of ultra-faint compact satellites ($r_{1/2} < 15$~pc; $M_V > -3.5$) with generally more ambiguous classifications. We find that both our measurement for \dtwo{} and upper limit for \booV{} fall near the lower envelope of the observed MW satellite galaxy population in the $\sigma_v$--$M_V$ plane, broadly consistent with the (noisy) empirical trend that fainter MW satellites have lower observed velocity dispersions \citep[e.g.,][]{2019ARA&A..57..375S,2022NatAs...6..659B,2026arXiv260210202G}. Our upper limit for \booV{} is notably among the tightest upper limits yet measured for an ultra-faint MW satellite and excludes the dispersions observed in $>80\%$ of the confirmed MW ultra-faint dwarfs; however, there are at least two faint dwarfs with resolved dispersions below this limit (Tucana V and Bo\"otes~II; \citealt{2024ApJ...968...21H,2026ApJ...999..140G}) and thus our limit alone does not make it an outlier. Finally, \lmi{} joins a small number of faint  MW satellites with particularly weak velocity dispersion upper limits owing to the very small member samples ($N_{\rm kin} \lesssim 5$). It is not possible to extract physical insight into these satellites' internal kinematics beyond that they host identifiably co-moving member stars.  

\subsubsection{Velocity Dispersion Systematics}
\label{sec:vdispsystematics}
Determining robust estimates of the velocity dispersions of the faintest galaxies is challenging. In the small-$N$, low-$\sigma_v$ regime, fit hyperparameters such as the stellar membership, binary rejection threshold, and the assumed Bayesian priors can have an outsized effect that exceeds formal statistical uncertainties. While fully addressing these limitations will require larger member samples and more extensive velocity monitoring,  we explored variations in these hyperparameters to contextualize our primary results. 

\par Beginning with stellar membership, we assessed the impact of re-including the most distant stars that we conservatively excluded from our primary sample for \booV{} and \dtwo{} in \secref{membership}. For \booV{}, adding back the two stars at $>8a_{1/2}$ into our kinematic sample resulted in a resolved dispersion of \BooVvdispwithDistantStars{}. This credible interval overlaps with that from our primary fit (95\% credible upper limit of \BooVvdisp{})  but is significantly inflated by the inclusion of one star that is offset by $\sim$4.5$\sigma$ from \booV{}'s mean velocity. Similarly, adding the two distant stars to our \dtwo{} kinematic sample inflated the dispersion to \DELVEtwovdispwithDistantStars{} -- larger than nearly all ultra-faint dwarfs at its luminosity (see \figref{masses}). Arguably, these results suggest that the distant stars are more likely non-members, as the simplest explanation for a larger dispersion in these outer-halo satellites' outskirts is contamination. Indeed, if we alternatively limit our  \booV{} kinematic sample to a \textit{stricter} radial selection of $r_{\rm ell} < 2a_{1/2}$, we find an upper limit of  \BooVvdispinnerOnlyStars; all confident \dtwo{} members fall within $r_{\rm ell} < 2a_{1/2}$. The fact that the dispersion for both of these satellites qualitatively changes only when the two most distant stars are included -- and not under conservative radial sample selections -- reiterates that these distant stars are outliers and are best excluded.
\input{AllResultsSeptember/AllTablesSeptember/T2_PropertiesTable_September}

\par We next considered the impact of raising our binary rejection threshold from $p_{\rm const} = 0.05$ to a more conservative $p_{\rm const} =0.1$ in light of the several stars at $0.05 < p_{\rm const} < 0.1$ identified in \secref{binaries}. In practice, this excluded one additional star in each of \booV{} and \lmi{} and two stars in \dtwo{}. Rerunning fits with these reduced samples resulted in new constraints of \BooVvdispstricterBinary{}, \LeoMinorIvdispstricterBinary{}, \DELVEtwovdispstricterBinary{} for \booV{}, \lmi{}, and \dtwo{}, with corresponding logarithmic Bayes factors of \BooVvdispbayesstricterBinary{}, \LeoMinorIvdispbayesstricterBinary{}, and \DELVEtwovdispbayesstricterBinary{}. These estimates are formally consistent with our primary results; however, the change in logarithmic Bayes factors would lead us to the opposite interpretation of the posterior for both \lmi{} and \dtwo, i.e., a marginally resolved dispersion for \lmi{} and a marginally unresolved dispersion for \dtwo{}. This sensitivity is to be expected given the particularly small kinematic samples for these latter two targets after additional stars have been removed, and leads us to give greater weight to our chemical constraints when determining these two satellites' classifications (see \secref{chemistries} and \secref{classifications}).

\par Finally, we re-ran our fits adopting a different velocity dispersion prior. Our primary fits assumed a uniform (flat) prior for $\sigma_v$ bounded by [0,10]~\kms; these prior boundaries were chosen to encompass the full empirical range of dispersions seen in the MW's ultra-faint dwarf galaxies and to yield conservative upper limits. Another common choice is a Jeffreys prior -- i.e., uniform in $\log(\sigma)$ -- that is uninformative for scale parameters. Taking such a prior bounded as  $-1 < \rm log_{10}(\sigma_v) < 1$ and rerunning the fits yielded alternative estimates of \BooVvdispLog{}, \LeoMinorIvdispLog{}, \DELVEtwovdispLog{} for \booV{}, \lmi{}, and \dtwo{}, respectively. These smaller dispersions or tighter limits arise because the log prior weights low dispersions far more heavily than the uniform prior: it assigns 50\% of its probability to $0.1~\kms < \sigma_v < 1~\kms$, versus 10\% over $0~\kms < \sigma_v < 1~\kms$ for the uniform prior. Nonetheless, they remain consistent with our primary fit results assuming the linear-uniform prior. For the remainder of this work, we adopt the more conservative limits from the linear-uniform prior fits given the small member samples available.

\subsection{Dynamical Masses and Mass-to-Light Ratios}
Using our new velocity dispersion measurements, we derived dynamical mass estimates or upper limits based on two distinct estimators. First, we computed masses using the \citet{2010MNRAS.406.1220W} estimator for dispersion-supported systems in dynamical equilibrium. This estimator relates the total enclosed mass within the 3D half-light radius, $M(r < r_{\rm 1/2,3D})$, to the velocity dispersion and the 2D (projected) half-light radius according to 
\begin{align}
M_{1/2} \equiv M(r < r_{\rm 1/2,3D}) \approx 930\left(\frac{\sigma_v}{\kms} \right)^2\left(\frac{r_{1/2}}{\rm pc}\right) M_\odot.
\end{align}
Next, we computed dynamical masses within $1.8r_{1/2}$ using the estimator from \citet{2018MNRAS.481.5073E},
\begin{align}
M(<1.8\,r_\mathrm{1/2}) \approx 3.5 \times 1.8\,r_\mathrm{1/2}\sigma_v^2  G^{-1}.
\end{align}
This latter estimator was derived to yield unbiased mass estimates assuming maximal ignorance about the underlying density profile and anisotropy. For both estimators, we propagated the posterior distributions for the velocity dispersion and half-light radius and computed credible intervals based on the median and 16th/84th percentiles or upper limits at the 95\% credible level. Finally, we also used these posteriors and the $L_V$ estimates from \tabref{propertiestable} to compute mass-to-light ratios within the half-light radius for each satellite (for the mass derived using the \citealt{2010MNRAS.406.1220W} estimator only).

\par We summarize our mass and mass-to-light ratio estimates in \tabref{propertiestable} and compare them to the known MW globular clusters, satellite galaxies, and ambiguous ultra-faint compact satellites in \figref{masses}. Most interestingly,  we find a mass-to-light ratio upper limit of \BooVmlratio{} for \booV{} that suggests it may be less dark-matter-dominated than the majority of known MW ultra-faint dwarfs. This low mass-to-light ratio arises not only because of the tight velocity dispersion limit that we derive but also because \booV{} is compact (making the dynamical mass, which scales as $M_{1/2} \propto r_{1/2}$, smaller at fixed $\sigma_v$) and somewhat overluminous compared to other ultra-faint dwarfs at its size (see \figref{sizeluminosity}). \dtwo{} (\DELVEtwomlratio{}), on the other hand, appears more typical and is consistent with the mean trend seen for ultra-faint dwarfs in \figref{masses}. Finally, our upper limit for \lmi{} (\LeoMinorImlratio{}) does not meaningfully constrain the system's dark matter content at this time.

\section{Stellar Metallicities}
\label{sec:chemistries}
To complement our kinematic constraints, we determined the mean iron abundance ($\rm \overline{[Fe/H]}$) and metallicity dispersion ($\sigma_{\rm [Fe/H]}$) of \booV{}, \lmi{}, and \dtwo{}. Using our samples of $N_{\rm [Fe/H]} = \BooVNfehnumberonly{}, \LeoMinorINfehnumberonly{}$, and \DELVEtwoNfehnumberonly{} member stars with CaT-derived metallicities in each satellite, respectively,  we performed a two-parameter Bayesian fit to each satellite's metallicity distribution assuming a Gaussian likelihood analogous to that used for the velocity dispersion fits. We adopted uniform priors of $-4 < \overline{\rm [Fe/H]} < -1$ and $0 \rm ~dex< \sigma_{\rm [Fe/H]} < 1~dex$ and derived posterior probability distributions for each model parameter through dynamic nested sampling with \texttt{dynesty}. This procedure was repeated for a no-dispersion model with $\sigma_{\rm [Fe/H]}$ fixed to zero, again enabling us to compute Bayes factors and thereby rigorously decide whether to quote resolved estimates vs. upper limits.
\par
\begin{figure*}
    \centering
    \includegraphics[width=0.32\linewidth]{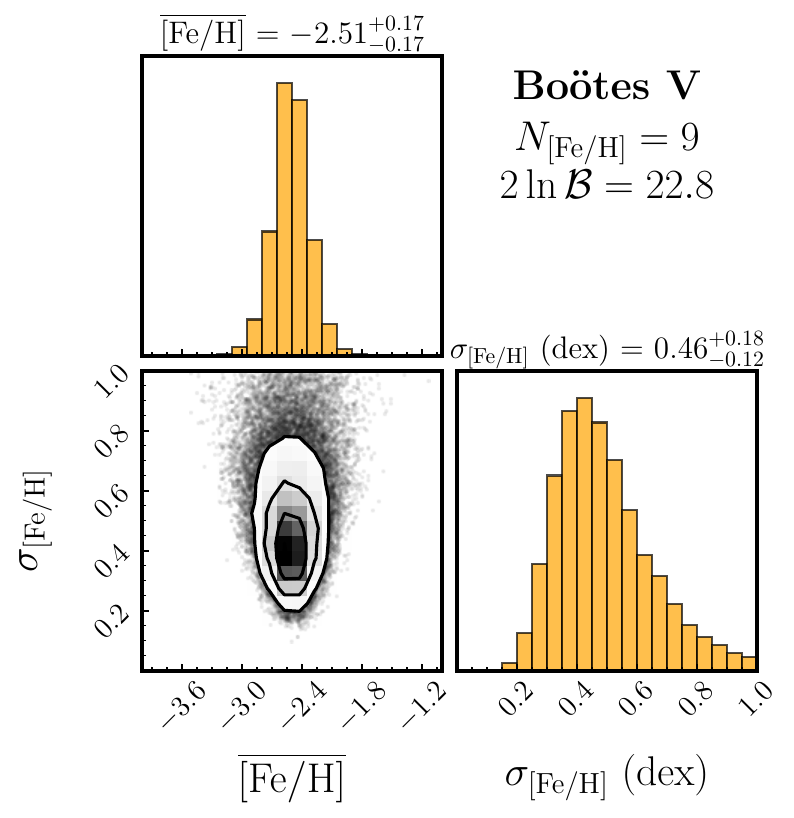}
    \includegraphics[width=0.32\linewidth]{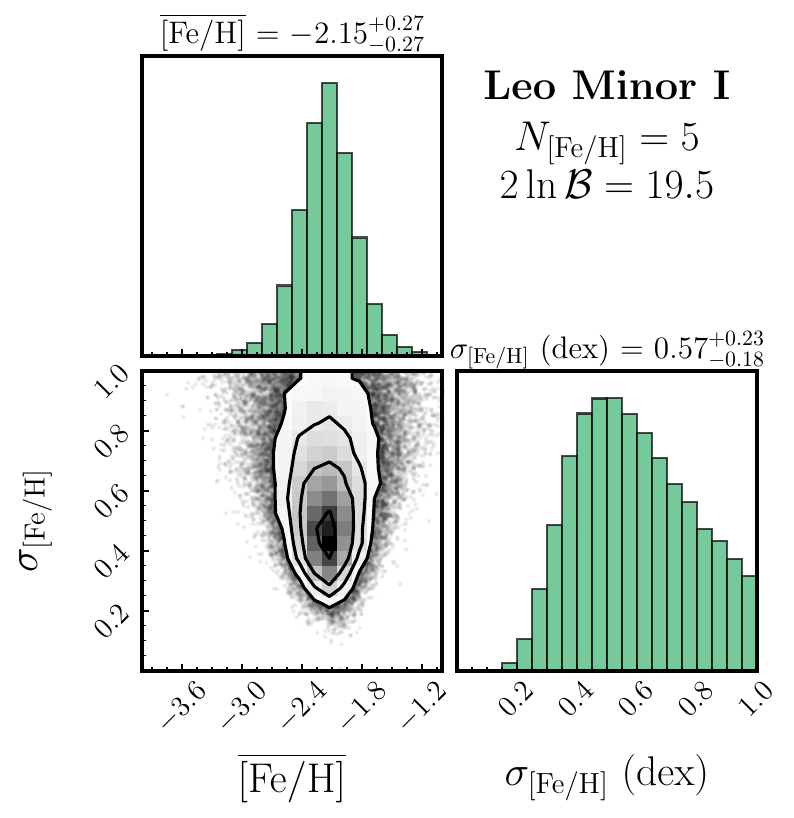}
        \includegraphics[width=0.32\linewidth]{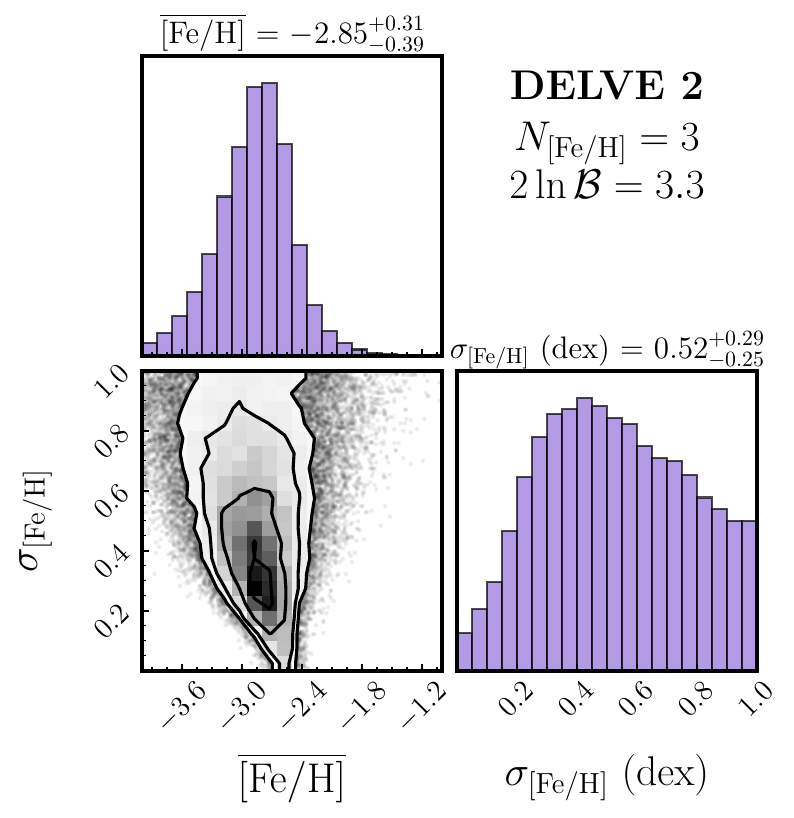}
    \caption{Mean metallicity and metallicity dispersion posteriors for \booV{} (left), \lmi{} (center), and \dtwo{} (right). In each panel, all axis limits are set to the boundaries of our fit priors, i.e., $[-4,-1]$ for [Fe/H] and [0,1]~dex for $\sigma_{\rm [Fe/H]}$. We clearly resolve the metallicity dispersion of both \booV{} and \lmi{} and find a modestly significant resolved dispersion for \dtwo{}.}
    \label{fig:metallicityposteriors}
\end{figure*}

 The resulting posterior probability distributions for the metallicity and metallicity dispersion of each satellite are shown in \figref{metallicityposteriors}, and we summarize our estimates in \tabref{propertiestable}. We find that all three satellites are very metal-poor, with mean metallicities \BooVfehmodel{} (\booV), \LeoMinorIfehmodel{} (\lmi{}), and \DELVEtwofehmodel{} (\dtwo). As seen in \figref{metallicities}, these metallicities are broadly consistent with the metallicities of confirmed ultra-faint MW satellite galaxies at similar luminosities; the few known globular clusters at similar luminosities are all significantly more metal rich ($\rm [Fe/H] > -2$). These metallicities thus provide some information about the classification of our targeted satellites; we return to this point in \secref{classifications}.
\par As for the satellites' metallicity dispersions, we estimate \BooVfehdisp{} (\booV{}), \LeoMinorIfehdisp{} (\lmi{}), and \DELVEtwofehdisp{} (\dtwo{}) from the same fits. The former two dispersions are clearly resolved, with Bayes factors of \BooVfehdispbayes{} and \LeoMinorIfehdispbayes{}. These dispersions persist even when the most metal-rich member in each satellite is excluded; in these cases, we recover \BooVfehdispExclMR{}
 (\booV) and  \LeoMinorIfehdispExclMR{} (\lmi). The presence of statistically significant internal metallicity spreads is further reinforced by the large [Fe/H] ranges seen across our sample of member stars in each of these two systems, \BooVfehrange{} and \LeoMinorIfehrange{}, both of which significantly exceed measurement uncertainties. On the other hand, the  metallicity spread that we identify in \dtwo{} is more modestly significant, with a logarithmic Bayes factor \DELVEtwofehdispbayes{} and an observed range of \DELVEtwofehrange{}. While the Bayesian evidence clearly favors a resolved dispersion for \dtwo{} and not an upper limit, the small number of metallicity members motivates follow-up to better characterize the system's metallicity distribution.

\section{Proper Motions and Orbital Histories}
\label{sec:orbits}
Our new radial velocity measurements allow us to study the orbital histories of the three satellites. To realize this possibility, we determined new \textit{Gaia}-based mean proper motions for each system (\secref{meanpms}), motivated by the pure member selections provided by our spectroscopy. We then integrated each satellite's orbit in the combined potential of the MW and LMC (\secref{integration}) and explored the connection between \dtwo{} and the LMC (\secref{delve2lmc}).

\subsection{Proper Motions}
\label{sec:meanpms}
Of the \BooVNmem{}, \LeoMinorINmem{}, and \DELVEtwoNmem{} spectroscopic members we identified in \booV,  \lmi, and \dtwo, respectively,  \BooVNpropermotion{}, \LeoMinorINpropermotion{}, and \DELVEtwoNpropermotion{} stars have \Gaia DR3 proper motions. All 16 of these stars pass basic astrometric quality cuts (\texttt{astrometric\_excess\_noise\_sig < 3} and \texttt{ruwe} $<$ 1.4). Using these measurements,  we 
determined updated mean proper motions $(\overline{\mu_{\alpha*}},\overline{\mu_\delta})$ for each satellite by fitting a bivariate (2D) Gaussian distribution, accounting for the covariance between components. Adopting the likelihood from \citet{2022ApJ...940..136P} and assuming no intrinsic proper motion dispersion, we performed MCMC sampling with \texttt{emcee} \citep{2013PASP..125..306F} to derive posteriors for each component. From the median and 16th/84th percentiles of the posteriors, we determine mean proper motions of 
\begin{align*}
    \text{\booV{}: } (\BooVpmranumberonly{},\BooVpmdecnumberonly{}) \rm \ mas \ yr^{-1} \\ 
    \text{\lmi{}: } (\LeoMinorIpmranumberonly{},\LeoMinorIpmdecnumberonly{}) \rm \ mas \ yr^{-1}   \\
    \text{\dtwo{}: } (\DELVEtwopmranumberonly{},\DELVEtwopmdecnumberonly{}) \rm \ mas \ yr^{-1}. 
\end{align*}
\par The uncertainties on these estimates include the formal covariance between the proper motion components for a given star but do not yet account for the systematic uncertainty associated with star-to-star spatial covariance discussed by \citet{2021A&A...649A...2L}. Following \citet{2022ApJ...940..136P}, we approximated the contribution of this systematic on our mean proper motion uncertainty based on the angular covariance function $V(\theta)$ provided in \citet{2021A&A...649A...2L} evaluated at the scale of each satellite's angular half-light radius (i.e., $\epsilon_{\rm\mu,sys} = \sqrt{V(\theta = r_{1/2})}$), finding  $\epsilon_{\rm \mu,sys} = 0.023 \rm \ mas \ yr^{-1}$ for each satellite. While subdominant, we accounted for this contribution during our orbit analyses by perturbing each posterior draw with a zero-mean Gaussian of standard deviation  $\epsilon_{\mu,\mathrm{sys}}$. 
\par Compared to independent prior determinations by \citet{2022A&A...657A..54B}, \citet{2023AJ....166...76S}, and \citet{2024MNRAS.527.4209J}, we generally find excellent ($\ll 1\sigma$) agreement. This reflects the highly overlapping \textit{Gaia} member samples identified in each study. The only (minor) exception to this trend was for \lmi{}, for which \citet{2024MNRAS.527.4209J} estimated $(-0.221 \rm \ mas \ yr^{-1}, -0.835 \rm \ mas \ yr^{-1})$ -- mildly offset in $\mu_\delta$ from our own result. This difference results from their inclusion of a fourth star (\textit{Gaia} DR3 732982829593106560), which their model assigns a membership probability of $p=0.3$. Our spectroscopy now clarifies that this star is an unambiguous velocity non-member.

\begin{figure*}
    \centering
    \includegraphics[width=\textwidth]{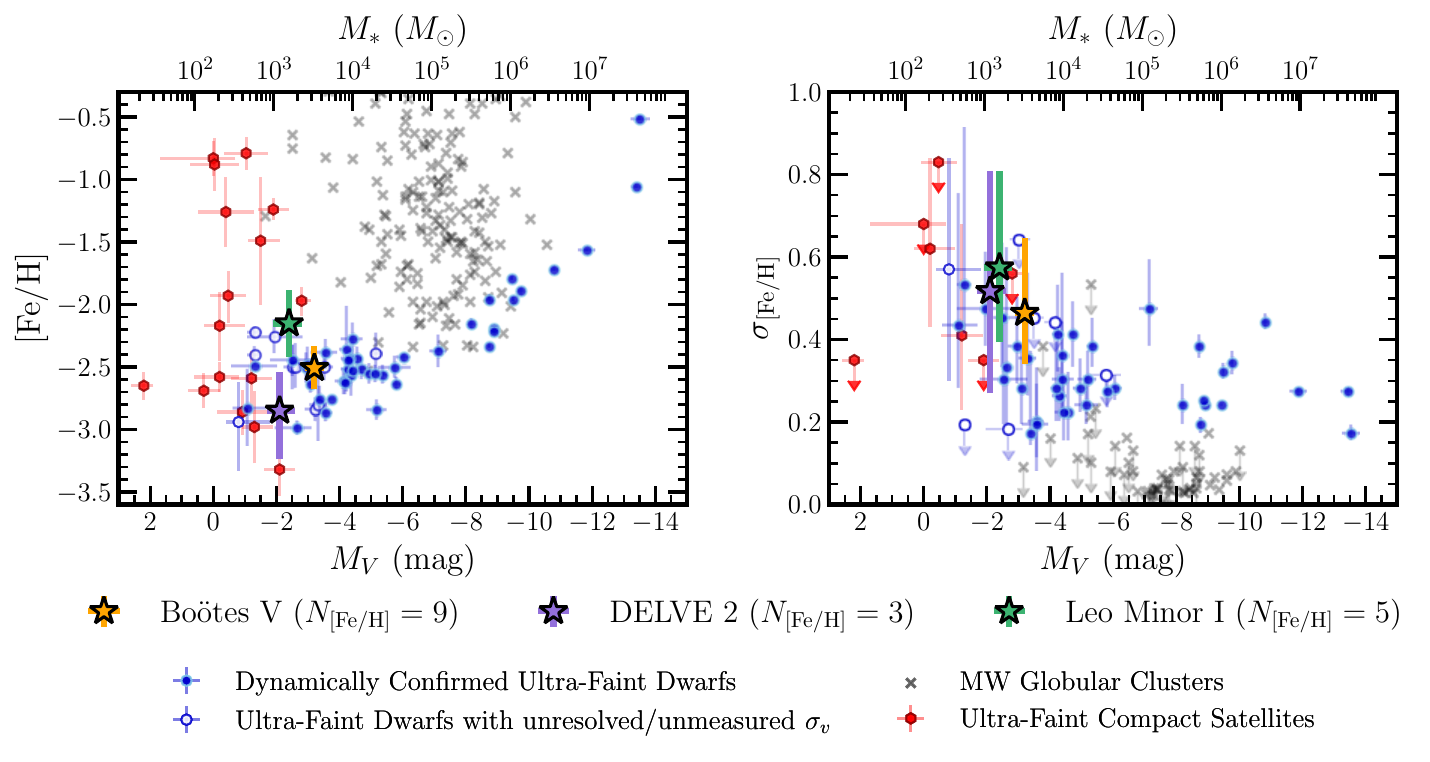}
    \caption{Comparing the metallicity properties of \booV{}, \lmi{}, and \dtwo{} to those of the MW globular cluster and satellite galaxy populations, as well as the population of ultra-faint compact satellites with relatively more ambiguous classifications. (Left) Spectroscopic mean metallicity vs. absolute magnitude. The three satellites lie on the empirical luminosity-metallicity relation for MW satellite galaxies. (Right) Metallicity dispersion vs. absolute magnitude for the same populations; we limit the comparison sample to satellites with well-measured spectroscopic metallicity dispersions (uncertainties $<0.4$~dex). All three satellites have metallicity dispersions consistent with the population of MW ultra-faint dwarfs and larger than those of the globular clusters. These metallicity dispersions provide the strongest evidence that \booV{}, \lmi{}, and \dtwo{} are  dwarf galaxies.}
    \label{fig:metallicities}
\end{figure*}

\subsection{Orbits in the Combined MW + LMC Potential}
\label{sec:integration}
We integrated the orbits of \booV{}, \lmi{}, and \dtwo{} in a time-dependent MW + LMC potential using the Action-based Galaxy Modeling Architecture (\texttt{AGAMA}; \citealt{2019MNRAS.482.1525V}) package. For the satellites' initial conditions, we drew from normal or split-normal distributions for the position components ($\alpha_{\rm 2000}$, $\delta_{\rm 2000}$, $D_\odot$) and directly sampled the motion components ($\mu_{\alpha*}$, $\mu_{\delta}$, $\overline{v_{\rm hel}}$) from the posteriors derived above, accounting for the covariance of the proper motion components. The MW was modeled with an \texttt{AGAMA} implementation of the \texttt{MilkyWayPotential\_v2} from the \texttt{gala} package \citep{2017JOSS....2..388P}, which is composed of an NFW dark matter halo \citep{1997ApJ...490..493N}, nucleus and bulge components modeled as Hernquist profiles \citep{1990ApJ...356..359H}, and the sum of three Miyamoto-Nagai disk models \citep{1975PASJ...27..533M} constructed to match the \textit{Gaia} rotation curve data from  \citet{2019ApJ...871..120E}. The LMC was modeled as a truncated NFW spheroid with mass $M_h = 1.5 \times 10^{11} \rm \ M_\odot$ (consistent with the estimates from, e.g., \citealt{2019MNRAS.487.2685E,2021MNRAS.501.2279V}) and scale radius $\sim$10.8~kpc experiencing Chandrasekhar dynamical friction \citep{1943ApJ....97..255C} with a spatially varying Coulomb logarithm calibrated against $N$-body simulations.\footnote{Our implementation of the LMC follows that found within the \texttt{AGAMA} script \texttt{example\_lmc\_mw\_interaction.py}; refer to this example for more details. } We adopted the centroid position and mean radial velocity of the LMC from \citet{2002AJ....124.2639V}, the LMC distance from \citet{2019Natur.567..200P}, and its mean proper motion from \citet{2021A&A...649A...7G}; explicitly, these are $(\alpha,\delta,D_\odot,\overline{\mu_{\alpha*}},\overline{\mu_\delta},\overline{v_{\rm hel}}) =$ (81.28\degree,$-$69.78\degree,49.6 kpc,  1.858~$\rm \ mas \ yr^{-1}$, 0.385~$\rm \ mas \ yr^{-1}$, 262.2~\kms). In our setup, the MW experiences reflex acceleration due to the gravitational influence of the LMC and thus defines a non-inertial frame; however, its density profile remains unperturbed.
\par Each satellite's orbit was integrated from 2.5 Gyr in the past to 0.5 Gyr into the future for 2500 randomly drawn sets of six-dimensional initial conditions, in addition to a single fiducial case where the satellites' phase space properties were set to match the median estimates presented in \tabref{propertiestable}. At the conclusion of all realizations, we computed basic properties including their apocenters ($r_{\rm apo}$), pericenters ($r_{\rm peri}$), and eccentricities ($e$). Owing to the infall of the LMC and the Milky Way's reflex response, these properties vary over time within each  realization. Given this, we specifically identified the apocenter/pericenter closest in time to $t=0$ for each realization -- whether in the past or in the future -- and computed the eccentricity from these quantities as $e \equiv (r_{\rm apo} - r_{\rm peri})/(r_{\rm apo} + r_{\rm peri})$. We summarize these properties in \tabref{propertiestable}.

\begin{figure*}
    \centering
    \includegraphics[width=\textwidth]{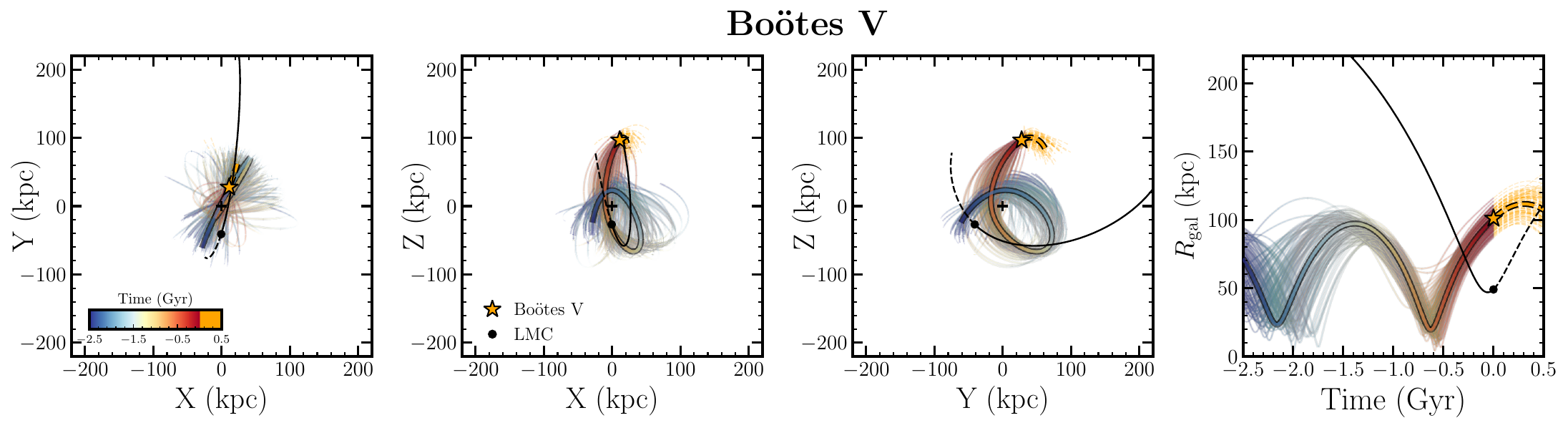}
    \caption{Four projections of the orbit of \booV{} in our MW + LMC potential. In all panels, we plot our fiducial orbit for \booV{} (thick, bordered line) as well as the orbits derived from 100 randomly sampled realizations of the satellite's initial conditions. Each of these 100 sampled trajectories is colored by time: times $[-2.5,0]$~Gyr are shown in a blue--yellow--red continuous colormap, whereas (future) times $[0,+0.5]$~Gyr are shown in orange. \booV{}'s present-day position is depicted with a large star symbol. The LMC is shown as a black dot, with a solid (dashed) black line tracing its past (future) orbit.  (Far Left) Galactocentric $Y$ vs $X$. (Center Left)  Galactocentric $Z$ vs $X$. (Center Right)  Galactocentric $Z$ vs $Y$. (Far Right) Galactocentric radius ($R_{\rm gal}$) vs time. \label{fig:booVorbit} }
     \includegraphics[width=\textwidth]{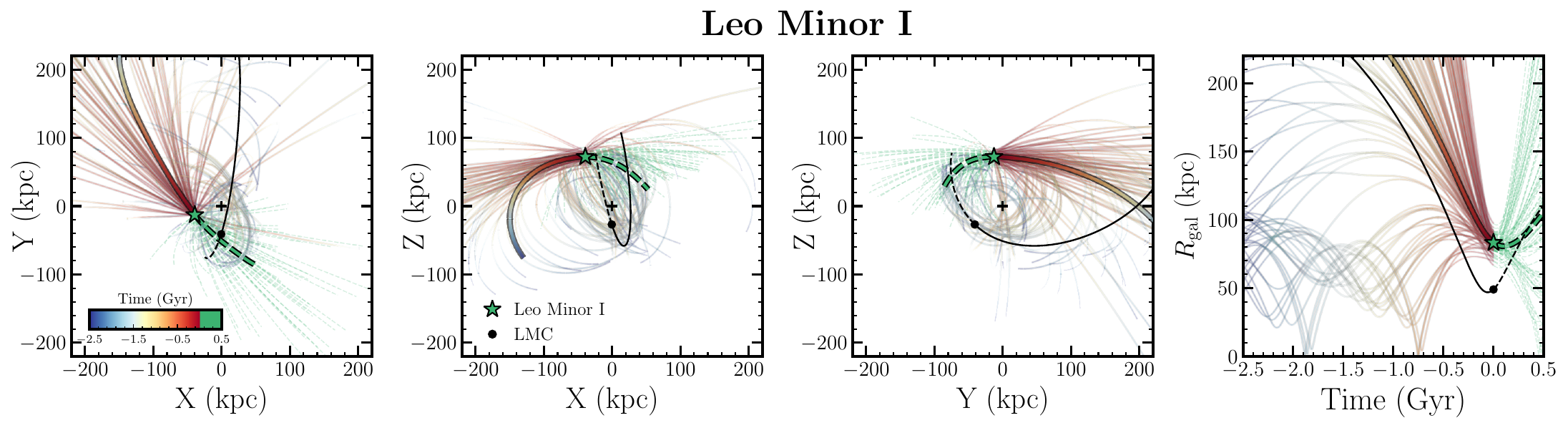}
    \caption{The same as \figref{booVorbit} except for \lmi{}; the system's present-day position and future trajectory are shown in green. \lmi{}'s orbit is considerably more uncertain than \booV{}'s due to the much poorer proper motion precision ($\sim$5--$9\times$ larger uncertainty for the two components). \label{fig:lmi1orbit} }   
         \includegraphics[width=\textwidth]{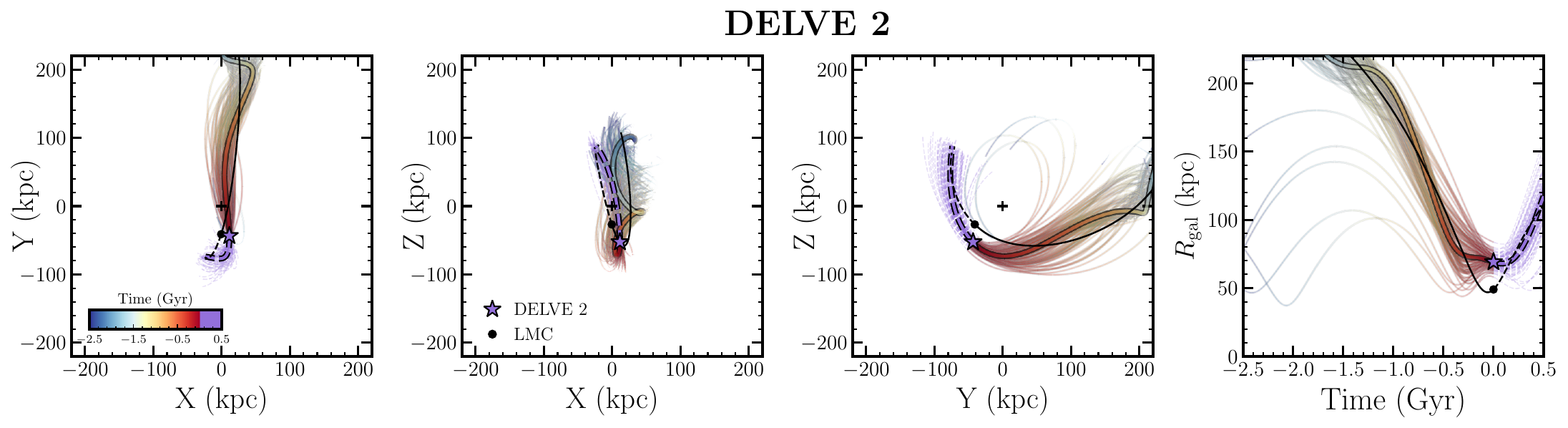}
   
     \caption{The same as Figures \ref{fig:booVorbit}--\ref{fig:lmi1orbit} except for \dtwo{}; the system's present-day position and future trajectory are shown in purple. \dtwo{} clearly traces the LMC's past orbit in all four panels and is energetically bound to the LMC at $t=0$ in \DELVEtwofboundLMCnow{} of realizations. \label{fig:delve2orbit}}   
\end{figure*}

\par In  Figures \ref{fig:booVorbit}--\ref{fig:delve2orbit}, we present the fiducial orbits of each satellite as well as 100 randomly drawn realizations from our set of 2500. Beginning with \booV{}, we find that the system occupies an eccentric, polar orbit (\BooVeccproximateAgama; \BooVLz) with an orbital period of $\sim$1.5~Gyr. The system is currently approaching its predicted apocenter (\BooVrapoproximateAgama{}) and has experienced a moderately close-in perigalacticon passage (\BooVrperiproximateAgama{}) $\sim$650 Myr ago. These results are in good agreement ($\lesssim$1$\sigma$) with those found by \citetalias{2023AJ....166...76S} in a MW-only potential. While a pericenter this close to the Galactic center suggests the possibility that \booV{} is vulnerable to tidal disruption, we find that its compact size and concomitantly higher density promote its resilience against Galactic tides. In fact, we find the ratio of \booV's central density ($\rho_{1/2} \equiv M_{1/2} \left(\frac{4}{3}\pi(1.305r_{1/2})^3\right)^{-1}$) to the density of the MW at the location of \booV{}'s pericenter in our fiducial model is $\rho_{1/2}/\rho_{\rm MW} > 10$ even if \booV{} were comprised entirely of stars. This strongly disfavors tidal disruption in \booV{}'s central regions, though does not rule out the possibility of disruption in its outskirts.

\par Unlike \booV{}, \lmi{} is not clearly a long-term bound satellite of the MW. Our orbit models, while highly uncertain, prefer a scenario in which \lmi{} is on its first infall onto the MW, plunging toward the Galactic center from a separation of $\gtrsim$300~kpc on a retrograde trajectory with respect to the disk (\LeoMinorILz). The system is currently situated at/near its orbital pericenter (\LeoMinorIrperiproximateAgama) and has not experienced an apocenter within the last 2.5 Gyr  in the majority of our realizations; given this, we do not report $r_{\rm apo}$ or $e$ for the system in \tabref{propertiestable}. A major caveat, though, is that these conditions (i.e., that \lmi{} is near its pericenter on an eccentric first-infall orbit) could arise purely from sampling the particularly large proper motion uncertainties for \lmi{} (see the discussion in Section 5.2 of \citealt{2022MNRAS.511.2610C}). More precise proper motion measurements -- such as those expected from the upcoming \textit{Gaia} DR4 -- will be necessary to break the degeneracy between a true first-infall orbit and the effect of the large uncertainties.
\par Lastly, we find that \dtwo{} is likely also on its first infall onto the MW and is currently situated near its pericenter (\DELVEtworperiproximateAgama); we therefore do not report $r_{\rm apo}$ or $e$  in \tabref{propertiestable}. These properties closely match those of the LMC -- which itself is believed to be on its first infall onto the MW \citep[e.g.,][]{2007ApJ...668..949B,2013ApJ...764..161K} -- strongly suggesting a physical association. We explore this likely connection between \dtwo{} and the LMC in detail in the following subsection.

\subsection{Confirmation of \dtwo{} as a Bound Satellite of the LMC}
\label{sec:delve2lmc}
At the time of discovery, \dtwo{} was identified as a promising candidate satellite of the Magellanic system based  on its proximity to the LMC and SMC (3D separations of $\sim$28~kpc and $\sim$12~kpc, respectively; \citetalias{2021ApJ...910...18C}) and consistency with the predicted distribution of LMC satellite debris from the simulations of \citet{2016MNRAS.461.2212J}. Using these simulations, \citetalias{2021ApJ...910...18C} predicted that the system would be  consistent with satellite debris of the LMC if its (then-unknown) radial velocity in the Galactic Standard of Rest (GSR) fell in the range $-80~\kms{} <   v_{\rm R,GSR} <  50~\kms{}$. In addition, \citetalias{2021ApJ...910...18C} rewound \dtwo{}'s orbit in the MW + LMC potential following the prescription described in \citet{2020MNRAS.495.2554E}, making a similar prediction of $-150~\kms{} <   v_{\rm R,GSR} <  80~\kms{}$ for  a $p > 0.5$ probability of association. Independently, both \citet{2022MNRAS.511.2610C} and \citet{2024MNRAS.527..437V} explored the possibility of an association between \dtwo{} and the LMC (again in the absence of a radial velocity measurement), predicting high association probabilities if the system's heliocentric radial velocity falls in the range $30~\kms{} \rm < \overline{v_{hel}} < 190$~\kms{}. 
\par Our new 6D kinematic measurements and reconstructed orbital history make abundantly clear that \dtwo{} is a bound LMC satellite. Specifically, our newly measured mean radial velocity of \DELVEtwomeanvhel{} (\DELVEtwomeanvgsr{}) falls within the predicted high-probability range of each of the above studies. In addition, \dtwo{} displays a highly negative angular momentum about the Galactocentric $X$ axis, \DELVEtwoLx, that is a defining kinematic characteristic of accreted debris from the Clouds \citep{2023ApJ...956..110C}. Finally, we find that \dtwo{} is instantaneously energetically bound to the LMC at $t=0$ in $\sim$\DELVEtwofboundLMCnow{} of realizations and bound 2.5 Gyr ago in $\sim$\DELVEtwofboundLMCpast{}; these can be considered rough association probabilities from our models.
\par
\begin{figure}
    \centering
    \includegraphics[width=0.9\linewidth]{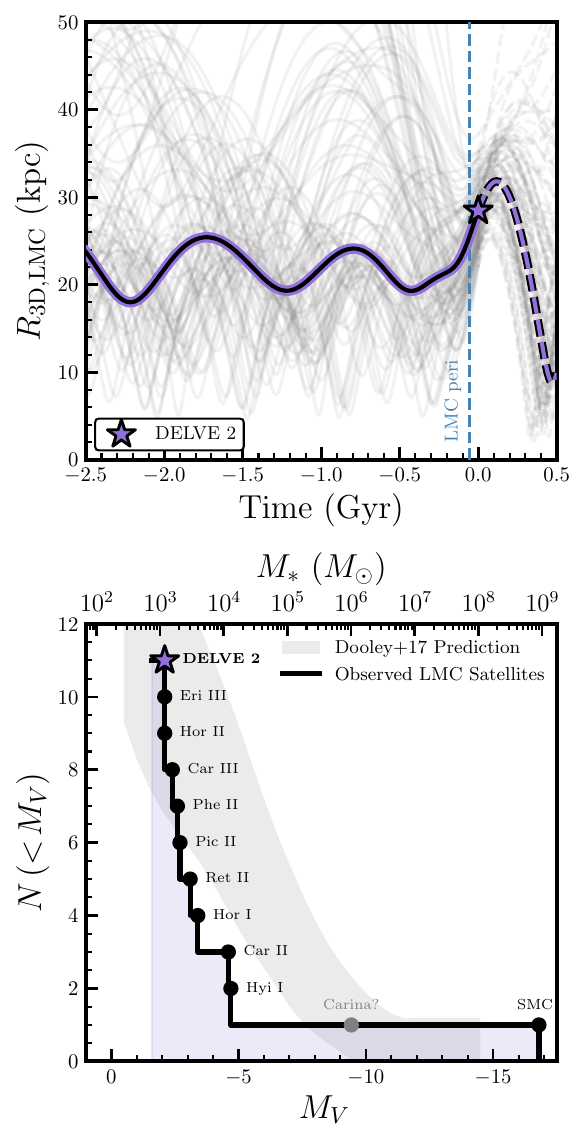}
    \caption{(Top) 3D separation between \dtwo{} and the LMC ($R_{3\rm D,LMC}$) from 2.5 Gyr ago to 0.5 Gyr in the future. \dtwo{} exhibits a stable, bound orbit about the LMC until the LMC's pericenter about the MW (dashed blue line), at which point its orbit becomes more eccentric and somewhat less bound. (Bottom) An updated view of the observed cumulative luminosity function (LF) of LMC satellites, adopting the list of high-probability LMC-associated dwarfs from \citet{2024MNRAS.527..437V}; no completeness corrections have been applied. The Carina dSph's status as a (former) LMC satellite is debated and we exclude it from the LF. Eridanus III is not yet confirmed to be a galaxy \citep{2024ApJ...976..256S,2026arXiv260217652C}, but we include it here. In gray, we overplot the abundance-matching prediction from \citet{2017MNRAS.472.1060D}. }
    \label{fig:lmcassociation}
\end{figure}

 Our fiducial orbital trajectory (purple line in \figref{lmcassociation}) describes that \dtwo{} has spent much of the last 2.5 Gyr on a relatively circular orbit taking it between $18 \rm \ kpc \lesssim \textit{R}_{\rm 3D,LMC} \lesssim 25 \ kpc$ with an orbital period about the LMC of $\sim 900$~Myr.  Interestingly, we observe that the system's orbit has been perturbed in the last $\sim$200~Myr, coinciding with the approach of the LMC toward its perigalacticon about the MW. This perturbation significantly increases the eccentricity of \dtwo{}'s orbit about the LMC, and our forward-integration predicts that \dtwo{} will come within $10$~kpc of the LMC center in $\sim$500~Myr. Despite this perturbation,  the system is predicted to remain energetically bound. 
\par  Presuming that it is indeed a dwarf galaxy (as appears likely; see \secref{disco}), \dtwo{} represents the eleventh MW satellite galaxy with a likely kinematic link to the Clouds (after the SMC, Hydrus I, Carina II, Horologium I, Horologium II, Reticulum II, Pictor II, Phoenix II, Carina III, Eridanus III; see e.g., \citealt{2018ApJ...867...19K,2020ApJ...893..121P,2022ApJ...940..136P,2025OJAp....8E.112P,2026arXiv260217652C,2024MNRAS.527..437V,2026arXiv260613787M}  for discussion of these associations and their caveats). Among this list, \dtwo{} is statistically tied for the faintest known LMC-associated dwarf galaxy, with an absolute magnitude matching that of Eridanus III and Horologium II ($M_V = -2.1 \pm 0.5$ and $-2.1\pm0.4$, respectively; \citealt{2018ApJ...852...68C,2026arXiv260625058P}, and see \figref{lmcassociation}).  It is also the closest (likely) ultra-faint dwarf to the SMC, at a distance of just $\sim$12~kpc. Echoing \citetalias{2021ApJ...910...18C}, we find the SMC's gravitational influence on \dtwo{} is still much weaker compared to the LMC's despite this proximity, making our omission of the SMC in our orbit models a tolerable approximation.
\par Motivated by this now-confirmed link between the LMC and \dtwo{}, we present an updated view of the Magellanic satellite luminosity function (LF) down to the ultra-faint regime in \figref{lmcassociation}. We compare the observed LF to predictions from \citet{2017MNRAS.472.1060D}, who applied an abundance-matching approach to the \textit{Caterpillar} cosmological simulations \citep{2016ApJ...818...10G} to generate satellite LFs for LMC-like hosts. The observed cumulative satellite luminosity function of the LMC -- with no completeness corrections applied --  is less gradual than the predicted LF, with both an apparent dearth of bright satellites ($M_V \lesssim -4$) and an excess at the faint end. These features were also observed by \citet{2017MNRAS.472.1060D} albeit from a smaller satellite sample that predated the availability of \textit{Gaia} proper motions. The authors found this LF shape challenging to reproduce in their models but argued that it may reflect the need for a lower critical mass threshold for galaxy formation, a later epoch of cosmic reionization for the Local Group ($z \lesssim 9$), a faint-end flattening in the halo stellar mass-halo mass relationship, and/or the significant tidal stripping of the Magellanic satellites. 

\section{Discussion}
\label{sec:disco}
\subsection{The Nature of \booV, \lmi, and \dtwo{}: Three Ultra-Faint Dwarf Galaxies}
\label{sec:classifications}
The most significant question surrounding \booV{}, \lmi{}, and \dtwo{} is that of their classifications: are they dwarf galaxies or star clusters? Based on our new spectroscopic velocity and metallicity measurements, we find that  \booV{} and \lmi{} are clearly dwarf galaxies. We further find that \dtwo{} is most likely a dwarf galaxy as well, though the evidence is less definitive.
\par To begin, our results confirm that all three satellites are very metal-poor: \BooVfehmodel{} (\booV), \LeoMinorIfehmodel{} (\lmi), \DELVEtwofehmodel{} (\dtwo{}). As seen in \figref{metallicities}, these low metallicities place each satellite on/near the empirical luminosity-metallicity sequence for dwarf galaxies \citep[see e.g.,][]{2013ApJ...779..102K,2026arXiv260210202G} and, for \booV{}  and \dtwo{}, place them at or below the low-metallicity edge of the classical MW globular cluster population ($\rm [Fe/H] \approx -2.5$).\footnote{Globular clusters below this empirical metallicity ``floor'' have been suspected and confirmed in at least one case, and thus a low metallicity alone is not a sufficient condition for a galaxy classification \citep{2018MNRAS.477.4565S,2020Sci...370..970L,2025OJAp....8E..86U,2026arXiv260217652C,2026arXiv260612517B}.} While \lmi{} falls toward the high-metallicity side of this relation, this very likely reflects the small sample size (\LeoMinorINfeh) and the presence of a single bright member at \LeoMinorIfehbrightest. In addition to their low metallicities, all three satellites are characterized  by ancient stellar populations ($\tau > 10~\Gyr$; see \citetalias{2021ApJ...910...18C}, \citetalias{2023ApJ...953....1C}, and \citetalias{2023AJ....166...76S}). While these ages and metallicities are consistent with both reionization-quenched ultra-faint dwarfs and many ancient MW globular clusters, these measurements distinguish \booV{}, \lmi{}, and \dtwo{} from systems such as the extended, low-luminosity satellite Crater / Laevens~1 -- one of very few MW globular clusters known at $r_{1/2} \gtrsim 20$~pc -- which displays a considerably younger stellar population ($\tau = 7.5$~Gyr; \citealt{2016ApJ...822...32W}).
\par Stronger evidence for dwarf galaxy classifications comes from these satellites' internal metallicity spreads: \BooVfehdisp{}, \LeoMinorIfehdisp{}, and \DELVEtwofehdisp{} for \booV{}, \lmi{}, and \dtwo{}, respectively. The former two satellites' metallicity dispersions are clearly resolved in our data, with logarithmic Bayes factors \BooVfehdispbayes{} and \LeoMinorIfehdispbayes{} (corresponding to approximate frequentist significances of \BooVmetallicitydispersionsignificance and \LeoMinorImetallicitydispersionsignificance). Moreover, these two dispersions remain resolved even when the most metal-rich member is exclude (see \secref{chemistries}), suggesting they are robust to membership uncertainty. On the other hand, \dtwo{}'s metallicity dispersion is only modestly resolved (\DELVEtwofehdispbayes{}) in our data and will require further confirmation from a larger member sample, though we highlight that the combination of a very low metallicity \textit{and} a tentative metallicity spread is highly suggestive of a dwarf galaxy classification.
\par Finally, we find that our observational constraints on the kinematics of the three satellites do not strongly inform us of their classifications, despite the relatively precise per-star velocity measurements available ($\sim$1--5~\kms{}). Our upper limits for the velocity dispersions, dynamical masses, and mass-to-light ratios of \booV{} and \lmi{} are formally consistent with both the globular cluster and dwarf galaxy scenarios, even while the chemical evidence suggests that both are galaxies. The only satellite with a resolved velocity dispersion in our data is \dtwo{} (\DELVEtwovdisp{}, with a modestly significant \DELVEtwovdispbayes{}). This dispersion supports the dark-matter-dominated dwarf galaxy scenario for \dtwo{} (with \DELVEtwomlratio{}), though we caution that it is sensitive to two stars with low-significance hints of velocity variability (see \secref{vdispsystematics}).

\subsubsection{Are \booV, \lmi, and \dtwo{}  ``Globular-Cluster-Like Dwarfs''?}
Recently, \citet{2025Natur.645..327T} identified a population of Globular-Cluster-Like Dwarfs (GCDs) in the EDGE suite of high-resolution hydrodynamical simulations that have intermediate physical sizes and luminosities ($r_{1/2} \approx 25$--50~pc; $-4.9 \lesssim  M_V \lesssim -2.6$) comparable to the three satellites studied here.\footnote{See also \citet{2016ApJ...831..204R}  and \citet{2025arXiv251121824B}, who identify GCD-like populations in different hydrodynamical simulations.} The GCDs are a proposed subclass of ultra-faint dwarf galaxies formed from a single, self-quenching star formation event in low-mass halos at high redshift. Owing to this formation channel, the defining features of these postulated systems -- beyond their morphologies -- are their low dark matter densities ($M/L_V\approx$~10--100~$\rm M_\odot L_\odot^{-1}$), short formation timescales (age spreads 10--20 Myr),  and small-but-nonzero metallicity spreads ($\sigma_{\rm [Fe/H]} \approx 0.1$--0.3~dex). \citet{2025Natur.645..327T} specifically identified \booV{} as a candidate GCD on the basis of its size and luminosity, though \dtwo{} and \lmi{} appear to be equally viable candidates based on their similar morphologies and masses.
\par Our new spectroscopic measurements somewhat disfavor the GCD scenario for at least two of our three satellites primarily on the basis of their larger metallicity dispersions. Specifically, our dispersion fits yielded 95\% credible lower limits of \BooVfehdispfive{}, \LeoMinorIfehdispfive{}, and \DELVEtwofehdispfive{} -- excluding much of the GCD parameter space for \booV{} and \lmi{}. The large metallicity \textit{ranges} of \BooVfehrange{} and \LeoMinorIfehrange{} that we observe for \booV{} and \lmi{} may also be challenging to explain with a single burst of star formation. In addition, our mass-to-light ratio estimate for \dtwo{}, \DELVEtwomlratio{}, is formally more dark-matter-dominated than the predicted GCDs; however, the uncertainty on this estimate is large and stellar binaries may be inflating this measurement (see \secref{vdispsystematics}).  We conclude that the GCD scenario is unlikely but not fully ruled out for the three satellites, with the greatest remaining prospects for \dtwo.

\subsection{Implications for Dark Matter Annihilation}
\label{sec:annihilation}
Owing to their proximity and high dark matter densities, the ultra-faint dwarf galaxies orbiting the MW are ideal sites to search for the gamma-ray emission from the annihilation of dark matter into Standard Model particles \citep[e.g.,][]{2008ApJ...678..614S,2015PhRvL.115w1301A,2015ApJ...801...74G,2020JCAP...02..012H,2024PhRvD.109f3024M}. For the simplest case of velocity-independent ($s$-wave) annihilation, the expected annihilation flux can be written as
\begin{align}
    \frac{d\Phi}{dE} = \underbrace{\frac{\langle\sigma v\rangle}{8\pi m_\chi^2} \frac{dN}{dE}}_{\text{particle physics}} \times \underbrace{ \int_{\Delta\Omega} \int_{\text{l.o.s}} \rho^2(l,\Omega)\, dl \, d\Omega.}_{\text{astrophysics}}
\end{align}
Here, the former term expresses the dependence on the underlying particle properties (the velocity-averaged cross section $\langle \sigma v\rangle$, mass $m_\chi$, and energy spectrum $dN/dE$) while the latter, astrophysical term is known as the $J$-factor, and corresponds to the integral of the dark matter density squared ($\rho^2$) over the line-of-sight in a fixed solid angle $\Omega$. The density $\rho$ is often inferred through Jeans modeling, but the small number of spectroscopic members in each of our target galaxies makes this challenging. Therefore, to estimate the $J$-factor within a fiducial solid angle with radius $\theta = 0.5^{\circ} $, hereafter $J(0.5^\circ)$, we instead used the scaling relation from \citet{2019MNRAS.482.3480P},
\begin{align}
J(0.5^\circ) &= 10^{17.87} 
\left( \frac{\sigma_{\mathrm{v}}}{5 \, \mathrm{km\,s^{-1}}} \right)^4
\left( \frac{D_\odot}{100 \, \mathrm{kpc}} \right)^{-2}
\left( \frac{r_{1/2}}{100 \, \mathrm{pc}} \right)^{-1}
\end{align}
directly propagating our posteriors for each input. This scaling relation was derived based on MW satellite galaxies as faint as $M_V = - 1.5$, and thus no extrapolation was necessary for our trio of galaxies.
\par For \dtwo{}, the sole satellite for which we measured a resolved velocity dispersion, we estimated a relatively modest \DELVEtwojfactor{}. This places the system in the lowest quartile of MW satellite galaxies by $J$-factor (see \citealt{2019MNRAS.482.3480P,2024ApJ...961..234H,2024PhRvD.109f3024M,2026arXiv260727326R} for recent compilations), but the system may nonetheless provide an incremental benefit in the context of stacked analyses of many local dwarfs. On the other hand, we determined only $J$-factor upper limits of \BooVjfactor{} and \LeoMinorIjfactor{} for \booV{} and \lmi{}, respectively. These estimates leave open the possibility that \lmi{} is a useful target in the search for the indirect detection of dark matter annihilation but clearly demonstrate that \booV{} is not. In fact, our relatively strong upper limit for \booV{} implies that the system's $J$-factor is among the lowest of all known ultra-faint dwarfs, owing to both the system's significant distance and stringent velocity dispersion upper limit.

\begin{figure*}
    \centering
    \includegraphics[width=0.95\textwidth]{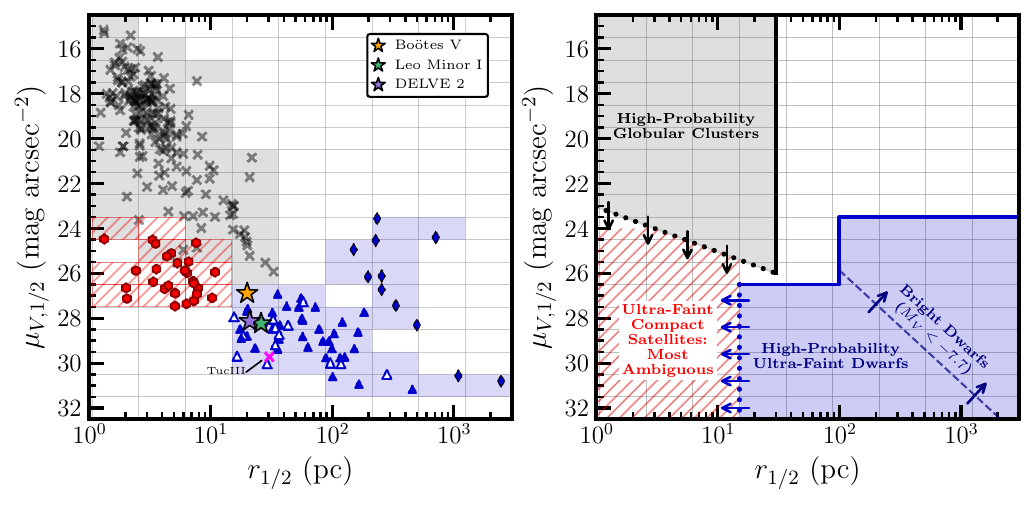}
    \caption{An updated perspective on the classifications of the MW satellite population in the surface brightness ($\mu_{V,1/2}$) vs. azimuthally averaged half-light radius ($r_{1/2}$) plane. (Left) The observed populations of MW globular clusters (black), bright dwarf galaxies (blue diamonds), ultra-faint dwarfs (blue triangles), and ultra-faint compact satellites (red hexagons). To guide the eye, we bin this parameter space and shade each bin based on the constituent objects; empty bins are shown in white. Notably, we find that for radii $r_{1/2} > 15$~pc and surface brightnesses $\mu_{V,1/2} > 26.5$ mag arcsec$^{-2}$, all but a single object (Tucana~III; shown as a pink $\times$) that has been spectroscopically characterized is more likely a galaxy.  (Right) A schematic representation of the classification landscape for MW satellites. At radii $r_{1/2}>15$~pc, we find that ultra-faint dwarf galaxies can be separated from extended star clusters based on a surface brightness threshold of $\mu_{V,1/2} > 26.5$ mag arcsec$^{-2}$. At smaller sizes ($r_{1/2} < 15$~pc) and low surface-brightnesses, classifications are most ambiguous and morphological separation of dwarfs and clusters is challenging.}
    \label{fig:landscape}
\end{figure*}

\subsection{An Updated Perspective on the Classifications of Faint Galactic Satellites}
Whether explicitly or implicitly, many studies of Local Group satellites have discussed the ``trough of uncertainty'' or ``valley of ambiguity'':  the slice of the size-luminosity plane at intermediate physical sizes for which the classifications of faint satellites are purportedly most uncertain \citep[see e.g.,][who propose $r_{1/2} \approx 10$--30~pc]{2018ApJ...852...68C}. Given the growing number of ultra-faint satellites in this regime with spectroscopic data, recent interest in the possibility of satellite galaxies at even smaller physical scales \citep{2020MNRAS.491.4591E,2024ApJ...976..256S,2024ApJ...961...92S,2026arXiv260217652C}, and the dawn of several major sky surveys poised to discover numerous new satellites with small physical sizes, it is worth revisiting the Milky Way satellite classification landscape.
\par In \figref{landscape}, we present the complete MW satellite population in the surface-brightness\footnote{Throughout, we define $\mu_{V,1/2}$ as the mean surface brightness within the azimuthally averaged half-light radius  $r_{1/2}$.} vs. azimuthally averaged half-light radius ($\mu_{\rm {V,1/2}}$ vs. $r_{1/2}$) plane, complementing the similar presentation in \figref{sizeluminosity}. In this parameter space, the classical MW globular clusters obey a relatively tight linear $\mu_{V,1/2}$--$r_{1/2}$ correlation that largely reflects a characteristic luminosity scale $\sim$10$^5 \rm \  L_\odot$ for globular clusters, whereas the population of known, faint MW satellite galaxies centers around a surface brightness scale of $\sim$28.5$\rm \ mag \rm \ arcsec^{-2}$ with scatter $\sim$1--2$\rm \, mag \, arcsec^{-2}$. At the smallest sizes ($r_{1/2} < 15$~pc) and low surface brightnesses ($>24\rm \ mag \rm \ arcsec^{-2}$) are the Ultra-Faint Compact Satellites (UFCSs); these systems lack reliable classifications in many cases but very likely include both the faintest galaxies and the faintest star clusters \citep[e.g.,][]{2011AJ....142...88F,2012ApJ...753L..15M,2019MNRAS.490.1498L,2024ApJ...976..256S,2026arXiv260612517B,2026arXiv260217652C}.
\par As this figure clearly reveals, \textit{there is no ambiguity in the classifications of MW satellites with azimuthally averaged radii $r_{1/2} >35$~pc}: echoing  \citet{2007ApJ...663..948G},  all spectroscopically characterized  MW satellites discovered in this regime are clearly dwarf galaxies.  Yet what recent spectroscopic observations -- including those presented here -- have made increasingly clear is that dwarf galaxies can be distinguished from their star cluster counterparts down to a factor of two smaller physical sizes \textit{based on their surface brightnesses}. More specifically,  \figref{landscape} shows that \textit{all satellites at $r_{1/2} > 15$~pc and $\mu_{V,1/2} > 26.5 \rm \ mag \ arcsec^{-2}$ that have been characterized spectroscopically are most likely dwarf galaxies, with at most one exception.} 
\par This latter claim relies heavily on recent observations that have clarified the nature of objects in this morphological regime. Of the 21 known MW satellites at $15\rm ~pc \leq r_{1/2} \leq 35$~pc and $\mu_{V,1/2} > 26.5 \rm \ mag \ arcsec^{-2}$, 14 now have published spectroscopic evidence suggesting dwarf galaxy classifications (Boötes II, Carina III, Cetus II, Draco~II, Horologium~I, Phoenix~II, Pictor II, Segue 1, Triangulum II, Tucana V, and Willman 1; \citealt{2014ApJ...786...74F,2015ApJ...814L...7K,2015ApJ...811...62K,2018ApJ...857..145L,2019A&A...623A.129F,2023ApJ...950..167B,2023ApJ...959..141W,2024ApJ...968...21H,2025OJAp....8E.112P,2026ApJ..1002...55C,2026arXiv260217652C}, plus the three satellites from this work) while a further six lack sufficient spectroscopic measurements (Horologium II, Pictor I, Phoenix III, Carina IV, Virgo I, and Virgo II; \citealt{2016ApJ...832...21H,2023ApJ...953....1C,2026ApJ..1000...46T}). The low surface brightnesses of these galaxies and galaxy candidates separate them from all confirmed MW globular clusters, including the most extended MW cluster discovered to date (Sagittarius II, at $r_{1/2} = 32$~pc; \citealt{2018ApJ...863...25M,2021MNRAS.503.2754L,2025ApJ...987..217Z}).
\par The final satellite in the critical $15\rm ~pc \leq r_{1/2} \leq 35$~pc size regime -- and the single possible case that challenges the empirical separation of ultra-faint dwarfs and extended clusters based on surface brightness  -- is the MW satellite Tucana III ($r_{1/2}= 30$~pc, $\mu_{V,1/2} \approx 29.7 \rm \, mag \, arcsec^{-2}$; \citealt{2018ApJ...863...25M}), shown as a pink cross in \figref{landscape}. At the time of discovery, Tucana~III was suspected to be a dwarf galaxy  based on its similar size and surface brightness to other known dwarfs \citep{2015ApJ...813..109D}. Subsequent chemical measurements \citep{2017ApJ...838...44H,2019ApJ...882..177M} and the detection of a tentative metallicity gradient \citep{2018ApJ...866...22L} appeared to support this scenario, though the system's significant $r$-process element enhancement complicated the interpretation of its full abundance pattern.   More recent re-interpretations of the same data, however, have suggested that the system is more likely to be a star cluster \citep{2025ApJ...987..217Z,2026ApJ...999L...8C} -- consistent with the initial velocity dispersion upper limit found for the system \citep{2017ApJ...838...11S}. If so, Tucana~III could suggest that the empirical separation of dwarfs/clusters highlighted here can be contaminated in rare cases by heavily disrupted star clusters.

\par While the case of Tucana~III remains unresolved, and while the classifications of some dwarf galaxy candidates have not yet been definitively confirmed (unfilled triangles in \figref{landscape}), it is clear that the simple combination of a size and surface brightness selection can isolate a pure sample of ultra-faint dwarfs that is free (or nearly free) of contamination from star clusters.  The value of this result is twofold. First, it provides an objective framework for assigning preliminary classifications for discoveries in current surveys and those  anticipated with \textit{Rubin}, \textit{Roman}, and \textit{Euclid}: satellites with $r_{1/2} > 15$~pc, $\mu_{V,1/2} > 26.5 \rm \, mag \, arcsec^{-2}$ are likely dwarf galaxies awaiting spectroscopic confirmation, those at $r_{1/2} > 15$~pc and $\mu_{V,1/2} < 26.5 \rm \, mag \, arcsec^{-2}$ are likely star clusters, and finally, satellites at $r_{1/2} < 15$~pc are the most ambiguous and cannot be classified even preliminarily based on their morphologies. Second, this result can be applied in the context of  abundance matching analyses that seek to infer the total MW satellite galaxy population and constrain the galaxy-halo connection. Recent state-of-the-art analyses have typically excluded satellites below a radius scale of $r_{1/2} \approx 10$--15~pc \citep[e.g.,][]{2020ApJ...893...48N,2026ApJ..1000...87T} to avoid contamination from star clusters. Our exploration here places this approach on a much firmer basis, but advocates for the additional inclusion of a surface-brightness threshold of $\mu_{V,1/2} > 26.5 \rm \ mag \ arcsec^{-2}$.

\par We stress that these conclusions are \textit{probabilistic empirical results}, and not physical predictions or exceptionless rules. Extended, low-surface-brightness clusters may yet exist in greater abundance than  currently recognized -- particularly in the outer Galactic halo where the sensitivity of wide-field surveys to diffuse systems is poor. Moreover, the classifications of the low-surface-brightness satellites at $r_{1/2} < 15$~pc -- the UFCSs -- remain poorly understood. Spectroscopy will therefore continue to play an essential role in determining reliable classifications for newly discovered satellites.

\section{Summary}
\label{sec:summary}
We have presented new multi-object Keck/DEIMOS and Magellan/IMACS spectroscopy  of stars in \booV, \lmi{}, and \dtwo{} -- three small, ultra-faint stellar systems in the MW halo with no prior multi-object spectroscopic measurements. From this data, our main findings are as follows: 
\begin{enumerate}
    \item Leveraging DELVE photometry, \textit{Gaia} astrometry, and our new radial velocity measurements, we identify \BooVNmemclear{}, \LeoMinorINmemclear{}, and \DELVEtwoNmemclear{} spectroscopic members in \booV, \lmi{}, and \dtwo{}, respectively. 

    \item We determine the velocity dispersion of each system, finding an upper limit of \BooVvdisp{} for \booV{} and a modestly significant resolved dispersion of \DELVEtwovdisp{} for \dtwo{} that implies a mass-to-light ratio of \DELVEtwomlratio{}. We place a non-constraining upper limit of \LeoMinorIvdisp{} for \lmi{}.

    \item We estimate the mean metallicity and metallicity dispersion of each satellite, finding that all three are very metal-poor ($-2.85 \leq \overline{\rm [Fe/H]} \leq -2.15$). \booV{} and \lmi{} exhibit clear, statistically significant metallicity spreads of \BooVfehdisp{} and \LeoMinorIfehdisp{}. For \dtwo{}, we find a modestly significant spread of \DELVEtwofehdisp{} across our sample of three members with metallicities.

    \item Based primarily on their metallicity spreads, we confidently classify \booV{} and \lmi{} as dwarf galaxies. \dtwo{}'s low metallicity, tentative velocity dispersion, and modestly significant metallicity dispersion strongly support a dwarf galaxy classification as well. 
    
    \item We integrate the orbits of the three satellites in a joint MW + LMC potential. Most notably, we confirm for the first time that \dtwo{} is a long-term bound satellite of the LMC, increasing the census of likely Magellanic satellite galaxies to 11. 
     
\end{enumerate}

Although we were ultimately able to classify \booV{}, \lmi{}, and \dtwo{}, our work emphasizes the significant challenge in making gold-standard \textit{kinematic} classifications for such faint satellites. Echoing other recent analyses of faint MW satellite galaxies, our classifications for the three targets rely nearly entirely on stellar metallicity measurements; this reflects the difficulty of robustly resolving velocity dispersions at the $\sim$1--3~\kms{} scale.  Efficiently achieving direct kinematic classifications will likely require the advent of new instrumentation capable of achieving $\ll 1$~\kms{} velocity precision for faint stars (e.g., the upcoming MOONS spectrograph at VLT or ViaSpec at MMT and Magellan; \citealt{2014SPIE.9147E..0NC,2026arXiv260618332T}) and/or the use of 30-m-class telescopes. \newline

\par In light of our three new satellite classifications and other recent spectroscopic studies from the literature, we also reviewed the empirical MW satellite classification landscape. We conclude that: 

\begin{enumerate}[label=\Alph*.]
    \item All spectroscopically characterized MW satellites with azimuthally averaged half-light radii $r_{1/2} > 35$~pc are dwarf galaxies.
    \item In the size regime $15\rm ~pc \leq r_{1/2} \leq 35\rm ~pc$,  every spectroscopically characterized satellite  with $\mu_{V,1/2} > 26.5 \rm \ mag \ arcsec^{-2}$ is likely a dwarf galaxy, while every satellite at $\mu_{V,1/2} < 26.5 \rm \ mag \ arcsec^{-2}$ is a globular cluster.  We identify at most one possible exception to this division (Tucana~III).
    \item At the smallest physical sizes ($r_{1/2} < 15\rm ~pc$) and low surface brightnesses ($\mu_{V,1/2} \gtrsim 24 \rm \ mag \ arcsec^{-2}$), satellite classifications are the most ambiguous and cannot be ascertained from morphology alone.
\end{enumerate}

Based on the semi-analytic models from \citet{2022MNRAS.516.3944M}, we find that nearly 50\% of the MW satellite galaxies that the Vera C. Rubin Observatory is predicted to discover will have half-light radii in the critical range $15~\rm \,pc \leq r_{1/2} \leq 35$~pc. We therefore anticipate that these criteria can provide valuable insights into the classifications of the deluge of satellites discovered by \textit{Rubin}, as well as those found by the \textit{Nancy Grace Roman Space Telescope} and \textit{Euclid}.

\section*{Data Availability}
\label{sec:dataavail}
This work is accompanied by a Zenodo repository that includes a machine-readable version of our final spectroscopic member catalogs: \url{https://doi.org/10.5281/zenodo.23045325}. A future update, anticipated at the time of journal publication, will include non-member catalogs and posterior chains from our analyses.

\section*{Acknowledgments}
\par We thank Jaclyn Jensen for helpful correspondence regarding the proper motion measurements for \lmi{} and \dtwo{} from \citet{2024MNRAS.527.4209J} and Simon Smith for providing timing information for the \citetalias{2023AJ....166...76S} GMOS velocity for \booV{}'s brightest star.

\par W.C. gratefully acknowledges support from a Gruber Science Fellowship at Yale University. This material is based upon work supported by the National Science Foundation Graduate
Research Fellowship Program under Grant No. DGE2139841. Any opinions, findings, and conclusions or recommendations expressed in this material are those of the author(s) and do not necessarily reflect the views of the National Science Foundation. D.J.S. acknowledges support from NSF grant AST-2508746.
\par The DELVE Collaboration gratefully acknowledges support from Fermilab LDRD L2019-011, the NASA Fermi Guest Investigator Program Cycle 9 No. 91201, and the National Science Foundation under Grant Nos. AST-2108168, AST-2108169, AST-2307126, and AST-2407526. This research is partially funded by a generous gift from Charles Simonyi to the NSF Division of Astronomical Sciences. The award is made in recognition of significant contributions to Rubin Observatory’s Legacy Survey of Space and Time.
\par Some of the data presented herein were obtained at Keck Observatory, which is a private 501(c)3 non-profit organization operated as a scientific partnership among the California Institute of Technology, the University of California, and the National Aeronautics and Space Administration. The Observatory was made possible by the generous financial support of the W. M. Keck Foundation. The authors wish to recognize and acknowledge the very significant cultural role and reverence that the summit of Maunakea has always had within the Native Hawaiian community. We are most fortunate to have the opportunity to conduct observations from this mountain. This research has also made use of the Keck Observatory Archive (KOA), which is operated by the W. M. Keck Observatory and the NASA Exoplanet Science Institute (NExScI), under contract with the National Aeronautics and Space Administration.

\par This paper includes data gathered with the 6.5-meter Magellan Telescopes located at Las Campanas Observatory, Chile. This work is based in part on Magellan Baade observations obtained as part of NOIRLab program 2023B-865674 (PI W. Cerny). 

\par This project used data obtained with the Dark Energy Camera, which was constructed by the Dark Energy Survey (DES) collaboration.
Funding for the DES Projects has been provided by 
the DOE and NSF (USA),   
MISE (Spain),   
STFC (UK), 
HEFCE (UK), 
NCSA (UIUC), 
KICP (U. Chicago), 
CCAPP (Ohio State), 
MIFPA (Texas A\&M University),  
CNPQ, 
FAPERJ, 
FINEP (Brazil), 
MINECO (Spain), 
DFG (Germany), 
and the collaborating institutions in the Dark Energy Survey, which are
Argonne Lab, 
UC Santa Cruz, 
University of Cambridge, 
CIEMAT-Madrid, 
University of Chicago, 
University College London, 
DES-Brazil Consortium, 
University of Edinburgh, 
ETH Z{\"u}rich, 
Fermilab, 
University of Illinois, 
ICE (IEEC-CSIC), 
IFAE Barcelona, 
Lawrence Berkeley Lab, 
LMU M{\"u}nchen, and the associated Excellence Cluster Universe, 
University of Michigan, 
NSF's National Optical-Infrared Astronomy Research Laboratory, 
University of Nottingham, 
Ohio State University, 
OzDES Membership Consortium,
University of Pennsylvania, 
University of Portsmouth, 
SLAC National Lab, 
Stanford University, 
University of Sussex, 
and Texas A\&M University.

Based on observations at NSF Cerro Tololo Inter-American Observatory, NSF NOIRLab (NOIRLab Prop. ID 2019A-0305; PI: Drlica-Wagner), which is managed by the Association of Universities for Research in Astronomy (AURA) under a cooperative agreement with the U.S. National Science Foundation.

This work has made use of data from the European Space Agency (ESA) mission {\it Gaia} (\url{https://www.cosmos.esa.int/gaia}), processed by the {\it Gaia} Data Processing and Analysis Consortium (DPAC, \url{https://www.cosmos.esa.int/web/gaia/dpac/consortium}).
Funding for the DPAC has been provided by national institutions, in particular the institutions participating in the {\it Gaia} Multilateral Agreement.

\par This research used data obtained with the Dark Energy Spectroscopic Instrument (DESI). DESI construction and operations is managed by the Lawrence Berkeley National Laboratory. This material is based upon work supported by the U.S. Department of Energy, Office of Science, Office of High-Energy Physics, under Contract No. DE–AC02–05CH11231, and by the National Energy Research Scientific Computing Center, a DOE Office of Science User Facility under the same contract. Additional support for DESI was provided by the U.S. National Science Foundation (NSF), Division of Astronomical Sciences under Contract No. AST-0950945 to the NSF’s National Optical-Infrared Astronomy Research Laboratory; the Science and Technology Facilities Council of the United Kingdom; the Gordon and Betty Moore Foundation; the Heising-Simons Foundation; the French Alternative Energies and Atomic Energy Commission (CEA); the National Council of Humanities, Science and Technology of Mexico (CONAHCYT); the Ministry of Science and Innovation of Spain (MICINN), and by the DESI Member Institutions: www.desi.lbl.gov/collaborating-institutions. The DESI collaboration is honored to be permitted to conduct scientific research on I’oligam Du’ag (Kitt Peak), a mountain with particular significance to the Tohono O’odham Nation. Any opinions, findings, and conclusions or recommendations expressed in this material are those of the author(s) and do not necessarily reflect the views of the U.S. National Science Foundation, the U.S. Department of Energy, or any of the listed funding agencies.

\par Guoshoujing Telescope (the Large Sky Area Multi-Object Fiber Spectroscopic Telescope LAMOST) is a National Major Scientific Project built by the Chinese Academy of Sciences. Funding for the project has been provided by the National Development and Reform Commission. LAMOST is operated and managed by the National Astronomical Observatories, Chinese Academy of Sciences.

\par This work made use of Astropy:\footnote{\url{http://www.astropy.org}} a community-developed core Python package and an ecosystem of tools and resources for astronomy.

\par This work has made use of the Local Volume Database \citep{2025OJAp....8E.142P}.\footnote{\url{https://github.com/apace7/local_volume_database}}

\par This research uses services or data provided by the Astro Data Lab \citep{2014SPIE.9149E..1TF,2020A&C....3300411N}, which is part of the Community Science and Data Center (CSDC) Program of NSF NOIRLab. NOIRLab is operated by the Association of Universities for Research in Astronomy (AURA), Inc. under a cooperative agreement with the U.S. National Science Foundation.
\par Contour plots were generated using corner.py \citep{2016JOSS....1...24F}.

\par Generative AI Usage Disclosure: the scientific conceptualization, data collection, analysis design, methodological decisions, interpretation of the results, and manuscript writing presented herein were conducted exclusively by the authors. Anthropic's \texttt{Claude Code} (models Opus 4.6--5.5 and Sonnet 4.6--5.5) was used to (1) develop and refine analysis code for this work, including the code used to generate manuscript figures and tables, (2) scan for typos, grammatical errors, and numerical inconsistencies in the manuscript text, (3) provide minor suggestions to enhance the clarity of existing text, which were adopted or rejected at the authors' discretion, and (4) prepare member tables and associated documentation for public release on Zenodo. No original manuscript text was generated by a large language model (LLM). All AI-assisted code contributions were reviewed and vetted for accuracy, and all scientific content and manuscript text are the sole responsibility of the authors. \\

 \large{\textit{Facilities:}} \normalsize
\facility{Blanco, \Gaia, Keck:II (DEIMOS), Magellan:Baade (IMACS)} 

\software{\code{numpy} \citep{2011CSE....13b..22V,2020Natur.585..357H}, \code{scipy} \citep{2020NatMe..17..261V}, \texttt{dynesty} \citep{2020MNRAS.493.3132S,2023zndo...3348367K}, \texttt{matplotlib} \citep{2007CSE.....9...90H}, \texttt{pandas} \citep{2024zndo..10957263T}, \texttt{astropy} \citep{2013A&A...558A..33A},   \texttt{emcee} \citep{2013PASP..125..306F}, \texttt{corner.py} \citep{2016JOSS....1...24F}, \texttt{AGAMA} \citep{2019MNRAS.482.1525V}}

\bibliography{main}{}
\bibliographystyle{aasjournal}


\appendix



\section{Member tables for \booV{}, \lmi{}, and \dtwo{}}

In Tables \ref{tab:members_Bootes_V}--\ref{tab:members_DELVE_2} below, we present catalogs of member stars in the three satellites.

\input{AllResultsSeptember/AllTablesSeptember/T3_MembersTable_BootesV_September}   
\input{AllResultsSeptember/AllTablesSeptember/T4_SupplementalTable_BootesV_September}

\input{AllResultsSeptember/AllTablesSeptember/T5_MembersTable_LeoMinorI_September}   
\input{AllResultsSeptember/AllTablesSeptember/T6_MembersTable_DELVE2_September}

\FloatBarrier


\section{References for Literature Comparison Samples}
\label{sec:datareferences}
In \figref{sizeluminosity}, \figref{masses}, and \figref{metallicities}, we compare \booV{}, \lmi{}, and \dtwo{} to the population of MW globular clusters, candidate and confirmed dwarf galaxies, and ultra-faint compact satellites. For the globular clusters and dwarfs, we adopted the measurements compiled in the Local Volume Database version 1.1.0 (LVDB; \citealt{2025OJAp....8E.142P}). For the ultra-faint compact satellites, properties were taken from the compilation and new spectroscopic measurements presented in \citet{2026arXiv260217652C}. Nearly all of these references are listed in Appendix B of \citet{2026arXiv260217652C}, which includes all references for the satellite samples dating to LVDB version 1.0.4. Measurements in LVDB 1.1.0 that are included in this work but not cited in the aforementioned Appendix include \citet{2009A&A...506..729K}, \citet{2018A&A...620A..96M}, \citet{2019MNRAS.490..741S}, \citet{2020MNRAS.492.1061V}, \citet{2021MNRAS.503.4336M}, \citet{2022MNRAS.514.1706B}, \citet{2023ApJ...951...17C}, \citet{2025A&A...694A.248L}, \citet{2025AJ....170...37B}, \citet{2025OJAp....8E..86U}, \citet{2025A&A...700A.201M}, \citet{2025OJAp....8E.112P}, \citet{2025A&A...703A.267G}, \citet{2026ApJ...998...47S}, \citet{2026arXiv260210202G}, \citet{2026A&A...706A.338G}, \citet{2026ApJ...999..140G}, \citet{2026A&A...707A..83P}, \citet{2025arXiv251201530R}, \citet{2026ApJ..1002...55C}, \citet{2026ApJ..1000...46T}, and \citet{2026arXiv260304507C}.
\par

\end{document}

%% file: commands.tex
\usepackage{amsmath}
\usepackage{amssymb}
\usepackage{xspace}
\usepackage{xifthen}
\usepackage{eso-pic}

\newcommand{\Gaia}{{\it Gaia}\xspace}

\definecolor{forestgreen}{HTML}{228B22}
\definecolor{urlblue}{HTML}{000000}

\mathchardef\mhyphen="2D

\newlength{\dhatheight}

\newcommand{\code}[1]{\texttt{#1}\xspace}

\newcommand{\unit}[1]{\ensuremath{\mathrm{\,#1}}\xspace}

\newcommand{\Gyr}{\unit{Gyr}}

\newcommand{\degree}{\ensuremath{{}^{\circ}}\xspace}

\newcommand{\km}{\unit{km}}

\newcommand{\secref}[1]{Section~\ref{sec:#1}}

\newcommand{\tabref}[1]{Table~\ref{tab:#1}}

\newcommand{\figref}[1]{Figure~\ref{fig:#1}}

\newcommand{\bandvar}[2][]{%
  \ifthenelse{\isempty{#1}}{\var{#2}}{\var{#2\_#1}}%
}

\newcommand{\var}[1]{\ensuremath{\texttt{\MakeUppercase{#1}}}\xspace}

\providecommand\physrep{\ref@jnl{Phys.~Rep.}}%
\providecommand\apjs{\ref@jnl{ApJS}}%
\providecommand{\jcap}{\ref@jnl{JCAP}}%

\newcommand\kms{\mbox{km\,s$^{\rm -1}$}}
\newcommand{\MLunit}{\ensuremath{\mathrm{M_{\odot}}\,\mathrm{L_{\odot}^{-1}}}}
\newcommand{\masyr}{\ensuremath{\rm mas\,yr^{-1}}\xspace}

%% file: AllResultsSeptember/AllTablesSeptember/CommandList_September.tex
\newcommand{\BooVNunique}{71\xspace}
\newcommand{\BooVNpassesinitial}{40\xspace}
\newcommand{\BooVNpassesInitialAndIso}{32\xspace}
\newcommand{\BooVNpassesbothiterative}{16\xspace}

\newcommand{\BooVNmem}{14\xspace}
\newcommand{\BooVNmemclear}{13\xspace}

\newcommand{\BooVNkin}{$N_{\rm kin}$ = 13\xspace}

\newcommand{\BooVNpropermotion}{7\xspace}

\newcommand{\BooVNfehnumberonly}{9\xspace}

\newcommand{\BooVvdisp}{\ensuremath{\sigma_v <\! 2.7~\kms}\xspace}
\newcommand{\BooVmlratio}{\ensuremath{M_{1/2}/L_{V,1/2} <\! 180~\MLunit}\xspace}
\newcommand{\BooVjfactor}{\ensuremath{\log_{10} J(0.5^\circ) <\! 17.52~\log_{10}(\rm GeV^2\,cm^{-5})}\xspace}
\newcommand{\BooVvdispbayes}{\ensuremath{2\ln\mathcal{B} = -2.3}\xspace}

\newcommand{\BooVpmranumberonly}{-0.23^{+0.04}_{-0.04}\xspace}
\newcommand{\BooVpmdecnumberonly}{-0.29^{+0.07}_{-0.07}\xspace}

\newcommand{\BooVitermeanvhel}{-9.3~\kms\xspace}
\newcommand{\BooVitermeanpmra}{-0.22~\masyr\xspace}
\newcommand{\BooVitermeanpmdec}{-0.29~\masyr\xspace}

\newcommand{\BooVvdispLog}{\ensuremath{\sigma_v <\! 2.0~\kms}\xspace}

\newcommand{\BooVfehbrightest}{[Fe/H]$_{\rm brightest} = -2.70^{+0.15}_{-0.15}$\xspace}

\newcommand{\BooVfehrange}{$\rm Range([Fe/H]) = 1.25 \pm 0.29$~dex\xspace}
\newcommand{\BooVfehmodel}{\ensuremath{\overline{\rm [Fe/H]} = -2.51^{+0.17}_{-0.17}}\xspace}
\newcommand{\BooVfehdisp}{$\sigma_{\rm [Fe/H]} = 0.46^{+0.18}_{-0.12}$~dex\xspace}
\newcommand{\BooVfehdispbayes}{\ensuremath{2\ln\mathcal{B} = 22.8}\xspace}
\newcommand{\BooVmetallicitydispersionsignificance}{\ensuremath{4.8\sigma}\xspace}
\newcommand{\BooVfehdispfive}{$\sigma_{\rm [Fe/H]} >\! 0.28$~dex\xspace}
\newcommand{\BooVfehdispExclMR}{$\sigma_{\rm [Fe/H]} = 0.39^{+0.17}_{-0.12}$~dex\xspace}

\newcommand{\BooVrperiproximateAgama}{$r_{\rm peri} = 23^{+12}_{-10}$~kpc\xspace}

\newcommand{\BooVrapoproximateAgama}{$r_{\rm apo} = 111^{+8}_{-9}$~kpc\xspace}

\newcommand{\BooVeccproximateAgama}{\ensuremath{e = 0.66^{+0.13}_{-0.14}}\xspace}

\newcommand{\BooVLz}{\ensuremath{L_z = -0.1^{+0.9}_{-1.0}\times 10^3~\rm kpc\,km\,s^{-1}}\xspace}

\newcommand{\LeoMinorINunique}{50\xspace}

\newcommand{\LeoMinorINpassesInitialAndIso}{26\xspace}
\newcommand{\LeoMinorINpassesbothiterative}{7\xspace}

\newcommand{\LeoMinorINmem}{7\xspace}
\newcommand{\LeoMinorINmemclear}{7\xspace}
\newcommand{\LeoMinorINmemclearWord}{seven\xspace}

\newcommand{\LeoMinorINkin}{$N_{\rm kin}$ = 4\xspace}

\newcommand{\LeoMinorINpropermotion}{3\xspace}
\newcommand{\LeoMinorINfeh}{$N_{\rm [Fe/H]}$ = 5\xspace}
\newcommand{\LeoMinorINfehnumberonly}{5\xspace}

\newcommand{\LeoMinorIvdisp}{\ensuremath{\sigma_v <\! 7.8~\kms}\xspace}
\newcommand{\LeoMinorImlratio}{\ensuremath{M_{1/2}/L_{V,1/2} <\! 4530~\MLunit}\xspace}
\newcommand{\LeoMinorIjfactor}{\ensuremath{\log_{10} J(0.5^\circ) <\! 19.48~\log_{10}(\rm GeV^2\,cm^{-5})}\xspace}
\newcommand{\LeoMinorIvdispbayes}{\ensuremath{2\ln\mathcal{B} = 0.0}\xspace}

\newcommand{\LeoMinorIpmranumberonly}{0.01^{+0.37}_{-0.37}\xspace}
\newcommand{\LeoMinorIpmdecnumberonly}{-1.29^{+0.35}_{-0.35}\xspace}

\newcommand{\LeoMinorIitermeanvhel}{3.5~\kms\xspace}
\newcommand{\LeoMinorIitermeanpmra}{-0.01~\masyr\xspace}
\newcommand{\LeoMinorIitermeanpmdec}{-1.29~\masyr\xspace}

\newcommand{\LeoMinorIvdispLog}{\ensuremath{\sigma_v <\! 5.1~\kms}\xspace}

\newcommand{\LeoMinorIfehbrightest}{[Fe/H]$_{\rm brightest} = -1.52^{+0.14}_{-0.14}$\xspace}

\newcommand{\LeoMinorIfehrange}{$\rm Range([Fe/H]) = 1.22 \pm 0.26$~dex\xspace}
\newcommand{\LeoMinorIfehmodel}{\ensuremath{\overline{\rm [Fe/H]} = -2.15^{+0.27}_{-0.27}}\xspace}
\newcommand{\LeoMinorIfehdisp}{$\sigma_{\rm [Fe/H]} = 0.57^{+0.23}_{-0.18}$~dex\xspace}
\newcommand{\LeoMinorIfehdispbayes}{\ensuremath{2\ln\mathcal{B} = 19.5}\xspace}
\newcommand{\LeoMinorImetallicitydispersionsignificance}{\ensuremath{4.4\sigma}\xspace}
\newcommand{\LeoMinorIfehdispfive}{$\sigma_{\rm [Fe/H]} >\! 0.31$~dex\xspace}
\newcommand{\LeoMinorIfehdispExclMR}{$\sigma_{\rm [Fe/H]} = 0.48^{+0.28}_{-0.22}$~dex\xspace}

\newcommand{\LeoMinorIrperiproximateAgama}{$r_{\rm peri} = 79^{+7}_{-28}$~kpc\xspace}

\newcommand{\LeoMinorILz}{\ensuremath{L_z = 10.5^{+6.2}_{-5.7}\times 10^3~\rm kpc\,km\,s^{-1}}\xspace}

\newcommand{\DELVEtwoNunique}{29\xspace}

\newcommand{\DELVEtwoNpassesInitialAndIso}{25\xspace}
\newcommand{\DELVEtwoNpassesbothiterative}{9\xspace}

\newcommand{\DELVEtwoNmem}{7\xspace}
\newcommand{\DELVEtwoNmemclear}{7\xspace}
\newcommand{\DELVEtwoNmemclearWord}{seven\xspace}
\newcommand{\DELVEtwoNmemclearWordUppercase}{Seven\xspace}
\newcommand{\DELVEtwoNkin}{$N_{\rm kin}$ = 6\xspace}

\newcommand{\DELVEtwoNpropermotion}{6\xspace}

\newcommand{\DELVEtwoNfehnumberonly}{3\xspace}

\newcommand{\DELVEtwomeanvhel}{\ensuremath{\overline{v_{\rm hel}} = 129.9^{+1.1}_{-1.2}~\kms}\xspace}
\newcommand{\DELVEtwomeanvgsr}{\ensuremath{\overline{v_{\rm GSR}} = -22.8^{+1.1}_{-1.2}~\kms}\xspace}

\newcommand{\DELVEtwovdisp}{\ensuremath{\sigma_v = 2.1^{+1.5}_{-0.9}~\kms}\xspace}
\newcommand{\DELVEtwomlratio}{\ensuremath{M_{1/2}/L_{V,1/2} = 280^{+640}_{-190}~\MLunit}\xspace}
\newcommand{\DELVEtwojfactor}{\ensuremath{\log_{10} J(0.5^\circ) = 17.34^{+0.95}_{-1.00}~\log_{10}(\rm GeV^2\,cm^{-5})}\xspace}
\newcommand{\DELVEtwovdispbayes}{\ensuremath{2\ln\mathcal{B} = 2.9}\xspace}

\newcommand{\DELVEtwopmranumberonly}{0.97^{+0.12}_{-0.12}\xspace}
\newcommand{\DELVEtwopmdecnumberonly}{-0.95^{+0.09}_{-0.09}\xspace}

\newcommand{\DELVEtwoitermeanvhel}{130.6~\kms\xspace}
\newcommand{\DELVEtwoitermeanpmra}{0.97~\masyr\xspace}
\newcommand{\DELVEtwoitermeanpmdec}{-0.95~\masyr\xspace}

\newcommand{\DELVEtwovdispLog}{\ensuremath{\sigma_v <\! 3.8~\kms}\xspace}

\newcommand{\DELVEtwofehrange}{$\rm Range([Fe/H]) = 0.98^{+0.78}_{-0.49}$~dex\xspace}
\newcommand{\DELVEtwofehmodel}{\ensuremath{\overline{\rm [Fe/H]} = -2.85^{+0.31}_{-0.39}}\xspace}
\newcommand{\DELVEtwofehdisp}{$\sigma_{\rm [Fe/H]} = 0.52^{+0.29}_{-0.25}$~dex\xspace}
\newcommand{\DELVEtwofehdispbayes}{\ensuremath{2\ln\mathcal{B} = 3.3}\xspace}

\newcommand{\DELVEtwofehdispfive}{$\sigma_{\rm [Fe/H]} >\! 0.15$~dex\xspace}

\newcommand{\DELVEtworperiproximateAgama}{$r_{\rm peri} = 68^{+4}_{-5}$~kpc\xspace}

\newcommand{\DELVEtwofboundLMCnow}{99\%\xspace}
\newcommand{\DELVEtwofboundLMCpast}{93\%\xspace}

\newcommand{\DELVEtwoLx}{\ensuremath{L_x = -18.3^{+3.2}_{-3.6}\times 10^3~\rm kpc\,km\,s^{-1}}\xspace}

\newcommand{\BooVvdispwithDistantStars}{\ensuremath{\sigma_v = 3.2^{+1.2}_{-1.0}~\kms}\xspace}

\newcommand{\BooVvdispinnerOnlyStars}{\ensuremath{\sigma_v <\! 3.6~\kms}\xspace}

\newcommand{\BooVvdispstricterBinary}{\ensuremath{\sigma_v <\! 2.6~\kms}\xspace}
\newcommand{\BooVvdispbayesstricterBinary}{\ensuremath{2\ln\mathcal{B} = -3.5}\xspace}

\newcommand{\DELVEtwovdispwithDistantStars}{\ensuremath{\sigma_v = 5.8^{+1.9}_{-1.4}~\kms}\xspace}

\newcommand{\DELVEtwovdispstricterBinary}{\ensuremath{\sigma_v <\! 7.0~\kms}\xspace}
\newcommand{\DELVEtwovdispbayesstricterBinary}{\ensuremath{2\ln\mathcal{B} = -1.5}\xspace}

\newcommand{\LeoMinorIvdispstricterBinary}{\ensuremath{\sigma_v = 3.6^{+3.3}_{-2.0}~\kms}\xspace}
\newcommand{\LeoMinorIvdispbayesstricterBinary}{\ensuremath{2\ln\mathcal{B} = 0.6}\xspace}


%% file: AllResultsSeptember/AllTablesSeptember/T1_ObsTable_September.tex
\begin{table*}[htbp]
\centering
\caption{Keck and Magellan spectroscopic observations of Leo Minor~I, Boötes~V, and DELVE~2.}
\label{tab:obstable}
\small
\newsavebox{\obstblbox}
\sbox{\obstblbox}{%
\begin{tabular}{lllllll}
\hline\hline
\multicolumn{1}{c}{Satellite} & \multicolumn{1}{c}{Instrument} & \multicolumn{1}{c}{Maskname} & \multicolumn{1}{c}{Local Date$^{a}$} & \multicolumn{1}{c}{MJD$^{a}$} & \multicolumn{1}{c}{$t_\mathrm{exp}$ (s)} & \multicolumn{1}{c}{$N_*$$^{c}$} \\
\midrule
Leo Minor I & DEIMOS & \texttt{LeoMinorI\_mask1} & 2022 June 4, 5 & 59735.53 & 6000 & 24 \\
 &  & \texttt{LeoMinorI\_mask2} & 2023 April 23 & 60058.37 & 7200 & 32 \\
\midrule
Bo\"{o}tes V & DEIMOS & \texttt{BootesV\_mask1} & 2023 April 23 & 60058.49 & 8400 & 40 \\
 &  & \texttt{BootesV\_mask2} & 2023 June 11 & 60107.29 & 8400 & 39 \\
\midrule
DELVE 2 & IMACS & \texttt{DELVE2\_mask1} & 2021 July 10, 11, 13 & 59407.53 & 27000$^{b}$ & 13 \\
 &  &  & 2021 October 9 & 59497.13 & 4800 & 5 \\
 &  &  & 2022 October 28, 31 & 59883.06 & 15600 & 18 \\
 &  &  & 2023 October 4 & 60222.13 & 12000$^{b}$ & 9 \\
 &  &  & 2024 December 11 & 60656.16 & 13020 & 27 \\
\bottomrule
\end{tabular}%
}
\usebox{\obstblbox}
\par\vspace{4pt}
\begin{minipage}{\wd\obstblbox}
\raggedright
\footnotesize
\noindent$^{a}$ The listed dates correspond to the start of each observing night in local time, and the listed Modified Julian Date (MJD) corresponds to the mean across exposures. For multi-night runs, this mean can fall during the daytime.
\par
\noindent$^{b}$ One exposure was excluded from our reductions due to very poor seeing (2021 July run) and two were excluded due to elevated pattern noise (2023 October run); these exposures are not included in the listed exposure times.
\par
\noindent$^{c}$ $N_{*}$ is the number of unique stars with spectra that yielded useful radial-velocity measurements.
\end{minipage}
\end{table*}

%% file: AllResultsSeptember/AllTablesSeptember/T2_PropertiesTable_September.tex
\begin{table*}[htbp]
\centering
\caption{Morphological, Kinematic, and Chemical Properties of the Three Satellites}
\label{tab:propertiestable}
\small
\newsavebox{\ptblbox}
\sbox{\ptblbox}{%
\begin{tabular}{l|l|ccc|l}
\hline\hline
\multicolumn{1}{c}{Property} & \multicolumn{1}{c}{Description} & \multicolumn{1}{c}{Bo\"otes V} & \multicolumn{1}{c}{Leo Minor I} & \multicolumn{1}{c}{DELVE 2} & \multicolumn{1}{c}{Units} \\
\hline
$\alpha_{2000}$ & Right Ascension (J2000) & $213.909^{+0.001}_{-0.001}$ & $164.261^{+0.005}_{-0.006}$ & $28.772^{+0.006}_{-0.005}$ & deg \\
$\delta_{2000}$ & Declination (J2000) & $32.914^{+0.001}_{-0.001}$ & $28.875^{+0.004}_{-0.003}$ & $-68.253^{+0.002}_{-0.002}$ & deg \\
$a_{1/2}$ & Angular semi-major axis of the half-light ellipse & $0.8^{+0.1}_{-0.1}$ & $1.1^{+0.4}_{-0.3}$ & $1.0^{+0.2}_{-0.2}$ & arcmin \\
$r_{1/2}$ & Azimuthally-averaged half-light radius & $20^{+2}_{-2}$ & $26^{+9}_{-9}$ & $21^{+4}_{-3}$ & pc \\
$D_\odot$ & Heliocentric distance & $102^{+7}_{-7}$ & $82^{+4}_{-7}$ & $71^{+4}_{-4}$ & kpc \\
$(m-M)_0$ & Distance modulus & $20.04^{+0.10}_{-0.10}$ & $19.56^{+0.11}_{-0.19}$ & $19.26^{+0.10}_{-0.10}$ & mag \\
$M_V$ & Absolute $V$-band magnitude & $-3.2^{+0.3}_{-0.3}$ & $-2.4^{+0.5}_{-0.4}$ & $-2.1^{+0.4}_{-0.5}$ & mag \\
$L_V$ & Luminosity in the $V$ band & $1630^{+510}_{-390}$ & $740^{+350}_{-260}$ & $620^{+340}_{-200}$ & $L_{\odot}$ \\
$M_*$ & Stellar mass (assuming $M_*/L_V = 2$) & $3260^{+1030}_{-780}$ & $1480^{+710}_{-520}$ & $1240^{+680}_{-400}$ & $M_{\odot}$ \\
\hline
$N_{\rm mem}$ & Number of spectroscopic members & 13\textsuperscript{a} & 7 & 7 & stars \\
$N_{\rm kin}$ & Number of kinematic members used for $\sigma_v$ fit & 13 & 4 & 6 & stars \\
$\overline{v_{\rm hel}}$ & Mean heliocentric radial velocity & $-9.7^{+0.7}_{-0.7}$ & $4.1^{+1.7}_{-1.7}$ & $129.9^{+1.1}_{-1.2}$ & $\kms{}$ \\
$\sigma_v$ & Velocity Dispersion & $<\! 2.7$ & $<\! 7.8$\textsuperscript{b} & $2.1^{+1.5}_{-0.9}$ & $\kms{}$ \\
$M_{1/2}$ & Dynamical mass within $r_{\rm 1/2,3D}$ (Wolf estimator) & $<\! 1.4$ & $<\! 14.8$\textsuperscript{b} & $0.9^{+1.8}_{-0.6}$ & $10^5\,M_{\odot}$ \\
$M_{1/2}/L_{V,1/2}$ & Mass-to-light ratio within $r_{1/2}$ & $<\! 180$ & $<\! 4530$\textsuperscript{b} & $280^{+640}_{-190}$ & $M_{\odot} \ L_{\odot}^{-1}$ \\
$M_{<1.8r_{1/2}}$ & Dynamical mass within $1.8\,r_{1/2}$ (Errani estimator) & $<\! 2.2$ & $<\! 23.1$\textsuperscript{b} & $1.4^{+2.8}_{-1.0}$ & $10^5\,M_{\odot}$ \\
\hline
$\log_{10} J_{0.5}$ & J-factor within $0.5^\circ$ & $<\! 17.52$ & $<\! 19.48$ & $17.34^{+0.95}_{-1.00}$ & $\log_{10}(\rm GeV^2\ cm^{-5})$ \\
$N_{\rm [Fe/H]}$ & Number of stars with metallicity measurements & 9 & 5 & 3 & stars \\
$\overline{\rm [Fe/H]}$ & Mean metallicity & $-2.51^{+0.17}_{-0.17}$ & $-2.15^{+0.27}_{-0.27}$ & $-2.85^{+0.31}_{-0.39}$ &  \\
$\sigma_{\rm [Fe/H]}$ & Metallicity dispersion & $0.46^{+0.18}_{-0.12}$ & $0.57^{+0.23}_{-0.18}$ & $0.52^{+0.29}_{-0.25}$ & dex \\
${\rm [Fe/H]}_{\rm brightest}$ & [Fe/H] of the brightest member & $-2.70^{+0.15}_{-0.15}$ & $-1.52^{+0.14}_{-0.14}$ & $-2.49^{+0.09}_{-0.09}$ &  \\
$N_{\rm 6D}$ & Number of RV-confirmed proper motion members & 7 & 3 & 6 & stars \\
\hline
$\overline{\mu_{\alpha*}}$ & Mean proper motion in R.A. & $-0.23^{+0.04}_{-0.04}$ & $0.01^{+0.37}_{-0.37}$ & $0.97^{+0.12}_{-0.12}$ & $\rm mas \ yr^{-1}$ \\
$\overline{\mu_{\delta}}$ & Mean proper motion in decl. & $-0.29^{+0.07}_{-0.07}$ & $-1.29^{+0.35}_{-0.35}$ & $-0.95^{+0.09}_{-0.09}$ & $\rm mas \ yr^{-1}$ \\
$\xi_{\mu}$ & Correlation coefficient between $\mu_{\alpha*}$ and $\mu_{\delta}$ & $0.17$ & $-0.40$ & $0.22$ &  \\
\hline
$R_{\rm gal}$ & Galactocentric distance & $101^{+7}_{-8}$ & $83^{+5}_{-6}$ & $69^{+4}_{-4}$ & kpc \\
$E_{\rm tot}$ & Total orbital energy & $-0.45^{+0.02}_{-0.02}$ & $-0.1^{+0.6}_{-0.3}$ & $-0.2^{+0.1}_{-0.1}$ & $10^5\ \rm km^2\ s^{-2}$ \\
$L_z$ & z-component of angular momentum & $-0.1^{+0.9}_{-1.0}$ & $10.5^{+6.2}_{-5.7}$ & $-2.0^{+1.9}_{-2.1}$ & $10^3\ \rm kpc\ km\ s^{-1}$ \\
$r_{\rm apo}$ & Orbital apocenter closest in time & $111^{+8}_{-9}$ & \ldots\textsuperscript{c} & \ldots\textsuperscript{c} & kpc \\
$r_{\rm peri}$ & Orbital pericenter closest in time & $23^{+12}_{-10}$ & $79^{+7}_{-28}$ & $68^{+4}_{-5}$ & kpc \\
$e$ & Orbital eccentricity & $0.66^{+0.13}_{-0.14}$ & \ldots\textsuperscript{c} & \ldots\textsuperscript{c} &  \\
$f_{\rm bound}$ & Fraction of orbit realizations that are bound ($E_{\rm tot}<0$) & $1.00$ & $0.59$ & $0.96$ &  \\
\hline
\end{tabular}%
}
\ifdim\wd\ptblbox>\textwidth
\resizebox{\textwidth}{!}{\usebox{\ptblbox}}
\setlength{\wd\ptblbox}{\textwidth}
\else
\usebox{\ptblbox}
\fi
\par\vspace{4pt}
\begin{minipage}{\wd\ptblbox}
\raggedright\footnotesize
\textit{Note} --- Quantities above the first horizontal divider were taken from \citetalias{2023ApJ...953....1C} for \booV{} and \lmi{} and \citetalias{2021ApJ...910...18C} for \dtwo{}.\par
$^{a}$ For \booV{}, we identified \BooVNmemclear{} clear members as well as one candidate member binary.\par
$^{b}$ \lmi{}'s velocity dispersion, dynamical mass, and mass-to-light ratio are poorly constrained due to the small sample size; use with caution.\par
$^{c}$ \lmi{} and \dtwo{} are likely on first infall and have no well-defined apocenter and eccentricity.
\end{minipage}
\end{table*}

%% file: AllResultsSeptember/AllTablesSeptember/T3_MembersTable_BootesV_September.tex
\begin{table}[htbp]
\centering
\caption{Spectroscopic Members of Bo\"{o}tes V}
\label{tab:members_Bootes_V}
\scriptsize
\setlength{\tabcolsep}{4pt}

\newsavebox{\memboxBooV}\savebox{\memboxBooV}{%
\begin{tabular}{lrrrrrrrrrrrr}
\hline\hline
\multicolumn{1}{c}{Name} & \multicolumn{1}{c}{RA} & \multicolumn{1}{c}{DEC} & \multicolumn{1}{c}{Gaia Source ID} & \multicolumn{1}{c}{$g_0$} & \multicolumn{1}{c}{Type} & \multicolumn{1}{c}{MJD} & \multicolumn{1}{c}{S/N} & \multicolumn{1}{c}{$v_{\rm hel}$} & \multicolumn{1}{c}{$\sum \rm EW_{\rm CaT}$} & \multicolumn{1}{c}{[Fe/H]} & \multicolumn{1}{c}{$p_{\rm const}$} & \multicolumn{1}{c}{Comment} \\
\multicolumn{1}{c}{} & \multicolumn{1}{c}{(deg)} & \multicolumn{1}{c}{(deg)} & \multicolumn{1}{c}{} & \multicolumn{1}{c}{(mag)} & \multicolumn{1}{c}{} & \multicolumn{1}{c}{(d)} & \multicolumn{1}{c}{(pixel$^{-1}$)} & \multicolumn{1}{c}{(\kms)} & \multicolumn{1}{c}{(\AA)} & \multicolumn{1}{c}{} & \multicolumn{1}{c}{} & \multicolumn{1}{c}{} \\
\midrule
\texttt{BooV\_1} & 213.912 & +32.914 & 1478121382994560256 & 18.14 & RGB & 60058.49 & 150.0 & $-8.1 \pm 1.1$ & $2.93 \pm 0.08$ & \multicolumn{1}{c}{\ldots} & \multicolumn{1}{c}{\ldots} & \multicolumn{1}{c}{\ldots} \\
 & & & & & & 60107.29 & 145.3 & $-8.2 \pm 1.1$ & $2.85 \pm 0.08$ & \multicolumn{1}{c}{\ldots} & \multicolumn{1}{c}{\ldots} & \multicolumn{1}{c}{\ldots} \\
 & & & & & & \textit{Combined} & \multicolumn{1}{c}{\ldots} & $-8.2 \pm 0.8$ & $2.89 \pm 0.07$ & $-2.70 \pm 0.15$ & 0.961 & Binary?$^{a}$ \\
\hline
\texttt{BooV\_2} & 213.931 & +32.940 & 1478121722292128768 & 20.48 & BHB & 60058.49 & 18.2 & $-10.1 \pm 3.4$ & \multicolumn{1}{c}{\ldots} & \multicolumn{1}{c}{\ldots} & \multicolumn{1}{c}{\ldots} & \multicolumn{1}{c}{\ldots} \\
 & & & & & & 60107.29 & 16.1 & $-13.3 \pm 4.8$ & \multicolumn{1}{c}{\ldots} & \multicolumn{1}{c}{\ldots} & \multicolumn{1}{c}{\ldots} & \multicolumn{1}{c}{\ldots} \\
 & & & & & & \textit{Combined} & \multicolumn{1}{c}{\ldots} & $-11.2 \pm 2.8$ & \multicolumn{1}{c}{\ldots} & \multicolumn{1}{c}{\ldots} & 0.595 & \multicolumn{1}{c}{\ldots} \\
\hline
\texttt{BooV\_3} & 213.907 & +32.912 & 1478121584853173888 & 20.49 & BHB & 60107.29 & 15.9 & $-9.1 \pm 3.0$ & \multicolumn{1}{c}{\ldots} & \multicolumn{1}{c}{\ldots} & \multicolumn{1}{c}{\ldots} & \multicolumn{1}{c}{\ldots} \\
\hline
\texttt{BooV\_4} & 213.910 & +32.917 & 1478121584853174144 & 20.75 & BHB & 60058.49 & 11.1 & $-11.6 \pm 5.9$ & \multicolumn{1}{c}{\ldots} & \multicolumn{1}{c}{\ldots} & \multicolumn{1}{c}{\ldots} & \multicolumn{1}{c}{\ldots} \\
\hline
\texttt{BooV\_5} & 213.910 & +32.907 & 1478121378694742912 & 20.87 & RGB & 60058.49 & 25.6 & $-9.6 \pm 1.6$ & $1.74 \pm 0.21$ & $-2.90 \pm 0.20$ & \multicolumn{1}{c}{\ldots} & \multicolumn{1}{c}{\ldots} \\
\hline
\texttt{BooV\_6} & 214.003 & +32.946 & 1478126743113866624 & 20.92 & RGB & 60107.29 & 19.9 & $-9.6 \pm 2.6$ & $1.53 \pm 0.23$ & $-3.06 \pm 0.24$ & \multicolumn{1}{c}{\ldots} & \multicolumn{1}{c}{\ldots} \\
\hline
\texttt{BooV\_7} & 213.903 & +32.920 & 1478121584853174272 & 21.15 & RGB & 60058.49 & 23.0 & $-8.7 \pm 1.6$ & $3.61 \pm 0.26$ & $-1.93 \pm 0.16$ & \multicolumn{1}{c}{\ldots} & \multicolumn{1}{c}{\ldots} \\
\hline
\texttt{BooV\_8} & 213.921 & +32.922 & 1478121413054482560 & 21.22 & RGB & 60107.29 & 19.1 & $-13.9 \pm 1.8$ & $1.58 \pm 0.21$ & $-2.97 \pm 0.22$ & \multicolumn{1}{c}{\ldots} & \multicolumn{1}{c}{\ldots} \\
\hline
\texttt{BooV\_9} & 213.911 & +32.909 & 1478121382991253888 & 21.43 & RGB & 60058.49 & 16.0 & $-11.0 \pm 2.0$ & $3.64 \pm 0.39$ & \multicolumn{1}{c}{\ldots} & \multicolumn{1}{c}{\ldots} & \multicolumn{1}{c}{\ldots} \\
 & & & & & & 60107.29 & 14.7 & $-9.6 \pm 2.2$ & $3.87 \pm 0.39$ & \multicolumn{1}{c}{\ldots} & \multicolumn{1}{c}{\ldots} & \multicolumn{1}{c}{\ldots} \\
 & & & & & & \textit{Combined} & \multicolumn{1}{c}{\ldots} & $-10.4 \pm 1.5$ & $3.75 \pm 0.28$ & $-1.81 \pm 0.17$ & 0.643 & \multicolumn{1}{c}{\ldots} \\
\hline
\texttt{BooV\_10} & 213.920 & +32.905 & \multicolumn{1}{c}{\ldots} & 21.97 & RGB & 60058.49 & 9.9 & $-9.6 \pm 8.0$ & \multicolumn{1}{c}{\ldots} & \multicolumn{1}{c}{\ldots} & \multicolumn{1}{c}{\ldots} & \multicolumn{1}{c}{\ldots} \\
 & & & & & & 60107.29 & 10.1 & $-9.3 \pm 3.0$ & $2.32 \pm 0.36$ & \multicolumn{1}{c}{\ldots} & \multicolumn{1}{c}{\ldots} & \multicolumn{1}{c}{\ldots} \\
 & & & & & & \textit{Combined} & \multicolumn{1}{c}{\ldots} & $-9.4 \pm 2.8$ & $2.32 \pm 0.36$ & $-2.40 \pm 0.24$ & 0.973 & \multicolumn{1}{c}{\ldots} \\
\hline
\texttt{BooV\_11} & 213.888 & +32.897 & \multicolumn{1}{c}{\ldots} & 22.25 & RGB & 60107.29 & 7.7 & $-9.2 \pm 4.6$ & $1.98 \pm 0.44$ & $-2.55 \pm 0.33$ & \multicolumn{1}{c}{\ldots} & \multicolumn{1}{c}{\ldots} \\
\hline
\texttt{BooV\_12} & 213.918 & +32.923 & \multicolumn{1}{c}{\ldots} & 22.75 & RGB & 60058.49 & 6.0 & $-13.9 \pm 6.5$ & \multicolumn{1}{c}{\ldots} & \multicolumn{1}{c}{\ldots} & \multicolumn{1}{c}{\ldots} & \multicolumn{1}{c}{\ldots} \\
 & & & & & & 60107.29 & 5.2 & $-3.2 \pm 6.9$ & \multicolumn{1}{c}{\ldots} & \multicolumn{1}{c}{\ldots} & \multicolumn{1}{c}{\ldots} & \multicolumn{1}{c}{\ldots} \\
 & & & & & & \textit{Combined} & \multicolumn{1}{c}{\ldots} & $-8.9 \pm 4.7$ & \multicolumn{1}{c}{\ldots} & \multicolumn{1}{c}{\ldots} & 0.259 & \multicolumn{1}{c}{\ldots} \\
\hline
\texttt{BooV\_13} & 213.934 & +32.950 & \multicolumn{1}{c}{\ldots} & 22.83 & RGB & 60058.49 & 5.9 & $-11.3 \pm 5.8$ & \multicolumn{1}{c}{\ldots} & \multicolumn{1}{c}{\ldots} & \multicolumn{1}{c}{\ldots} & \multicolumn{1}{c}{\ldots} \\
 & & & & & & 60107.29 & 4.6 & $-10.7 \pm 5.7$ & \multicolumn{1}{c}{\ldots} & \multicolumn{1}{c}{\ldots} & \multicolumn{1}{c}{\ldots} & \multicolumn{1}{c}{\ldots} \\
 & & & & & & \textit{Combined} & \multicolumn{1}{c}{\ldots} & $-11.0 \pm 4.1$ & \multicolumn{1}{c}{\ldots} & \multicolumn{1}{c}{\ldots} & 0.950 & \multicolumn{1}{c}{\ldots} \\
\hline
\multicolumn{1}{c}{\ldots} & 213.918 & +32.919 & \multicolumn{1}{c}{\ldots} & 22.25 & RGB & 60107.29 & 7.9 & $5.6 \pm 5.2$ & $2.40 \pm 0.49$ & $-2.31 \pm 0.30$ & \multicolumn{1}{c}{\ldots} & \begin{tabular}[t]{@{}r@{}}Member\\binary?\end{tabular} \\
\bottomrule
\end{tabular}
}%
\ifdim\wd\memboxBooV>\linewidth\resizebox{\linewidth}{!}{\usebox{\memboxBooV}}\else\usebox{\memboxBooV}\fi
\par\vspace{2pt}\parbox{\linewidth}{\raggedright $^{a}$ This star displays no evidence for radial velocity variations in our DEIMOS data; however, an independent DESI measurement suggests the possibility that this star is in a binary system (see \tabref{supplemental_Bootes_V}).}
\end{table}

%% file: AllResultsSeptember/AllTablesSeptember/T4_SupplementalTable_BootesV_September.tex
\begin{table}[htbp]
\centering
\caption{External Spectroscopic Measurements for Bo\"{o}tes V Stars}
\label{tab:supplemental_Bootes_V}
\scriptsize
\setlength{\tabcolsep}{4pt}

\newsavebox{\tblbox}
\sbox{\tblbox}{%
\begin{tabular}{lllrrl}
\hline\hline
\multicolumn{1}{c}{Name} & \multicolumn{1}{c}{Source/Survey} & \multicolumn{1}{c}{Identifier} & \multicolumn{1}{c}{MJD} & \multicolumn{1}{c}{$v_{\rm hel}$ (\kms)} & \multicolumn{1}{c}{[Fe/H]} \\
\midrule
\texttt{BooV\_1} &  \citet{2023AJ....166...76S} & \ldots  &  59798.76 & $5.1 \pm 13.4$ &  $-2.85 \pm 0.1$\\
\texttt{BooV\_1} &  LAMOST DR5 & HD141746N331518M01 &  56797.93 & $-5.5 \pm 22.7$ &  $-2.25 \pm 0.6$\\
\texttt{BooV\_1} &  DESI DR1 & 2305843020223749931 &  59696.30 & $-0.2 \pm 3.2$ &  $-2.50 \pm 0.19$\\
\hline
\texttt{BooV\_2} & DESI DR1 & 39632941176848589 & 59303.90 &  $-7.3 \pm 5.4$ & BHB$^{a}$  \\
 & \ldots & \ldots &  59698.30 & $-18.1 \pm 22.8$ &  BHB$^{a}$ \\
\hline
\texttt{BooV\_4} &  DESI DR1 & 39632941172657712 &  59679.36 & $-5.0 \pm 15.9$ & BHB$^{a}$ \\
\bottomrule
\end{tabular}%
}
\usebox{\tblbox}
\par\vspace{4pt}
\begin{minipage}{\wd\tblbox}
\raggedright\footnotesize
$^{a}$While DESI reports metallicities for two BHB stars, we omit them due to reliability concerns given the low signal-to-noise.
\end{minipage}
\end{table}

%% file: AllResultsSeptember/AllTablesSeptember/T5_MembersTable_LeoMinorI_September.tex
\begin{table}[htbp]
\centering
\caption{Spectroscopic Members of Leo Minor I}
\label{tab:members_Leo_Minor_I}
\scriptsize
\setlength{\tabcolsep}{2pt}

\newsavebox{\memboxLMiI}\savebox{\memboxLMiI}{%
\begin{tabular}{lrrrrrrrrrrrr}
\hline\hline
\multicolumn{1}{c}{Name} & \multicolumn{1}{c}{RA} & \multicolumn{1}{c}{DEC} & \multicolumn{1}{c}{Gaia Source ID} & \multicolumn{1}{c}{$g_0$} & \multicolumn{1}{c}{Type} & \multicolumn{1}{c}{MJD} & \multicolumn{1}{c}{S/N} & \multicolumn{1}{c}{$v_{\rm hel}$} & \multicolumn{1}{c}{$\sum \rm EW_{\rm CaT}$} & \multicolumn{1}{c}{[Fe/H]} & \multicolumn{1}{c}{$p_{\rm const}$} & \multicolumn{1}{c}{Comment} \\
\multicolumn{1}{c}{} & \multicolumn{1}{c}{(deg)} & \multicolumn{1}{c}{(deg)} & \multicolumn{1}{c}{} & \multicolumn{1}{c}{(mag)} & \multicolumn{1}{c}{} & \multicolumn{1}{c}{(d)} & \multicolumn{1}{c}{(pixel$^{-1}$)} & \multicolumn{1}{c}{(\kms)} & \multicolumn{1}{c}{(\AA)} & \multicolumn{1}{c}{} & \multicolumn{1}{c}{} & \multicolumn{1}{c}{} \\
\midrule
\texttt{LMiI\_1} & 164.272 & +28.855 & 731478564536994432 & 20.27 & RGB & 59735.53 & 30.6 & $5.7 \pm 1.3$ & $5.24 \pm 0.24$ & \multicolumn{1}{c}{\ldots} & \multicolumn{1}{c}{\ldots} & \multicolumn{1}{c}{\ldots} \\
 & & & & & & 60058.37 & 30.6 & $4.2 \pm 1.3$ & $4.42 \pm 0.23$ & \multicolumn{1}{c}{\ldots} & \multicolumn{1}{c}{\ldots} & \multicolumn{1}{c}{\ldots} \\
 & & & & & & \textit{Combined} & \multicolumn{1}{c}{\ldots} & $5.0 \pm 0.9$ & $4.81 \pm 0.16$ & $-1.52 \pm 0.14$ & 0.393 & \multicolumn{1}{c}{\ldots} \\
\hline
\texttt{LMiI\_2} & 164.285 & +28.894 & 732979913304939776 & 20.84 & RGB & 59735.53 & 19.9 & $0.9 \pm 1.6$ & $3.01 \pm 0.36$ & $-2.14 \pm 0.21$ & \multicolumn{1}{c}{\ldots} & \multicolumn{1}{c}{\ldots} \\
\hline
\texttt{LMiI\_3} & 164.274 & +28.877 & 731478663321185152 & 21.20 & RGB & 59735.53 & 15.0 & $2.7 \pm 2.2$ & $1.93 \pm 0.27$ & \multicolumn{1}{c}{\ldots} & \multicolumn{1}{c}{\ldots} & \multicolumn{1}{c}{\ldots} \\
 & & & & & & 60058.37 & 14.6 & $14.6 \pm 4.1$ & \multicolumn{1}{c}{\ldots} & \multicolumn{1}{c}{\ldots} & \multicolumn{1}{c}{\ldots} & \multicolumn{1}{c}{\ldots} \\
 & & & & & & \textit{Combined} & \multicolumn{1}{c}{\ldots} & $5.3 \pm 1.9$ & $1.93 \pm 0.27$ & $-2.65 \pm 0.22$ & 0.011 & Binary \\
\hline
\texttt{LMiI\_4} & 164.256 & +28.886 & \multicolumn{1}{c}{\ldots} & 21.35 & RGB & 59735.53 & 12.6 & $2.0 \pm 2.5$ & $1.87 \pm 0.34$ & \multicolumn{1}{c}{\ldots} & \multicolumn{1}{c}{\ldots} & \multicolumn{1}{c}{\ldots} \\
 & & & & & & 60058.37 & 12.3 & $-11.6 \pm 3.0$ & $1.66 \pm 0.34$ & \multicolumn{1}{c}{\ldots} & \multicolumn{1}{c}{\ldots} & \multicolumn{1}{c}{\ldots} \\
 & & & & & & \textit{Combined} & \multicolumn{1}{c}{\ldots} & $-3.7 \pm 1.9$ & $1.76 \pm 0.24$ & $-2.74 \pm 0.22$ & 0.001 & Binary \\
\hline
\texttt{LMiI\_5} & 164.266 & +28.862 & \multicolumn{1}{c}{\ldots} & 21.72 & RGB & 59735.53 & 9.3 & $10.0 \pm 2.9$ & $3.71 \pm 0.57$ & \multicolumn{1}{c}{\ldots} & \multicolumn{1}{c}{\ldots} & \multicolumn{1}{c}{\ldots} \\
 & & & & & & 60058.37 & 9.2 & $3.2 \pm 3.1$ & $3.31 \pm 0.49$ & \multicolumn{1}{c}{\ldots} & \multicolumn{1}{c}{\ldots} & \multicolumn{1}{c}{\ldots} \\
 & & & & & & \textit{Combined} & \multicolumn{1}{c}{\ldots} & $6.9 \pm 2.1$ & $3.48 \pm 0.37$ & $-1.79 \pm 0.22$ & 0.112 & \multicolumn{1}{c}{\ldots} \\
\hline
\texttt{LMiI\_6} & 164.266 & +28.876 & \multicolumn{1}{c}{\ldots} & 22.31 & RGB & 59735.53 & 5.8 & $-1.4 \pm 4.5$ & \multicolumn{1}{c}{\ldots} & \multicolumn{1}{c}{\ldots} & \multicolumn{1}{c}{\ldots} & \multicolumn{1}{c}{\ldots} \\
 & & & & & & 60058.37 & 5.5 & $11.3 \pm 5.3$ & \multicolumn{1}{c}{\ldots} & \multicolumn{1}{c}{\ldots} & \multicolumn{1}{c}{\ldots} & \multicolumn{1}{c}{\ldots} \\
 & & & & & & \textit{Combined} & \multicolumn{1}{c}{\ldots} & $3.9 \pm 3.4$ & \multicolumn{1}{c}{\ldots} & \multicolumn{1}{c}{\ldots} & 0.065 & Binary? \\
\hline
\texttt{LMiI\_7} & 164.270 & +28.874 & \multicolumn{1}{c}{\ldots} & 22.63 & RGB & 60058.37 & 4.1 & $10.1 \pm 16.1$ & \multicolumn{1}{c}{\ldots} & \multicolumn{1}{c}{\ldots} & \multicolumn{1}{c}{\ldots} & \multicolumn{1}{c}{\ldots} \\
\bottomrule
\end{tabular}
}%
\ifdim\wd\memboxLMiI>\linewidth\resizebox{\linewidth}{!}{\usebox{\memboxLMiI}}\else\usebox{\memboxLMiI}\fi
\end{table}

%% file: AllResultsSeptember/AllTablesSeptember/T6_MembersTable_DELVE2_September.tex
\begin{table}[htbp]
\centering
\caption{Spectroscopic Members of DELVE 2}
\label{tab:members_DELVE_2}
\scriptsize
\setlength{\tabcolsep}{4pt}

\newsavebox{\memboxDELTwo}\savebox{\memboxDELTwo}{%
\begin{tabular}{lrrrrrrrrrrrr}
\hline\hline
\multicolumn{1}{c}{Name} & \multicolumn{1}{c}{RA} & \multicolumn{1}{c}{DEC} & \multicolumn{1}{c}{Gaia Source ID} & \multicolumn{1}{c}{$g_0$} & \multicolumn{1}{c}{Type} & \multicolumn{1}{c}{MJD} & \multicolumn{1}{c}{S/N} & \multicolumn{1}{c}{$v_{\rm hel}$} & \multicolumn{1}{c}{$\sum \rm EW_{\rm CaT}$} & \multicolumn{1}{c}{[Fe/H]} & \multicolumn{1}{c}{$p_{\rm const}$} & \multicolumn{1}{c}{Comment} \\
\multicolumn{1}{c}{} & \multicolumn{1}{c}{(deg)} & \multicolumn{1}{c}{(deg)} & \multicolumn{1}{c}{} & \multicolumn{1}{c}{(mag)} & \multicolumn{1}{c}{} & \multicolumn{1}{c}{(d)} & \multicolumn{1}{c}{(pixel$^{-1}$)} & \multicolumn{1}{c}{(\kms)} & \multicolumn{1}{c}{(\AA)} & \multicolumn{1}{c}{} & \multicolumn{1}{c}{} & \multicolumn{1}{c}{} \\
\midrule
\texttt{DEL2\_1} & 28.790 & $-$68.245 & 4697408171177907968 & 18.60 & RGB & 59407.53 & 28.4 & $131.3 \pm 1.1$ & $3.21 \pm 0.24$\phantom{$^{a}$} & \multicolumn{1}{c}{\ldots} & \multicolumn{1}{c}{\ldots} & \multicolumn{1}{c}{\ldots} \\
 & & & & & & 59497.13 & 13.8 & $129.8 \pm 1.3$ & $2.68 \pm 0.52$\phantom{$^{a}$} & \multicolumn{1}{c}{\ldots} & \multicolumn{1}{c}{\ldots} & \multicolumn{1}{c}{\ldots} \\
 & & & & & & 59883.06 & 36.2 & $131.5 \pm 1.0$ & $2.75 \pm 0.26$\phantom{$^{a}$} & \multicolumn{1}{c}{\ldots} & \multicolumn{1}{c}{\ldots} & \multicolumn{1}{c}{\ldots} \\
 & & & & & & 60222.13 & 18.9 & $131.5 \pm 1.1$ & $3.05 \pm 0.28$\phantom{$^{a}$} & \multicolumn{1}{c}{\ldots} & \multicolumn{1}{c}{\ldots} & \multicolumn{1}{c}{\ldots} \\
 & & & & & & 60656.16 & 51.0 & $131.9 \pm 1.0$ & $2.80 \pm 0.22$\phantom{$^{a}$} & \multicolumn{1}{c}{\ldots} & \multicolumn{1}{c}{\ldots} & \multicolumn{1}{c}{\ldots} \\
 & & & & & & \textit{Combined} & \multicolumn{1}{c}{\ldots} & $131.3 \pm 0.5$ & $2.93 \pm 0.12$\phantom{$^{a}$} & $-2.49 \pm 0.09$ & 0.777 & \multicolumn{1}{c}{\ldots} \\
\hline
\texttt{DEL2\_2} & 28.807 & $-$68.246 & 4697408171177907712 & 19.65 & BHB & 59407.53 & 3.0 & $133.5 \pm 5.4$ & \multicolumn{1}{c}{\ldots} & \multicolumn{1}{c}{\ldots} & \multicolumn{1}{c}{\ldots} & \multicolumn{1}{c}{\ldots} \\
 & & & & & & 59883.06 & 4.8 & $139.8 \pm 4.5$ & \multicolumn{1}{c}{\ldots} & \multicolumn{1}{c}{\ldots} & \multicolumn{1}{c}{\ldots} & \multicolumn{1}{c}{\ldots} \\
 & & & & & & 60656.16 & 8.2 & $127.4 \pm 3.5$ & \multicolumn{1}{c}{\ldots} & \multicolumn{1}{c}{\ldots} & \multicolumn{1}{c}{\ldots} & \multicolumn{1}{c}{\ldots} \\
 & & & & & & \textit{Combined} & \multicolumn{1}{c}{\ldots} & $132.3 \pm 2.4$ & \multicolumn{1}{c}{\ldots} & \multicolumn{1}{c}{\ldots} & 0.089 & Binary? \\
\hline
\texttt{DEL2\_3} & 28.786 & $-$68.233 & 4697411126115411328 & 19.69 & BHB & 60656.16 & 7.3 & $132.0 \pm 3.1$ & \multicolumn{1}{c}{\ldots} & \multicolumn{1}{c}{\ldots} & \multicolumn{1}{c}{\ldots} & \multicolumn{1}{c}{\ldots} \\
\hline
\texttt{DEL2\_4} & 28.746 & $-$68.249 & 4697411091755668608 & 20.46 & RGB & 59407.53 & 4.1 & $128.1 \pm 2.6$ & \multicolumn{1}{c}{\ldots} & \multicolumn{1}{c}{\ldots} & \multicolumn{1}{c}{\ldots} & \multicolumn{1}{c}{\ldots} \\
 & & & & & & 59883.06 & 6.2 & $131.5 \pm 2.2$ & $1.39 \pm 0.41$\phantom{$^{a}$} & \multicolumn{1}{c}{\ldots} & \multicolumn{1}{c}{\ldots} & \multicolumn{1}{c}{\ldots} \\
 & & & & & & 60656.16 & 10.8 & $125.5 \pm 1.5$ & $1.39 \pm 0.26$$^{a}$ & \multicolumn{1}{c}{\ldots} & \multicolumn{1}{c}{\ldots} & \multicolumn{1}{c}{\ldots} \\
 & & & & & & \textit{Combined} & \multicolumn{1}{c}{\ldots} & $127.5 \pm 1.1$ & $1.39 \pm 0.22$\phantom{$^{a}$} & $-3.15 \pm 0.25$ & 0.069 & Binary? \\
\hline
\texttt{DEL2\_5} & 28.796 & $-$68.224 & 4697411233489609856 & 21.08 & RGB & 59883.06 & 3.6 & $128.5 \pm 3.7$ & \multicolumn{1}{c}{\ldots} & \multicolumn{1}{c}{\ldots} & \multicolumn{1}{c}{\ldots} & \multicolumn{1}{c}{\ldots} \\
 & & & & & & 60656.16 & 6.1 & $129.5 \pm 2.4$ & $1.06 \pm 0.40$\phantom{$^{a}$} & \multicolumn{1}{c}{\ldots} & \multicolumn{1}{c}{\ldots} & \multicolumn{1}{c}{\ldots} \\
 & & & & & & \textit{Combined} & \multicolumn{1}{c}{\ldots} & $129.2 \pm 2.0$ & $1.06 \pm 0.40$\phantom{$^{a}$} & $-3.45 \pm 0.64$ & 0.820 & \multicolumn{1}{c}{\ldots} \\
\hline
\texttt{DEL2\_6} & 28.733 & $-$68.260 & 4697411091755664384 & 21.24 & RGB & 60656.16 & 4.7 & $127.6 \pm 2.0$ & \multicolumn{1}{c}{\ldots} & \multicolumn{1}{c}{\ldots} & \multicolumn{1}{c}{\ldots} & \multicolumn{1}{c}{\ldots} \\
\hline
\texttt{DEL2\_7} & 28.802 & $-$68.263 & \multicolumn{1}{c}{\ldots} & 21.76 & RGB & 60656.16 & 3.2 & $129.9 \pm 14.2$ & \multicolumn{1}{c}{\ldots} & \multicolumn{1}{c}{\ldots} & \multicolumn{1}{c}{\ldots} & \multicolumn{1}{c}{\ldots} \\
\bottomrule
\end{tabular}
}%
\ifdim\wd\memboxDELTwo>\linewidth\resizebox{\linewidth}{!}{\usebox{\memboxDELTwo}}\else\usebox{\memboxDELTwo}\fi
\par\vspace{2pt}\parbox{\linewidth}{\raggedright $^{a}$ We fit the EW of the third CaT line (8662~\AA) of this star in the December 2024 epoch with a Gaussian-only profile instead of the Gaussian-plus-Lorentzian profile used for all other IMACS CaT lines.}
\end{table}